\documentclass[longauth]{aa} 
\usepackage{graphicx}
\usepackage{txfonts}
\usepackage[dvipsnames]{xcolor}
\usepackage[breaklinks, colorlinks, citecolor=blue]{hyperref}
\newcommand{\SII}{[S~{\sc ii}]}
\newcommand{\OIII}{[O~{\sc iii}]}
\newcommand{\NII}{[N~{\sc ii}]}
\newcommand{\HII}{H~{\sc ii}}

\newcommand{\HI}{H~{\sc i}}
\newcommand{\Ha}{H$\alpha$}
\newcommand{\Hb}{H$\beta$}
\newcommand{\kms}{\,\mbox{km}\,\mbox{s}^{-1}}
\newcommand{\ergs}{\,\mbox{erg}\,\mbox{s}^{-1}}
\newcommand{\ergscm}{\,\mbox{erg}\,\mbox{s}^{-1}\,\mbox{cm}^{-2}}
\newcommand{\ergskpc}{\,\mbox{erg}\,\mbox{s}^{-1}\,\mbox{kpc}^{-2}}
\newcommand{\HST}{\textit{HST}}
\newcommand{\SIIHa}{[S~{\sc ii}]/H$\alpha$}

\newcommand{\OIIIHb}{[O~{\sc iii}]/H$\beta$}
\newcommand{\IS}{$I - \sigma$}
\newcommand{\SBDIG}{$\mathrm{\Sigma(H\alpha)_{DIG}}$}
\newcommand{\sigmaHa}{$\sigma$(H$\alpha$)}
\newcommand{\nregs}{1484}
\newcommand{\nregsnohst}{188}

\newcommand{\nregsnooldcls}{364}
\newcommand{\nregshaveyoungcls}{264}
\newcommand{\nshells}{171}
\newcommand{\ncentralpeak}{572}
\newcommand{\nregswithcls}{511}
\newcommand{\be}{\begin{equation}}
\newcommand{\ee}{\end{equation}}

\def\revone{}
\def\mathrevone{}
\begin{document}

   \title{Quantifying the energy balance between the turbulent ionised gas and young stars}


    
   \author{Oleg~V.~Egorov
          \inst{1}\fnmsep\thanks{\email{oleg.egorov@uni-heidelberg.de}}
          \and
          Kathryn Kreckel\inst{1}
          \and Simon~C.~O.~Glover\inst{2}
          \and Brent Groves\inst{3}
          \and
          Francesco Belfiore\inst{4} 
          \and
          Eric Emsellem\inst{5,6} 
          \and
          Ralf~S.~Klessen\inst{2,7}
          \and
          Adam K. Leroy\inst{8,9}
          \and
          Sharon E. Meidt\inst{10}
          \and
          Sumit K. Sarbadhicary\inst{8,9}
          \and
          Eva Schinnerer\inst{11}
          \and
          Elizabeth J. Watkins\inst{1}
          \and
          Brad C. Whitmore\inst{12}
          \and
          Ashley T. Barnes\inst{5,13}
          \and
          Enrico Congiu\inst{14}
          \and
          Daniel~A.~Dale\inst{15}
          \and
          Kathryn Grasha\inst{16}
          \and
          Kirsten L. Larson\inst{17}
          \and
          Janice C. Lee\inst{18,19}
          \and
          J. Eduardo Méndez-Delgado\inst{1}
          \and
          David A. Thilker\inst{20}
          \and
          Thomas G. Williams\inst{21}
          }

   \institute{Astronomisches Rechen-Institut, Zentrum f\"{u}r Astronomie der Universit\"{a}t Heidelberg, M\"{o}nchhofstr. 12-14, D-69120 Heidelberg, Germany
   \and
   Instit\"ut  f\"{u}r Theoretische Astrophysik, Zentrum f\"{u}r Astronomie der Universit\"{a}t Heidelberg, Albert-Ueberle-Strasse 2, 69120 Heidelberg, Germany
   \and 
   International Centre for Radio Astronomy Research, University of Western Australia, 7 Fairway, Crawley, 6009 WA, Australia
   \and
   INAF -- Osservatorio Astrofisico di Arcetri, Largo E. Fermi 5, I-50157 Firenze, Italy 
   \and
   European Southern Observatory, Karl-Schwarzschild Stra{\ss}e 2, D-85748 Garching bei M\"{u}nchen, Germany 
   \and
    Univ Lyon, Univ Lyon1, ENS de Lyon, CNRS, Centre de Recherche Astrophysique de Lyon UMR5574, F-69230 Saint-Genis-Laval France
    \and
     Interdisziplin\"{a}res Zentrum f\"{u}r Wissenschaftliches Rechnen der Universit\"{a}t Heidelberg, Im Neuenheimer Feld 205, D-69120 Heidelberg, Germany
     \and 
   Department of Astronomy, The Ohio State University, 140 West 18th Avenue, Columbus, Ohio 43210, USA
    \and
    Center for Cosmology and Astroparticle Physics, 191 West Woodruff Avenue, Columbus, OH 43210, USA
    \and 
    Sterrenkundig Observatorium, Universiteit Gent, Krijgslaan 281 S9, B-9000 Gent, Belgium
    \and
    Max-Planck-Institut f\"{u}r Astronomie, K\"{o}nigstuhl 17, D-69117, Heidelberg, Germany 
    \and
    Space Telescope Science Institute, 3700 San Martin Drive, Baltimore, MD, 21218, USA
    \and
    Argelander-Institut für Astronomie, Universität Bonn, Auf dem Hügel 71, 53121 Bonn, Germany
    \and
    European Southern Observatory (ESO), Alonso de C\'ordova 3107, Casilla 19, Santiago 19001, Chile
    \and
    Department of Physics \& Astronomy, University of Wyoming, Laramie, WY 82071, USA
    \and
    Research School of Astronomy and Astrophysics, Australian National University, Canberra, ACT 2611, Australia
    \and
    AURA for the European Space Agency (ESA), Space Telescope Science Institute, 3700 San Martin Drive, Baltimore, MD 21218, USA
    \and
    Gemini Observatory/NSF’s NOIRLab, 950 N. Cherry Avenue, Tucson, AZ, USA
        \and
    Steward Observatory, University of Arizona, 933 N Cherry Ave, Tucson, AZ 85721, USA
    \and
    Department of Physics \& Astronomy, Bloomberg Center for Physics and Astronomy, Johns Hopkins University, 3400 N. Charles Street, Baltimore, MD 21218
    \and
    Sub-department of Astrophysics, Department of Physics, University of Oxford, Keble Road, Oxford OX1 3RH, UK
             }

   \date{Received ????? ??, 2023; accepted ????? ??, ????}

 
  \abstract
   {Stellar feedback is a key contributor to 
   the morphology and dynamics of the interstellar medium in star-forming galaxies. In particular, energy and momentum input from massive stars can drive the turbulent motions in the gas, but the dominance and efficiency of this process are unclear. The study of ionised superbubbles enables quantitative constraints to be placed on the energetics of stellar feedback.}
   {We directly compare the kinetic energy of 
   expanding superbubbles and the turbulent motions in the interstellar medium with the mechanical energy deposited by massive stars in the form of winds and supernovae. With such a comparison we aim to answer whether the stellar feedback is responsible for 
   the observed turbulent motions and to quantify the fraction of mechanical energy retained in the superbubbles.}
   {We investigate the ionised gas morphology, excitation properties, and kinematics in 19 nearby star-forming galaxies from the PHANGS-MUSE survey. Based on the distribution of the flux and velocity dispersion in the \Ha\ line, we select \nregs\ regions of locally elevated velocity dispersion (\sigmaHa$>45\kms$), including at least \nshells\ expanding superbubbles. We analyse these regions and relate their properties to those of the young stellar associations and star clusters identified in PHANGS-HST data.}
   {We find a good correlation between the kinetic energy of the ionised gas and the total mechanical energy input from supernovae and stellar winds from the stellar associations. At the same time, the contribution of mechanical energy injected by the supernovae alone is not sufficient to explain the measured kinetic energy of the ionised gas, which implies that pre-supernova feedback in the form of radiation/thermal pressure and winds is necessary. We find that the gas kinetic energy decreases with metallicity for our sample covering $Z=0.5-1.0 \, Z_\odot$, reflecting the lower impact of stellar feedback. For the sample of \revone{well-resolved} superbubbles, we find that about 40\% of the young stellar associations are preferentially located in their rims. 
   We also find a slightly higher (by $\sim 15$\%) fraction of the youngest (\revone{< 3}~Myr) stellar associations in the rims of the superbubbles than in the centres, and the opposite trend for older associations, which implies possible propagation or triggering of star formation.}
   {Stellar feedback is the dominant source for powering the ionised gas in regions of locally (on 50--500~pc scale) elevated velocity dispersion, with a typical \revone{coupling} efficiency of $10-20$\%. Accounting for pre-supernovae feedback is required to set up the energy balance between gas and stars. }


   \maketitle
%

\section{Introduction}
\label{sec:intro}

Massive stars ($M>8~{\rm M}_\odot$) inject a large amount of energy and momentum into the interstellar medium (ISM) through several channels -- ionising radiation, stellar winds, and supernova explosions \cite[see, e.g.,][]{MacLow2004, Krumholz2014, Klessen2016, Girichidis2020}. This stellar output leads to the formation of expanding bubbles and superbubbles in the surrounding ISM with sizes from a few parsecs to 1--2~kpc \citep[e.g.][]{Oey1997, Nath2020, Watkins2022jwst}. These bubbles are seen in many star-forming galaxies, including our own, and have been observed in multiple ISM phases: atomic hydrogen, as traced by the \HI~21~cm line \citep[e.g.][]{Ehlerova2005, Bagetakos2011, Pokhrel2020}, ionised gas \cite[e.g.][]{ValdezGutierrez2001, Lozinskaya2003, AmbrocioCruz2016, Egorov2014, Egorov2017, Egorov2018, Gerasimov2022, Barnes2022}, warm dust \cite[e.g.][]{Churchwell2006, Watkins2022jwst, Barnes2022jwst}, and even in molecular gas \citep{Xu2020, Watkins2023}. Observationally, expanding superbubbles can be identified by their shell-like morphology and their emission-line profiles showing multiple components, corresponding to the approaching and receding sides of the bubble. When observed at low spectral and spatial resolution, the shell morphology can be lost, and the line profile broadened due to the overlapping of the individual unresolved components \citep[e.g.][]{TT1996, SP2021}. In such circumstances, the superbubbles can be observed as regions of locally elevated line-of-sight velocity dispersion (including individual supernovae remnants, see e.g. \citealt{Vasiliev2015}).

Outside of the expansion of the superbubbles, the ISM itself has broad kinematics, likely turbulent in nature, though this is not easy to quantify from observations \citep[see][for reviews]{Elmegreen2004, Burkhart2021}. The turbulent motions in nearby galaxies usually manifest themselves in the supersonic velocity dispersion of different tracers \citep[e.g.][]{Melnick1999, Green2014, MoiseevKlypin2015, Cosens2022, Law2022}. It is still not clear what mechanisms are responsible for the observed supersonic velocity dispersion of the ISM in galaxies, although stellar feedback and the gravitational potential of the galaxy are usually considered the main probable drivers. On the one hand, the mean velocity dispersion in the discs of the galaxies correlates with the total star formation rate (SFR) and its surface density ($\Sigma_\mathrm{SFR}$) \citep[see, e.g.,][]{Yu2019, Law2022}. It has also been shown for several samples of star-forming galaxies that at the current level of SFR, supernovae can produce sufficient energy to drive the turbulent motions in the atomic gas \revone{\citep{Dib2006, Tamburro2009, Bacchini2019, Bacchini2020}}. At high SFR, supernovae produce even more energy and momentum than is required to explain the observed velocity dispersion of \HI\ \citep{Utomo2019, Sarbadhicary2022}, with the excess of the energy likely deposited in the ionised gas or carried away by superbubble outflows. \revone{\citet{Girard2021} showed that supernova feedback alone can explain the observed velocity dispersion of molecular gas in their sample of galaxies with high specific SFR, although it is not sufficient to explain ionised gas kinematics there.}

In some cases, however, supernovae alone \revone{do not explain} the observed mean gas velocity dispersion, for example, in several gas-rich galaxies (e.g. \citealt{Fisher2019}  \revone{for ionised gas}, \citealt{Hunter2021, Elmegreen2022} \revone{for \HI}) and in the outer regions of some galaxies \citep{Tamburro2009, Koch2018}. In their analysis of four dwarf galaxies, \cite{Hunter2022} found that the \HI\  velocity dispersion on scales of $\sim 400$~pc correlates with the SFR 100--200~Myr ago, but no correlation was present between the previous star formation activity and the \Ha\ velocity dispersion. \cite{Colman2022} concluded that stellar feedback alone is not enough to explain the ISM structure in the LMC on scales larger than 60~pc -- if the ISM structure is generated by turbulence, then another large-scale driving mechanism is needed. 
\cite{Krumholz2016} tested models of feedback and gravity-driven turbulence and concluded that the gravity-driven model better explains the underlying observational measurements of the velocity dispersion and SFR, at least for the rapidly star-forming and high-velocity dispersion galaxies. 

Several theoretical models and simulations suggest that both stellar feedback and gravitational instabilities are responsible for the turbulence of the ISM, but at different scales. According to \cite{Nusser2022}, turbulence is driven mostly by stellar feedback in normal star-forming disc galaxies, while gravitational instabilities become important at high gas surface densities ($\Sigma_\mathrm{g} \gtrsim 50 \, {\rm M}_\odot \mathrm{pc}^{-2}$).  
\cite{Ejdetjarn2022} find that a difference in the velocity dispersion in warm (ionised) and cold (atomic, molecular) gas can be explained if stellar feedback is responsible for the high mean velocity dispersion in H$\alpha$, while the gravitational potential is a probable driver of that for cold gas. Gas accretion is also a possible driver of the turbulence in galaxies \revone{\citep[e.g.][]{Klessen2010, Forbes2022, Ginzburg2022, Sharda2023}}. In fact, an increase of the gas velocity dispersion and the non-circular motions, caused by the recent accretion of external gas from the intergalactic medium or nearby companions, has been observed at small ($<1$~kpc) scales \cite[e.g.][]{Silchenko2019, Egorova2019, Pilyugin2021}. 

In modern cosmological simulations, stellar feedback is often introduced as a sub-grid model \cite[e.g.,][]{Hopkins2011, Nelson2015, Pillepich2018}, and a proper understanding of what fraction of the energy injected by massive stars into the ISM is radiated away and what fraction is retained in the ISM in the form of kinetic energy is important for calibrating these models.
Different models predict different values of this mechanical energy \revone{coupling} efficiency -- from $\eta = E_\mathrm{kin}^\mathrm{(gas)}/E_\mathrm{mech}^\mathrm{(stars)}  \sim 1$\% to $\sim40$\% \citep[e.g.][]{Krause2014, Sharma2014, Yadav2017, Lancaster2021}, where 100\% means that all mechanical energy injected by stellar sources is converting to kinetic and turbulent energy of the gas. In particular, \cite{Sharma2014} demonstrated that about 40\% of the energy can be retained in superbubbles during a few tens of Myr, while single supernovae lose almost all their energy in $\sim 1$~Myr. The particular value of $\eta$ inferred from the models changes with the ISM gas density, age of the cluster, and number of OB stars, but also with the resolution of simulations \citep{Yadav2017}. It is therefore especially important to quantify this parameter from observations.

The main goal of the present paper is to establish the connection between the small-scale ($\sim 30 - 500$~pc) supersonic motions of the ionised gas (in the form of turbulence or expanding superbubbles) and the stellar population for a sample of nearby ($D<20$~Mpc) star-forming spiral galaxies. We are investigating 19 galaxies from the PHANGS survey\footnote{\url{http://phangs.org}} for which both integral-field spectroscopy (IFS) with Multi Unit Spectroscopic Explorer (MUSE) at the Very Large Telescope (VLT; \citealt{Emsellem2022}) and multi-band imaging with  Hubble Space Telescope (HST; \citealt{Lee2022}) are available. With this data set in hand, we can simultaneously analyse the properties of the ionised gas and the young stellar associations, and link their properties to each other. In particular, after identifying the regions dominated by non-circular supersonic motions of the ionised gas via their elevated velocity dispersion, we can compare their kinetic energy with the energy input produced by massive stars to judge the importance of stellar feedback in powering the regions of high velocity dispersion and estimate the efficiency of the mechanical stellar feedback.

The paper is organised as follows. Section~\ref{sec:obs} briefly describes the PHANGS-MUSE and PHANGS-HST observations and their data reduction. Section~\ref{sec:data_analysis} presents the data analysis, including the procedure for identifying the regions of locally increased velocity dispersion in the ionised gas, isolating the young stellar association linked to them, and measuring the parameters used in the present paper. In Section~\ref{sec:statistics} we consider the derived properties of the selected regions of high velocity dispersion. In Section~\ref{sec:discussion} we discuss the results, and Section~\ref{sec:summary} summarises our main findings.

\section{Observational data}    
\label{sec:obs}
\subsection{PHANGS-MUSE}
\label{sec:obs:muse}

\begin{figure*}
    \centering
    \includegraphics[width=\linewidth]{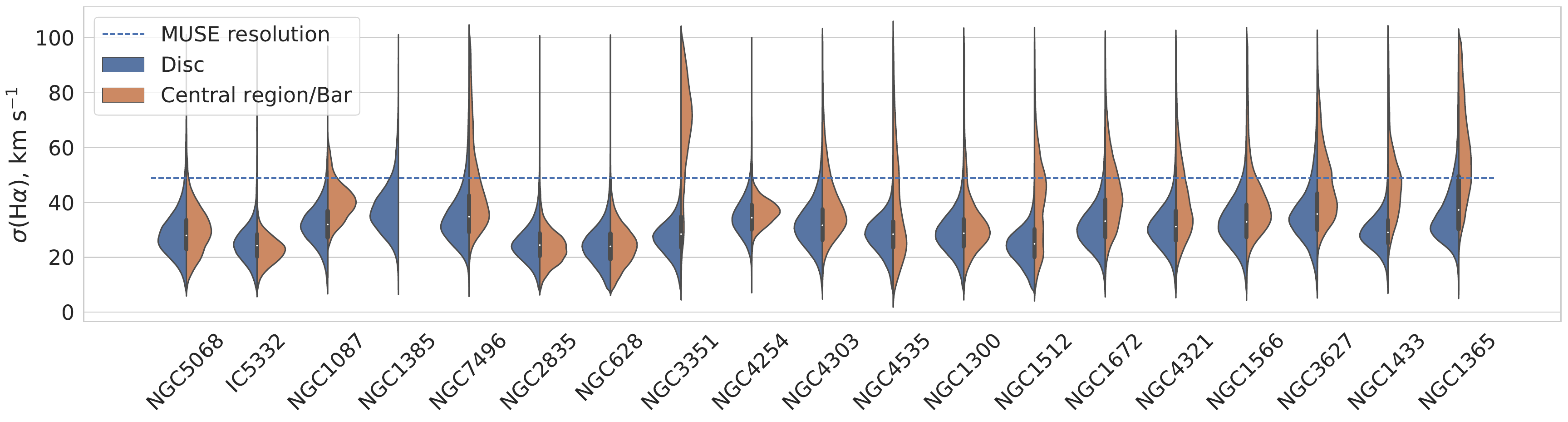}
    \caption{Statistics of the intrinsic \Ha\ velocity dispersion in PHANGS-MUSE galaxies (sorted by their stellar mass, see Table~\ref{tab:sample}) for different environments (blue colour is for the disc, orange is for the central regions and bars, according to the classification from \citealt{Querejeta2021}). Pixels with $S/N < 30$ in \Ha\ are excluded. The blue dashed line corresponds to the mean MUSE spectral resolution (velocity dispersion) at the wavelength of \Ha. The values of \sigmaHa\ are corrected for instrumental broadening by subtracting in quadrature the MUSE instrumental velocity dispersion for each galaxy.}
    \label{fig:sigma_stat}
\end{figure*}

This paper analyses IFS data from the MUSE/VLT \citep{Bacon2010} instrument as part of the PHANGS-MUSE survey (PI: Schinnerer). For details of the survey strategy, log of observations, and description of the data reduction process, we refer the reader to \cite{Emsellem2022}, while here we briefly describe those aspects important for the present analysis.

\subsubsection{Description of the data cubes and their analysis}

The important advantage of MUSE is the simultaneous combination of high angular resolution (PSF is given in Table~\ref{tab:sample} for each galaxy), small pixel size (0\farcs2), and large field of view ($1\arcmin \times 1\arcmin$). PHANGS-MUSE data cubes represent the mosaics of 3 to 15 MUSE fields for 19 nearby star-forming galaxies and cover a large part of their discs (extending to at least half of their optical radii, $R_{25}$). In the present paper, we use the publicly available\footnote{\url{https://archive.eso.org/scienceportal/home?data_collection=PHANGS}} data cubes and maps (`copt' dataset with homogenised PSFs) with the distribution of properties of the ionised gas. The maps were obtained with a data analysis pipeline (\textsc{dap}\footnote{\url{https://gitlab.com/francbelf/ifu-pipeline}}; based on the \textsc{gist}\footnote{\url{https://abittner.gitlab.io/thegistpipeline}} code by \citealt{Bittner2019}), which we also apply further to analyse the extracted spectra of the regions of interest. We derive the reddening-corrected fluxes (based on the observed Balmer decrement and a \cite{Cardelli1989} extinction curve), line-of-sight velocity, and velocity dispersion of the emission lines by fitting a single-component Gaussian model to the stellar population subtracted spectra. Details of the data reduction, mosaicking, and \textsc{dap} are given in \cite{Emsellem2022}. The typical spatial resolution of the PHANGS-MUSE data is 50 pc, and the 3$\sigma$ flux sensitivity in \Ha\ is \revone{$0.7-1.7 \times 10^{-17} \ergscm\ \mathrm{arcsec^{-2}}$ ($4-7 \times 10^{37} \ergskpc$)}.

In the initial steps of the present analysis, we used the spatial distribution of the \Ha\ flux, $F(\mathrm{H\alpha})$, and the measured velocity dispersion, $\sigma_\mathrm{obs}\mathrm{(H\alpha)}$, provided in the PHANGS-MUSE public data release DR1 (internal data release DR2.2). To remove the contribution of instrumental broadening, we derived the intrinsic velocity dispersion as $\sigma(\mathrm{H}\alpha) \approx \sqrt{\sigma_\mathrm{obs}\mathrm{(H\alpha)}^2 - \sigma_\mathrm{LSF}\mathrm{(H\alpha)}^2}$, where $\sigma_\mathrm{LSF}\mathrm{(H\alpha)}$ corresponds to the width of the MUSE instrumental profile at the observed wavelength of the \Ha\ line using the parametrisation of the line-spread function (LSF) as a function of wavelength from \cite{Bacon2017}. The mean value of $\sigma_\mathrm{LSF}\mathrm{(H\alpha)}$ is 48.9 $\kms$ after averaging over all sample galaxies. We do not correct \sigmaHa\ for thermal and fine structure broadening, as their contribution is negligible compared to the values measured in this paper. The maps of the stellar mass surface density $\Sigma_\star$ measured from the stellar continuum spectra and provided in the internal data release DR2.2 are used for the analysis in Sec.~\ref{sec:disc_turbulence}. We also analysed the integrated spectrum of each region with locally elevated \sigmaHa. For this purpose, we ran the \textsc{dap} on the extracted integrated spectra for each of these regions with the same settings as for the main PHANGS-MUSE products. As a result, we obtained information on the fluxes of the main diagnostic emission lines used in our later analysis (H$\beta$, [O~\textsc{iii}] 5007~\AA, [S~\textsc{ii}] 6717, 6731~\AA, H$\alpha$, [N~\textsc{ii}] 6584~\AA), and their velocity dispersion (corrected for instrumental broadening in the same way as described above).

Our analysis further relies on the catalogue of nebulae detected within PHANGS-MUSE galaxies (hereinafter -- the PHANGS-MUSE nebular catalogue; \citealt{Groves2023}). It was created by identifying \HII\ regions using the \textsc{HIIphot} code \citep{Thilker2000}, and includes derived physical properties  of these regions. In the present paper, we use the information on the electron density (derived from the ratio of the [S~\textsc{ii}]~6717/6731~\AA\, lines using \textsc{pyneb}; \citealt{pyneb}), size,  total reddening-corrected \Ha\ flux of the region, and  gas-phase metallicity (traced by oxygen abundance $\mathrm{12+\log(O/H)}$ measured using the S-calibration from \citealt{Pilyugin2016}). We consider only those nebulae classified as photoionised \HII\ regions (based on the diagnostic BPT diagrams; \citealt*{BPT}). Details of the PHANGS-MUSE nebular catalogue compilation and the parameters are provided in \cite{Groves2023}.

\subsubsection{On the reliability of the MUSE velocity dispersion measurements}

\begin{figure}
    \centering
    \includegraphics[width=\linewidth]{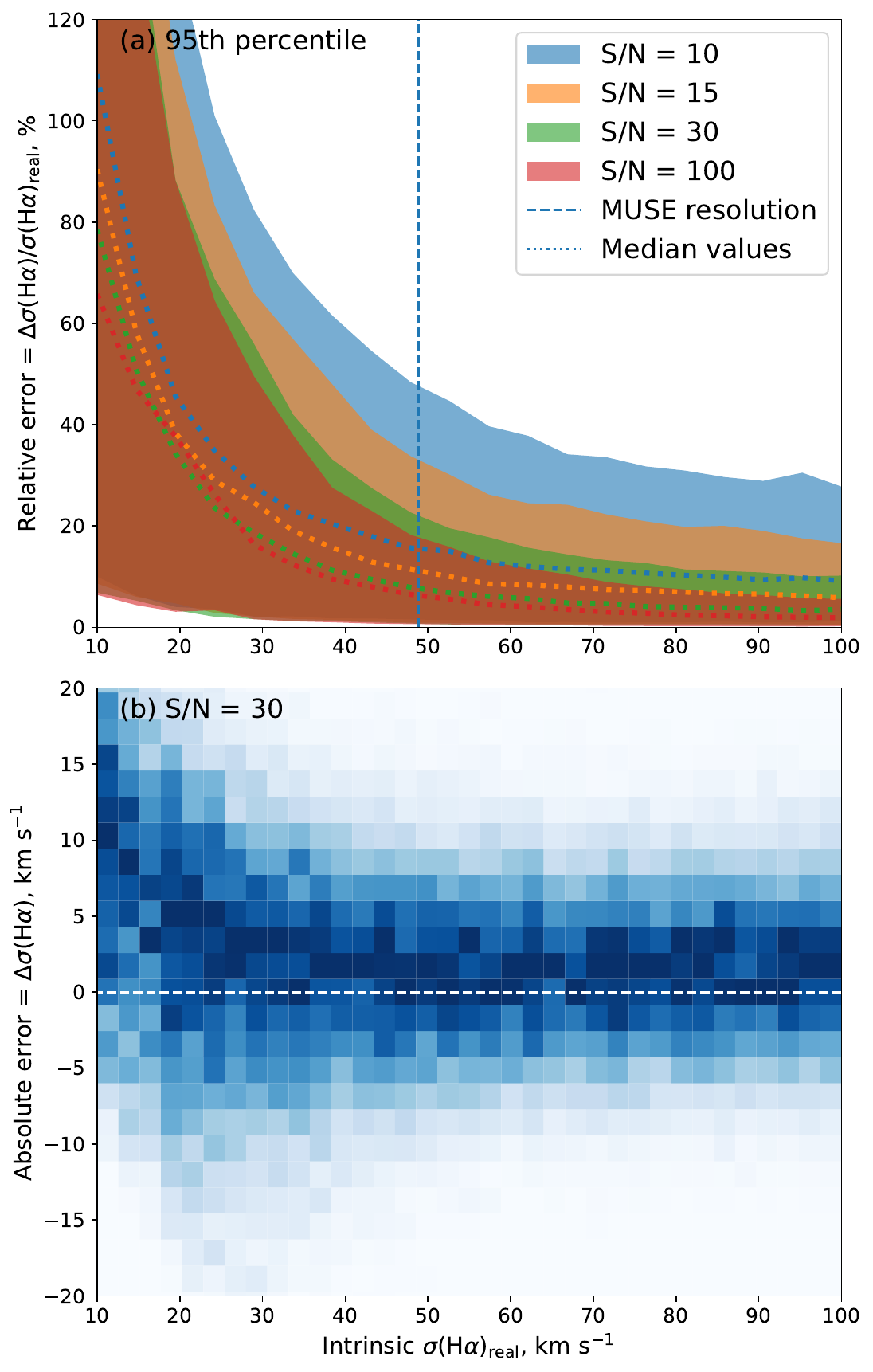}
    \caption{Relation between the recovered $\sigma\mathrm{(H\alpha)}$ and the true $\sigma\mathrm{(H\alpha)}$ depending on the $S/N$ ratio. Panel (a): Relative errors for various $S/N$ ratios. Coloured areas encompass the 5th-95th percentile interval and dotted lines trace the median values. Panel (b): Probability density of absolute errors for $S/N = 30$. The adopted MUSE resolution at the wavelength of \Ha\ is \mbox{$\mathrm{FWHM} = 2.54\pm0.1$~\AA\,} (according to \citealt{Bacon2017}, but with twice larger scatter; vertical dashed line on panel a).}
    \label{fig:sigma_test}
\end{figure}

Studies of the velocity dispersion of ionised gas with MUSE are limited due to its relatively modest spectral resolution -- the MUSE instrumental broadening is significantly larger than the typical values of $\sigma\mathrm{(H\alpha)} \sim 15-30\kms$ observed in the ISM of nearby star-forming galaxies \citep[e.g.][]{Terlevich1981, MoiseevKlypin2015, Green2014, Yu2019, Law2022}. This is also true for the PHANGS-MUSE galaxies -- the histograms in Fig.~\ref{fig:sigma_stat} show the distribution of $\sigma\mathrm{(H\alpha)}$. Only in some environments, such as bars and galaxy centres, is $\sigma\mathrm{(H\alpha)}$ above the MUSE resolution. 
 Our analysis is focused, however, on areas of locally elevated intrinsic velocity dispersion (exceeding $45~\kms$, see Sections~\ref{sec:isigma}, \ref{sec:disc_turbulence}), and thus the limited MUSE spectral resolution does not strongly impact our results. Nevertheless, it is important to estimate how precisely we can measure the relative variations of $\sigma\mathrm{(H\alpha)}$. 

To test the precision with which we can trace the relative changes of the \Ha\ velocity dispersion with MUSE, we performed a Monte Carlo (MC) simulation. We generated 50,000 synthetic Gaussian line profiles having the same sampling as the MUSE data ($1.25$ \AA\, per pixel) and a uniformly distributed line width (corresponding to $\sigma = 10 - 100 \kms$). These profiles were then convolved with a Gaussian with the width of the MUSE instrumental profile around its mean value at \Ha\ ($\mathrm{FWHM} = 2.54\pm0.05$~\AA) according to the parameterization from \citealt{Bacon2017}. We consider a two times larger scatter in this profile (0.1~\AA) given the higher spatial variation of the LSF in our data, likely due to the smaller number of rotations per pointing\footnote{The mosaics in PHANGS data were obtained from individual exposures obtained at 4 different orientations in order to minimize the instrumental artefacts and spatial variations of the LSF. The individual exposures in the deep field from \cite{Bacon2017} were obtained at 8 different orientations.}. 
We then added noise (both Gaussian and Poisson) according to the adopted signal-to-noise ($S/N$) ratio and fitted a Gaussian\footnote{\cite{Law2021sigma} compared the results of fitting the simple Gaussian and more realistic inverse gamma error distribution to the modelled line profiles and found a small ($\lesssim 10$\%) difference in \sigmaHa\ when both \sigmaHa\ and S/N are low, and negligible difference at high S/N.} to the obtained line profile. We repeated this procedure five times for different $S/N$ ratios and compared the measured LSF-subtracted velocity dispersion from the MC-sampled profiles to the true values. Fig.~\ref{fig:sigma_test} shows how the relative error of the measured $\sigma\mathrm{(H\alpha)}$ changes with velocity dispersion and with $S/N$ ratio (and also that for the absolute error for $S/N=30$). Qualitatively, these distributions agree with what was obtained by \cite{Law2021sigma} from a similar test for MaNGA data. As can be seen, we can trace the velocity dispersion in the \Ha\ line with MUSE with a precision better than 30\% (median uncertainty is $\sim12$\%) starting from $\sigma\mathrm{(H\alpha)} \sim 40 \kms$ for a $S/N$ = 30 or higher, and increasing the $S/N$ above 30 does not improve the precision significantly even for much higher values of $\sigma\mathrm{(H\alpha)}$. Panel (b) shows that for the regions having \sigmaHa$>45\kms$ and $S/N > 30$, we can expect the precision of the \sigmaHa\ measurements to be better than $\sim 5 \kms$, while the velocity dispersion is typically significantly overestimated with MUSE for \sigmaHa$<25\kms$ (typical values for \HII\ regions). 
Note also that, while the MUSE instrumental profile is quite well described by a Gaussian, the true instrumental LSF is slightly more square in shape according to \citet{Bacon2017}. This effect is not considered here but \revone{in principle, can produce slight additional} 
systematic offsets of the measurements.

\begin{figure*}
    \centering
    \includegraphics[width=\linewidth]{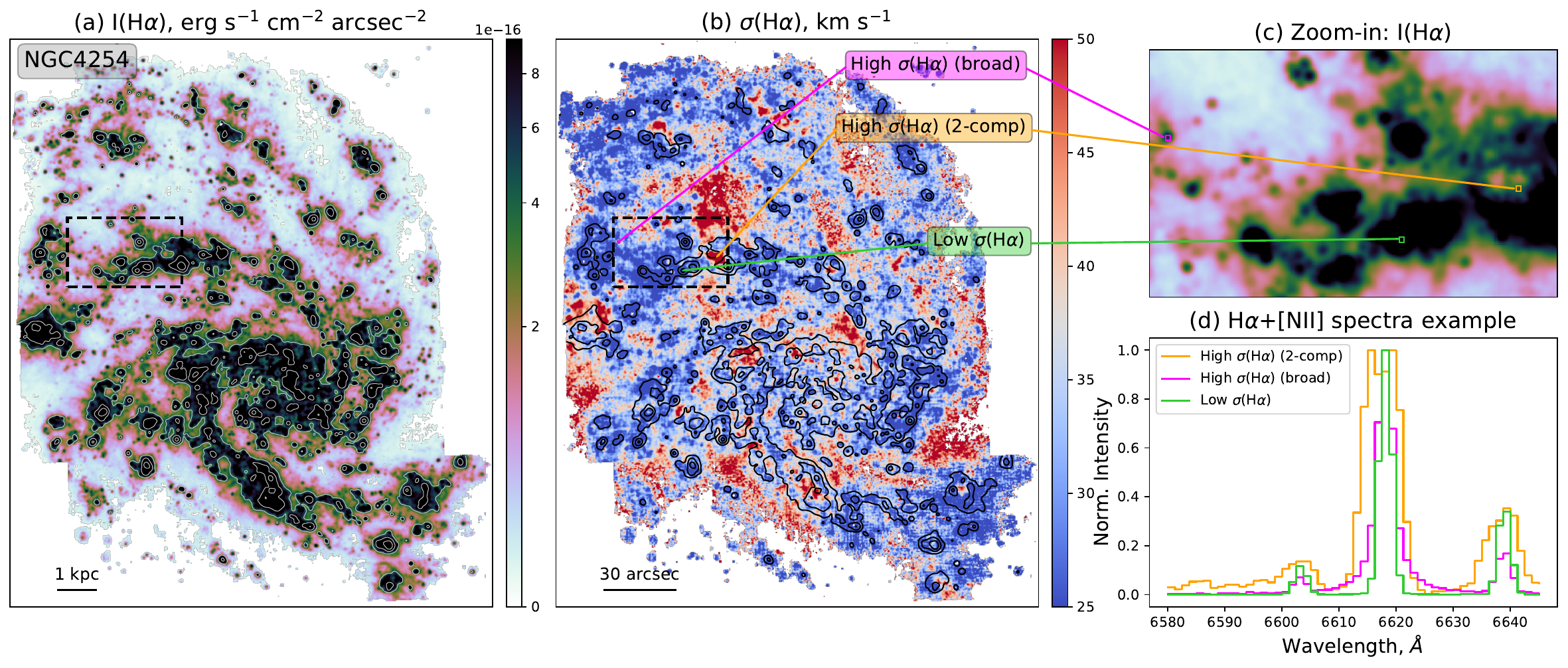}
    \caption{Distribution of the \Ha\ surface brightness (Panel a) and velocity dispersion (Panel b) for NGC~4254 shown as an example. Contours on these panels are the same and correspond to the lines of a constant \Ha\ brightness. Panel (c) demonstrates a zoom-in \Ha\ map of the area shown by dashed rectangle in Panels (a, b) with the same brightness limits as in Panel (a). As follows from Panel (b), velocity dispersion tends to be higher in the low surface brightness regions (dominated by DIG), but there are also areas of elevated \sigmaHa\ towards the bright regions. Panel (d) shows the examples of the \Ha\ and \NII\ $\lambda 6548,\ 6584$~\AA\, line profiles extracted from three regions with low (green line) and high (orange and magenta lines) \Ha\ velocity dispersion. \revone{These line profiles corresponding to high \sigmaHa\ are either broadened (magenta line) or can be decomposed by two components (orange line)}. Exact areas of the spectra extraction ($3\times3$ pixels) are shown on Panel (c) by squares of the same colour as the spectra. }
    \label{fig:data_example}
\end{figure*}

\subsection{PHANGS-HST}
\label{sec:obs:hst}

All 19 galaxies considered in this paper have publicly-available Hubble Space Telescope (\HST) observations obtained as part of the PHANGS-HST survey (PI: Lee). Some of the data\footnote{NGC~628, NGC~1433, NGC~1512, NGC~1566 and part of the data for NGC~3351 and NGC~3627} were obtained within the LEGUS (Legacy  ExtraGalactic  Ultraviolet  Survey) program \citep{Calzetti2015}. Images in 5 broad-band filters were obtained with Wide Field Camera 3 (WFC3) or Advanced Camera for Surveys (ACS) in bands roughly corresponding to the \textit{NUV}, \textit{U}, \textit{B}, \textit{V}, and \textit{I} bandpasses. A complete description of the PHANGS-HST survey and details on the data reduction are provided in \cite{Lee2022}.

At the HST resolution (PSF$\sim$0.08\arcsec = 2.0--7.6 pc at the distance of our galaxies), stellar associations can be identified by morphologically associating neighbouring young ($NUV$-bright) sources. \cite{Larson2022} have constructed a catalogue of stellar associations, which were defined as local peaks in the $NUV$ band, together with location masks derived by applying a watershed algorithm. Multi-scale catalogues were produced for fixed scales of 8, 16, 32 and 64~pc. Here we use the catalogue corresponding to the 32~pc spatial scale, which is typically a factor of two better than the MUSE spatial resolution, and thus allows us to cleanly associate the ionised gas with individual young stellar associations (the choice of spatial scale does not affect the results presented below). 

In addition to the young stellar associations, we also use the catalogue\footnote{\url{https://archive.stsci.edu/hlsp/phangs-cat}} of compact star clusters described in \cite{Whitmore2021} and \citet{Thilker2022}. From that catalogue, we selected only those human-classified clusters of Class 1 (symmetric, centrally concentrated), Class 2 (asymmetric, centrally concentrated) and Class 3 (compact stellar associations) that are not part of any of the young stellar associations from \cite{Larson2022}.

The catalogues of stellar associations and compact star clusters provide information on their mass and age, computed from fitting the observed spectral energy distribution (SED) in all 5 bands with \textsc{cigale} \citep{Boquien2019} assuming  theoretical models appropriate for a single stellar population from \cite{Bruzual2003}. The details of the SED fitting of the PHANGS-HST data are given in \cite{Turner2021}. In the present analysis, we use the measurements of the total stellar mass and the luminosity-weighted age of each association (or star cluster), and the corresponding stellar association masks. We note that both catalogues (stellar associations and compact star clusters) are used for the analysis, although for brevity we sometimes refer to stellar associations or star clusters only. 
The completeness limit for the young clusters and associations catalogues is about $3-5\times10^3\ M_\odot$ at the typical distance of PHANGS galaxies \citep{Turner2021}.

\section{Data analysis}
\label{sec:data_analysis}

We observe that the velocity dispersion of the ionised gas is not homogeneously distributed across the discs of our galaxies. Multiple areas of spatially coherent and locally elevated \sigmaHa\ are observed in all 19 PHANGS-MUSE galaxies. Those areas with small angular size and high \Ha\ surface brightness can be indicative of the presence of SNRs, Wolf-Rayet (WR) or Luminous Blue Variable (LBV) stars, or other high energy stellar sources \citep[e.g.][]{Moiseev2012, Yarovova2023}. More extended areas of elevated velocity dispersion are often observed in the inter-arm regions and are associated with the diffuse ionised gas (DIG, \citealt{Belfiore2022}). However, some of these regions are clearly connected to nearby \HII\ regions and identify the locations of expanding ionised superbubbles. Fig~\ref{fig:data_example} shows the distribution of \Ha\ flux and velocity dispersion for one galaxy from our sample: NGC~4254. The \Ha\ and \NII\ line profiles in panel (d) demonstrate the reliability of the observed variations in \sigmaHa. The presence of a broad underlying component or double-peaked emission lines (from the approaching and receding sides of a superbubble) \revone{is evident} in the emission line profiles extracted from regions of high \Ha\ velocity dispersion.

In this paper, we aim to isolate the regions of correlated supersonic motions in the ionised gas (caused by either turbulence or expanding superbubbles) within the PHANGS-MUSE data and to compare them with the presence and properties of the massive stars identified within the PHANGS-HST data. The analysis of both data sets can be briefly summarized in the following three major steps: 
\begin{enumerate}
    \item Identifying high \sigmaHa\ regions by their local excess of ionised gas velocity dispersion; 
    \item Searching for stellar associations residing within each identified region; 
    \item Deriving the physical properties of each region and of the stellar associations and calculating the kinetic energy of the ionised gas and the total mechanical energy input produced by the stellar associations. 
\end{enumerate}
These steps are described in detail in the rest of this Section.

\begin{figure*}
    \centering
    \includegraphics[width=\linewidth]{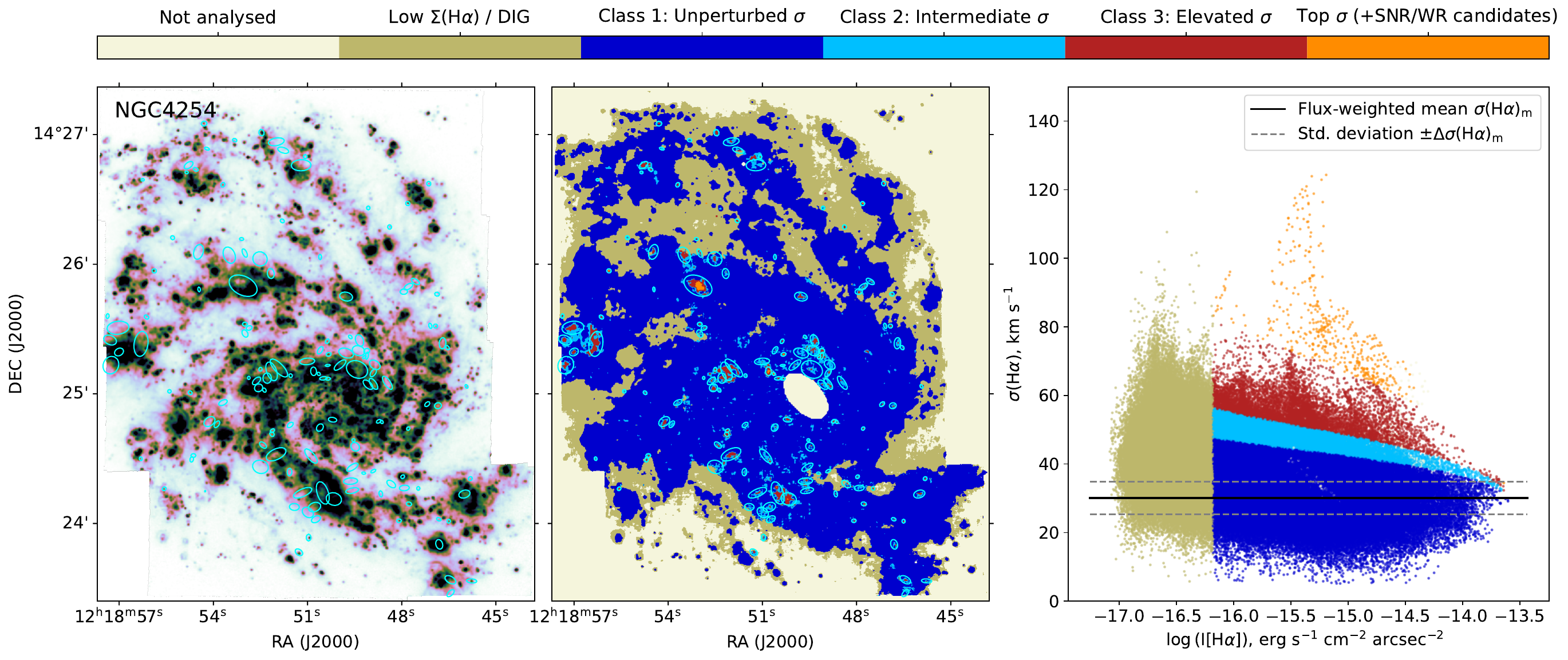}
    \caption{Localization of the regions of locally elevated \Ha\ velocity dispersion (cyan ellipses) in NGC~4254 galaxy, identified based on the `intensity -- velocity dispersion' (\IS) diagnostics, overlaid on the \Ha\ surface brightness map (left panel) and the classification map (central panel; see text). The \IS\ diagram is shown \revone{in the right panel}. The black solid line shows the mean value of \sigmaHa\ in the galaxy, and \revone{the dashed grey} lines show its 1$\sigma$ uncertainty. 
    See Fig~\ref{fig:isigma_all} for the rest of the PHANGS-MUSE galaxies.}
    \label{fig:isigma}
\end{figure*}

\subsection{Identification of expanding superbubbles and turbulent motions in the ionised ISM with \IS\ diagnostics}
\label{sec:isigma}

\begin{table*}
\centering
\caption{Sample of analysed galaxies, their properties and the number of identified regions with elevated velocity dispersion. }
\label{tab:sample}
\begin{small}\begin{tabular}{cccccccccccc} 
\hline
Galaxy & $D^1$ & $R_{25}^2$ & $\log M_*^3$  & $\log$(SFR)$^3$ & $12+\log(\mathrm{O/H})^4$  & PSF$^5$ &  $\mathrm{\sigma(H\alpha)_m}^6$ & $\mathrevone{\log \Sigma(H\alpha)_{DIG}}^6$  & \multicolumn{3}{c}{Number of regions$^6$}  \\ 
       & (Mpc) & (kpc) & ($M_\odot$) &  ($M_\odot\ \mathrm{yr}^{-1}$) & (dex at $R_\mathrm{eff}$)  & (\arcsec) & ($\kms$) & (erg s$^{-1}$ cm$^{-2}$ \revone{arcsec}$^{-2}$) & $N_\mathrm{tot}$ & $N_\mathrm{HST}$ & $N_\mathrm{cls}$ \\
\hline 
IC5332 & 9.01 & 8.0 & 9.67 & -0.39 & 8.30 & 0.87 & $23.0\pm5.2$ & -16.68 & 44 & 41 & 16 \\ 
NGC628 & 9.84 & 14.1 & 10.34 & 0.24 & 8.48 & 0.92 & $23.4\pm5.6$ & -16.28 & 116 & 69 & 30 \\ 
NGC1087 & 15.85 & 6.9 & 9.93 & 0.12 & 8.41 & 0.92 & $29.4\pm4.9$ & -16.18 & 28 & 28 & 19 \\ 
NGC1300 & 18.99 & 16.4 & 10.62 & 0.07 & 8.54 & 0.89 & $27.3\pm5.3$ & -16.58 & 64 & 63 & 10 \\ 
NGC1365 & 19.57 & 34.2 & 10.99 & 1.23 & 8.48 & 1.15 & $29.4\pm4.8$ & -16.48 & 55 & 20 & 2 \\ 
NGC1385 & 17.22 & 8.5 & 9.98 & 0.32 & 8.42 & 0.77 & $31.9\pm4.6$ & -16.38 & 108 & 108 & 33 \\ 
NGC1433 & 18.63 & 16.8 & 10.87 & 0.05 & 8.56 & 0.91 & $26.8\pm4.7$ & -16.78 & 103 & 50 & 6 \\ 
NGC1512 & 18.83 & 23.1 & 10.71 & 0.11 & 8.57 & 1.25 & $22.6\pm5.0$ & -16.98 & 51 & 51 & 10 \\ 
NGC1566 & 17.69 & 18.6 & 10.78 & 0.66 & 8.58 & 0.80 & $30.1\pm5.5$ & -16.23 & 107 & 102 & 38 \\ 
NGC1672 & 19.40 & 17.4 & 10.73 & 0.88 & 8.55 & 0.96 & $26.7\pm4.8$ & -16.38 & 93 & 93 & 19 \\ 
NGC2835 & 12.22 & 11.4 & 10.00 & 0.09 & 8.40 & 1.15 & $23.1\pm4.6$ & -16.28 & 58 & 48 & 28 \\ 
NGC3351 & 9.96 & 10.5 & 10.36 & 0.12 & 8.59 & 1.05 & $24.6\pm5.1$ & -16.38 & 43 & 37 & 13 \\ 
NGC3627 & 11.32 & 16.9 & 10.83 & 0.58 & 8.54 & 1.05 & $30.6\pm5.2$ & -15.98 & 102 & 101 & 47 \\ 
NGC4254 & 13.10 & 9.6 & 10.42 & 0.49 & 8.56 &  0.89 & $30.1\pm4.8$ & -16.18 & 137 & 137 & 61 \\ 
NGC4303 & 16.99 & 17.0 & 10.52 & 0.73 & 8.58 & 0.78 & $28.5\pm5.5$ & -15.98 & 108 & 108 & 58 \\ 
NGC4321 & 15.21 & 13.5 & 10.75 & 0.55 & 8.56 & 1.16 & $26.8\pm5.0$ & -16.28 & 58 & 46 & 27 \\ 
NGC4535 & 15.77 & 18.7 & 10.53 & 0.33 & 8.54& 0.56 & $26.9\pm5.7$ & -16.48 & 63 & 48 & 15 \\ 
NGC5068 & 5.20 & 5.7 & 9.40 & -0.56 & 8.32 & 1.04 & $24.3\pm5.5$ & -16.18 & 89 & 89 & 38 \\ 
NGC7496 & 18.72 & 9.1 & 10.00 & 0.35 & 8.51 & 0.89 & $28.7\pm4.8$ & -16.48 & 57 & 57 & 16 \\ 
\hline
\end{tabular}
\end{small}

\begin{footnotesize}\raggedright
$N_\mathrm{tot}$, $N_\mathrm{HST}$ and $N_\mathrm{cls}$ denote the total number of regions with elevated velocity dispersion, the number of those with available \HST\ data, and the number of regions containing young star clusters or associations, respectively.\\
$^{1}$\cite{Anand2021}, includes measurements from \cite{Jacobs2009} (NGC~628, NGC~1365, NGC~3351, NGC~3627), \cite{Kourkchi2017} (NGC~1087, NGC~1566, NGC~4303), \cite{Shaya2017, Kourkchi2020} (NGC~1300, NGC~1385, NGC~1672, NGC~7496), \cite{Scheuermann2022} (NGC~1433, NGC~1512), \cite{Nugent2006} (NGC~4254) and \cite{Freedman2001} (NGC~4321, NGC4535) \\
$^{2}$ From HyperLEDA \cite{Makarov2014}; \\
$^3$ \cite{Leroy2021}; \\
$^{4}$\cite{Groves2023};\\
$^{5}$\cite{Emsellem2022}; \\
$^{6}$This work.\\
\end{footnotesize}

\end{table*}

The first step in this analysis is the automated identification of regions with spatially-correlated \sigmaHa\ excess. For that, we rely on an analysis of the \IS\ diagrams representing the pixel-by-pixel variation of the velocity dispersion $\sigma$(H$\alpha$) versus the logarithm of the \Ha\ intensity,  $\log I\mathrm{(H\alpha)}$. 
This approach was introduced by \cite{MunozTunon1996} and further developed by \cite{Moiseev2012}. 
As was shown in these papers, an expanding superbubble appears on the \IS\ diagram as a triangle-shaped region of intermediate $I\mathrm{(H\alpha)}$ and elevated $\sigma$(H$\alpha$) in comparison to \HII\ regions, which tend to have a relatively uniform distribution in their velocity dispersion across almost the whole range of \Ha\ fluxes. \citet[][see also \citealt{Yarovova2023}]{Moiseev2012} demonstrated that stellar-like objects (e.g. SNRs, WR, and LBV stars) that are strongly impacting the ISM can also be easily identified on these diagrams as elongated diagonal stripes. 
The \IS\ diagnostics were successfully implemented for the detection and analysis of supersonic ionised gas motions in individual star-forming complexes of nearby galaxies and in irregular and blue compact dwarf galaxies based on high spectral resolution data from Fabry-Perot interferometer (FPI) and IFU (GMOS, KCWI) observations \citep[e.g.][]{MD2007, Bordalo2009, Egorov2018, Bresolin2020, Egorov2021, Gerasimov2022}. In this paper, we apply the \IS\ diagnostics in a form similar to that described in \cite{Egorov2021}. The \IS\ diagram constructed for NGC~4254 is shown in Fig.~\ref{fig:isigma} as an example (see Appendix~\ref{app:isigma} for other PHANGS-MUSE galaxies). The interpretation of the different zones and the analysis of these diagrams are described in detail in Appendix~\ref{app:isigma}, while here we outline the main steps in the analysis.

Compared with previous studies, the application of \IS\ diagnostics to the PHANGS-MUSE data has one obvious limitation -- the spectral resolution, which can be insufficient to resolve the complex kinematics of the ionised gas in star-forming complexes. 
As we demonstrated in Sec.~\ref{sec:obs:muse}, with $S/N \gtrsim 30$ and better we are able to reliably measure the relative variations of the intrinsic velocity dispersion at the level of $\sigma$(H$\alpha) \gtrsim 30 \kms$ with a typical uncertainty better than $5 \kms$. To ensure those measurement uncertainties do not bias our analysis, we masked out all pixels with $S/N < 30$ in \Ha. The majority of the areas of elevated velocity dispersion that we investigate in this paper have $\sigma$(H$\alpha) > 45 \kms$ and thus provide reliable measurements of $\sigma$(H$\alpha$) and even allow for the identification of kinematically distinct components for some of the regions (Fig.~\ref{fig:data_example}). Moreover, thanks to the high angular resolution of the MUSE data, the local peaks in the velocity dispersion are well distinguished, making it easier to identify expanding structures on the \IS\ plot.  

In contrast to the analysis of individual star-forming complexes and dwarf galaxies, massive galaxies provide a more diverse set of environments, with large variations in the properties of the ISM (i.e. temperature, density and metallicity) that affect the gas pressure and potentially the observed mean velocity dispersion of the \HII\ regions \cite[e.g.,][]{Barnes2021}. 
In galaxies with strong bars, massive bulges, or inner discs, each morphological component produces its own unique distribution of the data on the \IS\ plot, making the uniform application of this diagnostic across the whole galaxy very difficult. In general, the measured \sigmaHa\ tends to be higher in the galactic centres (including nuclear rings, lenses etc.) and bars of the PHANGS-MUSE galaxies than in their discs (Fig.~\ref{fig:sigma_stat}). To remove this bias, we exclude from further analysis all the regions related to these morphological components according to the Spitzer 3.6$\mu$m based environmental masks defined by \citet{Querejeta2021}. It is possible to extend the current study to these environments by analysing them separately, but for this work we focus only on regions in the discs. 
We skip the masking procedure for NGC~628 and for the two lowest mass galaxies (IC~5332 and NGC~5068) where the \sigmaHa\ distribution looks similar in all environments. Additional, manually defined masks were applied to NGC~1365, NGC~1672, and NGC~7496 to remove very extended regions of high \sigmaHa\ associated with \revone{AGN outflows \citep{Stuber2021}}. 

After applying these S/N and environmental masks, we constructed the \IS\ diagrams for each galaxy and analysed them to select the regions with spatially-correlated elevated velocity dispersion. As a short summary, we performed the following steps (see Appendix~\ref{app:isigma} for more details): 

\begin{enumerate}
    \item Isolated the peaks of high velocity dispersion corresponding to the DIG with surface brightness below \SBDIG\ (given in Table~\ref{tab:sample}), which were excluded from the analysis; 
    \item Derived the mean velocity dispersion (\sigmaHa$_\mathrm{m}$) and its \revone{standard deviation and defined the} normal distribution \revone{of \sigmaHa\ around \sigmaHa$_\mathrm{m}$} as a function of \Ha\ intensity for each galaxy; 
    \item Divided individual pixels into three major classes based on their \Ha\ intensity and \sigmaHa\ relative to the \revone{defined normal distribution around \sigmaHa$_\mathrm{m}$} (these classes are shown by different colours in Fig.~\ref{fig:isigma}, and particular criteria are summarized in Table~\ref{tab:method_summary});
    \item Applied \textsc{astrodendro} \citep{astrodendro} to the obtained classification map, isolating and tracing the variations of \sigmaHa\ for pixels with elevated velocity dispersion (Class 2 and Class 3 and SNR/WR candidates in Fig.~\ref{fig:isigma}). Defined the ellipses encircling the selected regions;
    \item Refined the estimates of the sizes for those regions showing clear peaks (either central or associated with the swept-up shells) in the radial distribution of their \Ha\ flux. If no such peak is found, we adopt $6.2\times\sigma_A$ as size of the regions along their major or minor axes, where $\sigma_A$ is the standard deviation along these axes of the
2D distribution resulting from the statistics computed by \textsc{astrodendro}. Significantly overlapping or elongated regions are excluded.
\end{enumerate}

As a result, we selected $\sim30-130$ regions of locally elevated \sigmaHa\ in each of the 19 PHANGS-MUSE galaxies (exact numbers are given in Table~\ref{tab:sample}), and \nregs\, regions in total.
This sample is not complete (some candidates were excluded as they do not pass all our criteria, and also better resolution data would allow us to detect more regions), however the number of false-detections is negligible ($<3$~regions per galaxy or $\sim$1\% in total\revone{, see Appendix~\ref{app:isigma}}).

\subsection{Connecting the identified regions with the stellar associations}
\label{sec:stars_identifications}

As a next step, we searched for young stellar associations 
 and compact star clusters which might be connected to the identified regions of locally elevated  ionised gas velocity dispersion. 
 For each region, we identified the stellar associations where the masks (or position of the star cluster) overlap with the elliptical borders calculated in the previous section. We also calculated the overlapping fraction as the ratio of the area of the stellar mask within the region to the total area of the association. Those associations residing mostly outside the elliptical borders (overlapping by less than 20\%; considering a higher or lower percentage within $\pm 10$\% does not change our results) were excluded from further analysis as they are unlikely to be physically associated, but rather just chance projections (this criterion was skipped, however, for the centrally concentrated regions). 
We also assume that the turbulent motions decay within a given timescale, which we set here equal to $10$~Myr. The exact timescale depends on the spatial scale, 
and velocity of the gas motions: a typical value is $1-10$~Myr, where the longest value corresponds to the characteristic crossing timescale for galaxy discs in the vertical direction \citep{Agertz2013, Ostriker2001}. Therefore we considered only the mechanical energy injection from the star clusters during the last 10~Myr (see Section~\ref{sec:physical_parameters}), and excluded from the calculations any stellar associations and star clusters older than 50~Myr ($\sim$10~Myr higher than the lifetime of an $8 M_\odot$ star which still can explode as a Type II SN). 

For \nregsnohst\, regions with high \sigmaHa\ we are unable to establish a link with a star cluster because they are \revone{outside} the area covered by \HST\ observations. For the rest of the regions, we find that \nregswithcls\, ($\sim39$\% of the sample covered by \HST) are associated with at least one stellar association or star cluster. This is almost twice what would be expected if the same stellar associations were randomly distributed across the discs. About 70\% of these regions (\nregsnooldcls) contain only relatively young stellar associations (younger than 10~Myr), and \nregshaveyoungcls\, contain stellar associations younger than 4~Myr.

\subsection{Estimation of physical parameters for the identified regions and stellar associations}
\label{sec:physical_parameters}

For each identified region with high \sigmaHa\ we extracted an integrated spectrum across an elliptical aperture covering $1/3$ of the region's effective radius. In choosing the central aperture we tried to minimise any contamination from nearby \HII\ regions or the denser rims in cases where the region is a superbubble (to isolate the emission from its approaching and receding sides). The integrated spectra were analysed with the \textsc{dap} in the same way as was done for the individual pixels in the PHANGS-MUSE datacubes (see Section~\ref{sec:obs:muse} and \citealt{Emsellem2022}). From these spectra, we extracted the \Ha\ surface brightness and emission line fluxes to characterise each region. We corrected their values for interstellar extinction using the observed Balmer decrement and the \citep{Cardelli1989} reddening curve with $R_V = 3.1$ as for our Galaxy, and assuming an intrinsic ratio of $I(\mathrm{H}\alpha)/I(\mathrm{H}\beta)=2.86$ for case B recombination \citep{Storey1995}.

The oxygen abundance $12+\log\mathrm{(O/H)}$ is taken from the smoothed two-dimensional distributions published by \cite{Williams2022}. These were derived by applying  Gaussian Process Regression to the metallicity measurements made for individual \HII\ regions for all 19 PHANGS-MUSE galaxies. These values rely on the S calibration from \citet{Pilyugin2016}, requiring measurements of the \SII~$\lambda$6717, 6731~\AA, \NII~$\lambda$6584~\AA, \OIII~$\lambda$5007~\AA, \Ha\ and \Hb\ emission lines. We use these estimates instead of obtaining metallicity measurements from the integrated region spectra, as the \HII\ regions are less biased by the DIG, although both methods gave us consistent results.

\subsubsection{Mass of the ionised gas}

We consider two ways to estimate the total mass of the ionised gas, $M_{\rm ion}$, in the regions. The first one assumes a thin spherical shell geometry for all of them. Then 
\begin{equation}
    \label{eq:mass_ini}
    M_{\rm ion}^{(1)} \simeq 4\pi R_{\rm eff}^2 \, \Delta R \, n_{\rm e} \, \mu m_{\rm H} = 4 \Omega_{\rm reg} D^2 \Delta R \, n_{\rm e} \, \mu m_{\rm H},
\end{equation}
where $R_{\rm eff} = \sqrt{R_{\rm min}R_{\rm maj}}$ is the effective radius of the region, computed as the geometric mean of the minor and major semi-axes of the ellipses encircling the region (see Sec.~\ref{sec:isigma}), $\Delta R$ is the thickness of the shell, $n_{\rm e}$ is the electron density, $m_{\rm H}$ is the mass of a hydrogen atom, the coefficient $\mu=1.27$ accounts for a singly-ionised helium contribution, $\Omega_{\rm reg}$ is the total area of the region in the plane of the sky, 
and $D$ is the distance to the galaxy. 
To calculate $\Delta R$, we can relate the emission measure $EM$ to the surface brightness of \Ha,  $I(\mathrm{H\alpha})$ as \citep{Reynolds1977, Pengelly1964}:
\begin{equation}
    EM = 2.75 \left(\frac{T_{\rm e}}{10^4 {\rm K}}\right)^{0.9} I(\mathrm{H}\alpha)\ \mathrevone{ pc\ cm^{-6}},
    \end{equation} 
where $I(\mathrm{H}\alpha)$ is in units of Rayleighs ($1\,\mathrm{Rayleigh} \simeq 5.69 \times 10^{-18}\, \mathrm{ergs\ s^{-1}\ cm^{-2}\ arcsec^{-2}}$ for the \Ha\ line), and $T_{\rm e}$ is the electron temperature. Assuming a homogeneous density distribution (that represents the average conditions), the emission measure is related to the thickness of the nebula (equal to $2\Delta R$ for a spherical shell geometry) as $EM = \int{n_{\rm e} n_{\rm H^+} ds} \simeq \int{n_{\rm e}^2 ds} \simeq 2\Delta R n_{\rm e}^2$. Thus, we obtain 
\begin{equation}
\label{eq:dr_flux}
    \Delta R \simeq 24.2 \left(\frac{T_{\rm e}}{10^4 {\rm K}}\right)^{0.9} \frac{F(\mathrm{H}\alpha) \Omega_{\rm c}^{-1}}{10^{-16}\ \mathrm{erg\  s^{-1}\ cm^{-2}\ arcsec^{-2}}} \left(\frac{n_{\rm e}}{\mathrm{cm}^{-3}}\right)^{-2}\, \mathrm{pc},
\end{equation}
where $F(\mathrm{H}\alpha) = I(\mathrm{H}\alpha) \Omega_{\rm c}$ is the total reddening-corrected \Ha\ flux from the central aperture used in the spectral extraction, which has an area of $\Omega_{\rm c}$ in the plane of the sky in squared arcseconds. Given that in our analysis we choose the central aperture for spectra extraction to have a size of 1/3 of the full size of the region (see Sec.~\ref{sec:isigma}), $\Omega_{\rm reg}/\Omega_c = 9$ for every region. Assuming that $T_{\rm e} = 8000$~K (typical for our sample of mostly slightly sub-solar metallicity \HII\ regions, but can be higher in shocked or in lower metallicity regions; \citealt{Kreckel2022}), we can combine Eqs.~(\ref{eq:mass_ini}) and  (\ref{eq:dr_flux}) to obtain
\begin{equation}
 \label{eq:mass_flux}
     M_{\rm ion}^{(1)} \simeq \mathrevone{522} \frac{F(\mathrm{H}\alpha)}{10^{-16}\ \mathrm{erg\ s^{-1}\ cm^{-2}}}\left(\frac{D}{\mathrm{Mpc}}\right)^2\left(\frac{n_e}{\mathrm{cm^{-3}}}\right)^{-1} M_\odot.
 \end{equation}

Note that this method does not depend on how well we can measure $R_{\rm eff}$ for our regions, which can be especially uncertain for small, poorly resolved regions \citep[see][]{Barnes2021}. However, it assumes a spherical shell geometry, with a uniform distribution of density (and brightness) within the shell. As previously mentioned in Sec.~\ref{sec:isigma}, at least \nshells\ of the regions demonstrate a shell-like morphology and thus are likely superbubbles, for which this approximation is arguably valid. This is not necessarily true for all regions. 

Given the spatial resolution of our data, we cannot provide better constraints on the shape of the high velocity dispersion regions. Also, the thin-shell approximation might not work for the small regions in our sample (i.e. where the size of the aperture for spectra extraction is smaller than the PSF). Therefore, we also consider a second method that does not depend on the geometry of the region but instead requires us to measure the total \Ha\ flux from the whole region, $F{(\rm H\alpha)_{tot}}$ (measured from the spectra extracted \revone{from} the elliptical apertures corresponding to the borders of a region). From atomic physics, we know that the number of hydrogen-ionising photons consumed within a region of volume $V$ is given by $Q({\rm H^0}) = \alpha_{\rm B} \langle n_{\rm H}^2 \rangle V \simeq \alpha_B \langle n_{\rm e}^2 \rangle V$ when the volume is in ionisation equilibrium, where $\alpha_{\rm B}$ is the case B recombination coefficient and $\langle n_{\rm e}^{2} \rangle$ is the mean squared electron density within $V$.\footnote{Accounting for helium makes only a small difference to this number.} If we assume, in the absence of evidence to the contrary, that the density distribution is uniform, this relation simplifies to $Q({\rm H^0}) = \alpha_B n_{\rm e}^2 V$. For our adopted $T_{\rm e}=8000$~K, interpolation of the values in \citet{Storey1995} yields $\alpha_B \simeq 3.1\times10^{-13}\ {\rm cm^3\ s^{-1}}$, and the ionising photon flux is related to the H$\alpha$ luminosity by $Q({\rm H^0}) \simeq 7.1\times10^{11} L({\rm H\alpha})$ \citep{Osterbrock2006}. \revone{Given that $L({\rm H\alpha}) = 4 \pi D^2 F{(\rm H\alpha)_{tot}}$, we then have}
\begin{equation}
    \label{eq:mass_flux_2}
    M_{\rm ion}^{(2)} = \mu m_{\rm H} n_{\rm e} V \simeq \mathrevone{29.3} \frac{F{(\rm H\alpha)_{tot}}}{\rm 10^{-16}\ erg\ s^{-1}\ cm^{-2}} \left(\frac{D}{\rm Mpc}\right)^2\left(\frac{n_{\rm e}}{\rm cm^{-3}}\right)^{-1} M_\odot,
\end{equation}

As our adopted value of $M_{\rm ion}$, we use $M_{\rm ion}^{(1)}$ for the regions having a size along the minor axis greater than $3\times$PSF, and $M_{\rm ion}^{(2)}$ for smaller regions. This allows us to minimize contamination by overlapping \HII\ regions in the case of large regions, and to correctly measure $M_{\rm ion}$ for small regions where the thin-shell approximation does not work due to insufficient resolution.

\subsubsection{Estimates of the electron density}

\begin{figure}
    \centering
    \includegraphics[width=\linewidth]{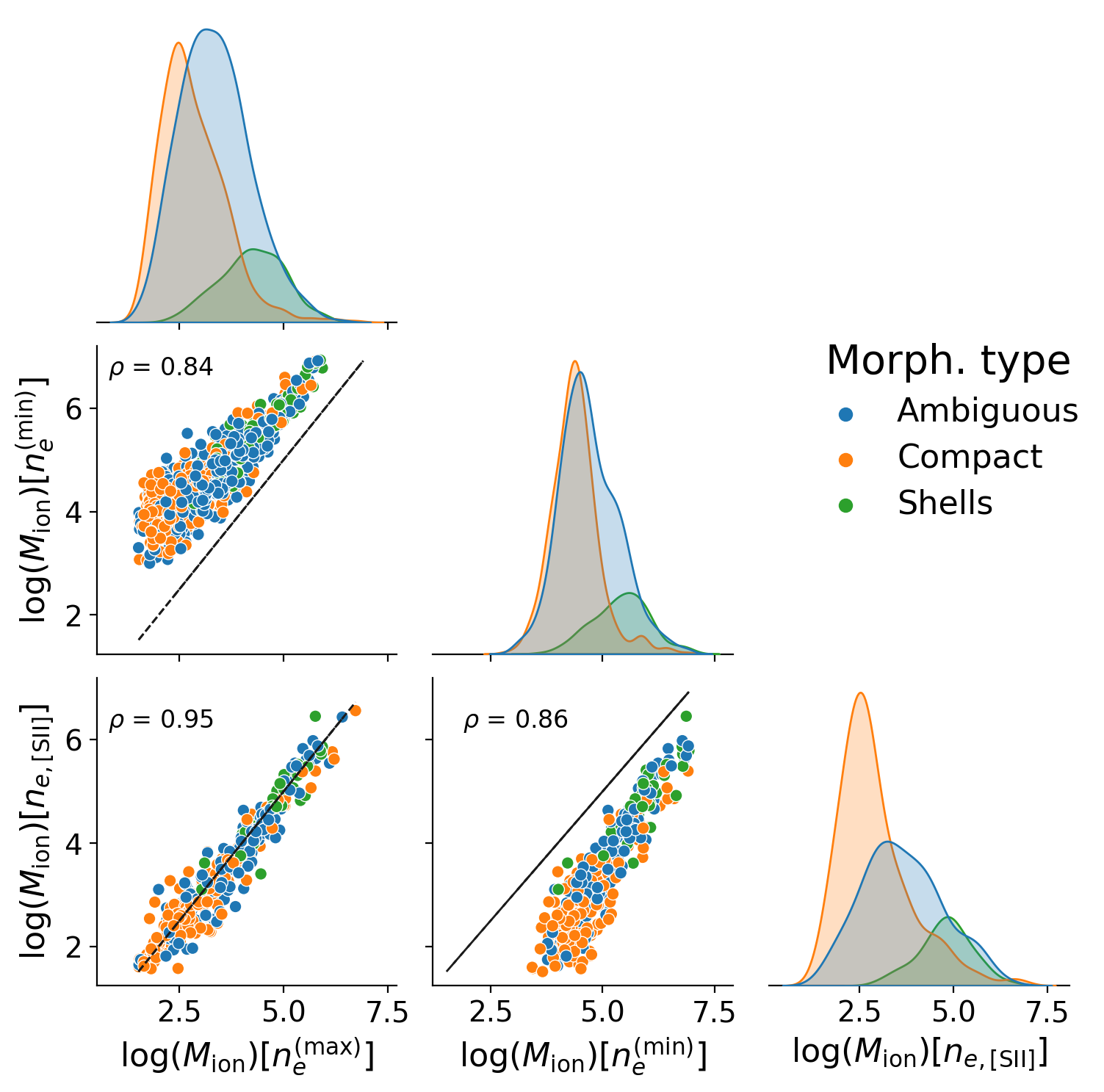}
    \caption{Mass of the ionised gas in the selected regions assuming different prescriptions for electron density: lower limit ($n_{\rm e}^{\rm (min)}$) and upper limit ($n_{\rm e}^{\rm (max)}$) derived from the reference \HII\ regions using Eq.~(\ref{eq:dens_hii_vol}) and from $R_\mathrm{[S II]}$ = [S~\textsc{ii}]$\lambda$6717\AA/[S~\textsc{ii}]$\lambda$6731\AA\, flux ratio, respectively, and $n_{{\rm e}, {\rm SII}}$ derived from $R_\mathrm{[S II]}$ of the observed spectra of \revone{the region} itself. Different morphological types introduced in Sec.~\ref{sec:regions_morphology} are shown by different colours. Diagonal lines show 1-to-1 relation. Spearman correlation coefficients $\rho$ are given on the plots.}
    \label{fig:gas_masses}
\end{figure}

Precise measurements of the electron density are crucial for deriving $M_{\rm ion}$, but difficult because of several observational and physical limitations. It is typically measured 
using density-sensitive line ratios like $R_\mathrm{[S II]}$ = [S~\textsc{ii}]$\lambda$6717\AA/[S~\textsc{ii}]$\lambda$6731\AA\ \citep[e.g.][]{Osterbrock2006, Kewley2019}. Such diagnostics have low-density limits (e.g.\ $n_e\sim30$~cm$^{-3}$ for $R_\mathrm{[S II]}$), below which changes in $n_e$ have no effect on the line ratios.
The densities typically measured in extragalactic \HII\ regions are close to this low-density limit \citep[e.g.][]{Kennicutt1984, Hunt2009} making the calculations uncertain. 
In addition, density inhomogeneities in the nebulae can introduce \revone{systematic} biases in the measurements of $n_{\rm e}$ and other physical properties \citep[e.g.][]{Peimbert1971, Rubin1989, Viegas1994, Cedres2013}. Here we use two different density diagnostics that we consider as lower and upper limits.

For both calculations of $M_{\rm ion}$ (Eqs.~\ref{eq:mass_flux}, \ref{eq:mass_flux_2}), we estimate the electron density $n_{\rm e,\mathrm{SII}}$ from the $R_\mathrm{[S II]}$ using the \textsc{pyneb} package \citep{pyneb} and assuming $T_{\rm e} = 8000$~K. The electron density of the ionised gas in our regions is close to the low-density limit. Because of this, we are able to reliably measure it only for a limited sample of $\sim20$\% of regions in which the observed $R_\mathrm{[S II]}$ falls below the value of 1.46\footnote{This value corresponds to the low-density limit at $T_e = 8000$~K \citep{Barnes2021}} by at least 3$\sigma$ (chosen to minimize the bias towards higher densities if $R_\mathrm{[S II]}$ is underestimated because of the observational uncertainties).
To use all the identified regions in the \revone{remainder of the} analysis, we assume they have in general the same $n_{\rm e}$ as \revone{their nearest-neighbour \HII\ regions}. We selected reference \HII\ regions from the nebular catalogue located within two effective radii of each region centre (this distance was increased up to four radii if none were found at the smaller distance), calculated $n_{\rm e,\mathrm{SII}}$ for each of these \HII\ regions in the same way as above, and adopted our final $n_{\rm e}^{\rm (max)}$ as the median value among them. We are able to measure $n_{\rm e,\mathrm{SII}}$ of at least one reference \HII\ region for all but 16 \revone{of our regions}, and of at least 3 for 36\% of our sample. Fig.~\ref{fig:gas_masses} compares the estimates of $M_{\rm ion}$ made assuming $n_{\rm e}$ calculated in different ways. For the regions where direct measurements of $n_{\rm e,\mathrm{SII}}$ from their spectra are possible, we obtain very similar $M_{\rm ion}$ as from the density adopted from the reference \HII\ regions. Therefore, for consistency, we consider \revone{hereafter} the electron density $n_{\rm e}^{\rm (max)}$ derived in the latter way. 

The estimates of $n_{\rm e}^{\rm (max)}$ obtained from $R_\mathrm{[S II]}$ are related to dense clumps in the nebulae where the emission of the \SII\ lines \revone{originates. Thus,} the derived values are not necessarily equal to the mean volume electron density $n_{\rm e}$, but represent its upper limit because of clumping of the ionised gas. Assuming a spherical shape for the reference \HII\ regions, we can measure the volume-averaged electron density as 

\begin{equation}
\label{eq:dens_hii_vol}
    n_{\rm e}^{\rm (min)} = \sqrt{\frac{3Q({\rm H^0})}{4\pi R_{\rm eff}^3 \alpha_B}} \simeq \mathrevone{1.36}\times10^{-16}\left(\frac{L({\rm H\alpha})}{\rm erg\ s^{-1}}\right)^{0.5}\left(\frac{R_{\rm eff}}{\rm pc}\right)^{-1.5} {\rm cm^{-3}}. 
\end{equation}
Applying this to the selected reference \HII\ regions in the same way as above, we estimate $n_{\rm e}^{\rm (min)}$ for all regions in our \revone{sample. 53\% have at least} 3 reference \HII\ regions with reliable estimates (\HII\ regions with an effective radius $R_{\rm eff}<0.5\times {\rm PSF}$ are excluded). In the ideal case, $n_{\rm e}^{\rm (min)}$ yields the true value of the mean volume density, however, only about 25\% of the \HII\ regions in PHANGS-MUSE galaxies are well-resolved and have reliable measurements of $R_{\rm eff}$ \citep{Barnes2021}. Thus many of the reference \HII\ regions in our sample are under-resolved, in which case $R_{\rm eff}$ is likely overestimated \citep{Barnes2022}. Moreover, the validity of Eq.~\ref{eq:dens_hii_vol} depends on the actual distribution of the ionised gas in \HII\ regions and can underestimate the result if the ionised gas does not occupy the entire volume of the nebula. Thus the value of $n_{\rm e}^{\rm (min)}$ corresponds to the lower limit of the electron density. 

We use the estimates of $n_{\rm e}^{\rm (min)}$ and $n_{\rm e}^{\rm (max)}$ for each region to measure the upper and lower limits of $M_{\rm ion}$, respectively. Fig.~\ref{fig:gas_masses} shows the comparison between different methods. While the measurements are correlated, we find an order of magnitude difference between $M_{\rm ion}$ calculated with these two prescriptions for $n_{\rm e}$.  In the case of the well-resolved reference \HII\ regions, $n_{\rm e}^{\rm (min)}$ should yield the volume-averaged density, less biased by the presence of dense clumps. We thus consider it as a better proxy of the true density of the ionised gas \revone{in our} selected regions of high velocity dispersion. Meanwhile, $n_{\rm e}^{\rm (max)}$ demands a knowledge of the filling factor to be converted to a volume-averaged density, which cannot be properly derived for our sample. At the same time, $n_{\rm e}^{\rm (max)}$ correlates extremely well with $n_{{\rm e},\mathrm{SII}}$ derived directly from the observed spectra and unaffected by the limited angular resolution. Thus, we expect that the upper limit of $M_{\rm ion}$ (based on $n_{\rm e}^{\rm (min)}$) is closer to the real value, while its lower limit (based on $n_{\rm e}^{\rm (max)}$) better traces the relative changes between the regions. The median formal uncertainties of $\log(M_{\rm ion}) [n_{\rm e}^{\rm (min)}]$ and $\log(M_{\rm ion}) [n_e^{\rm (max)}]$ are equal to 0.09~dex and 0.29~dex, respectively. Therefore our further analysis relies mainly on $M_{\rm ion}\ [n_{\rm e}^{\rm (min)}]$, but we compare the results with measurements made assuming  $M_{\rm ion}\ [n_{\rm e}^{\rm (max)}]$.

\subsubsection{Kinetic energy of the ionised gas and mechanical energy input from stars}

\begin{figure}
     \centering
     \includegraphics[width=\linewidth]{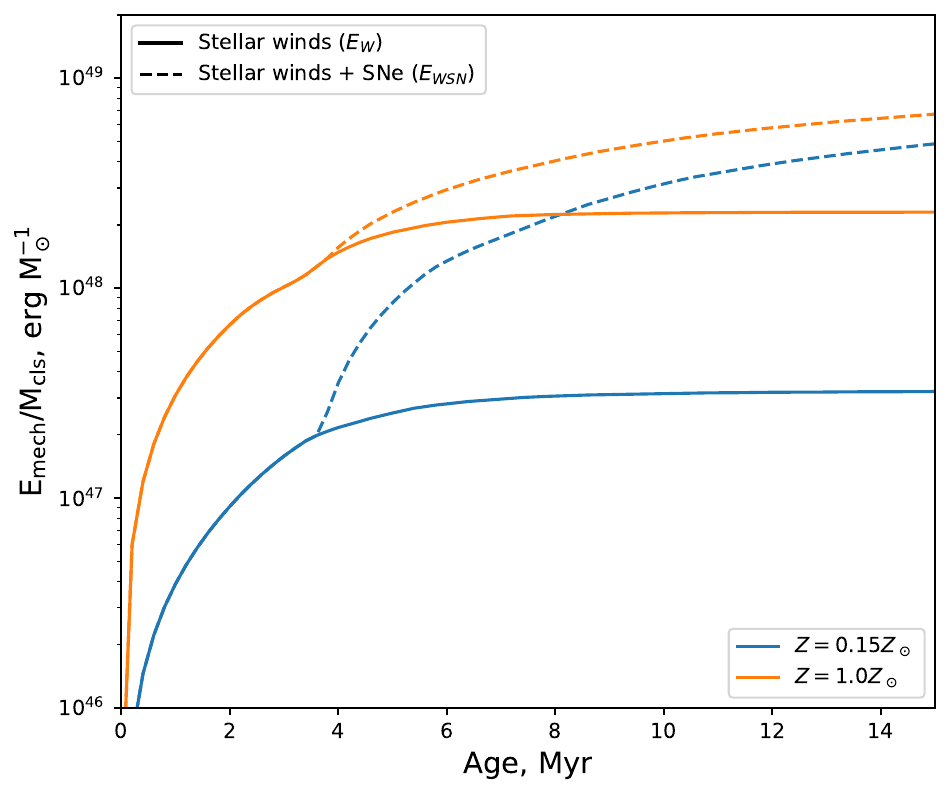}
     \caption{Evolution of the cumulative mechanical energy input to the ISM ($E_{\rm mech}$) normalized by the mass of star cluster ($M_{\rm cls}$)  for different metallicity (traced by different colours) according to \textsc{starburst99} \citep{Leitherer1999, Leitherer2014} models. The contribution produced by stellar winds alone is shown by the solid line, while the dashed line corresponds to the impact of both stellar winds and supernovae.}
     \label{fig:sb99}
 \end{figure}

\begin{figure*}
    \centering
    \includegraphics[width=\linewidth]{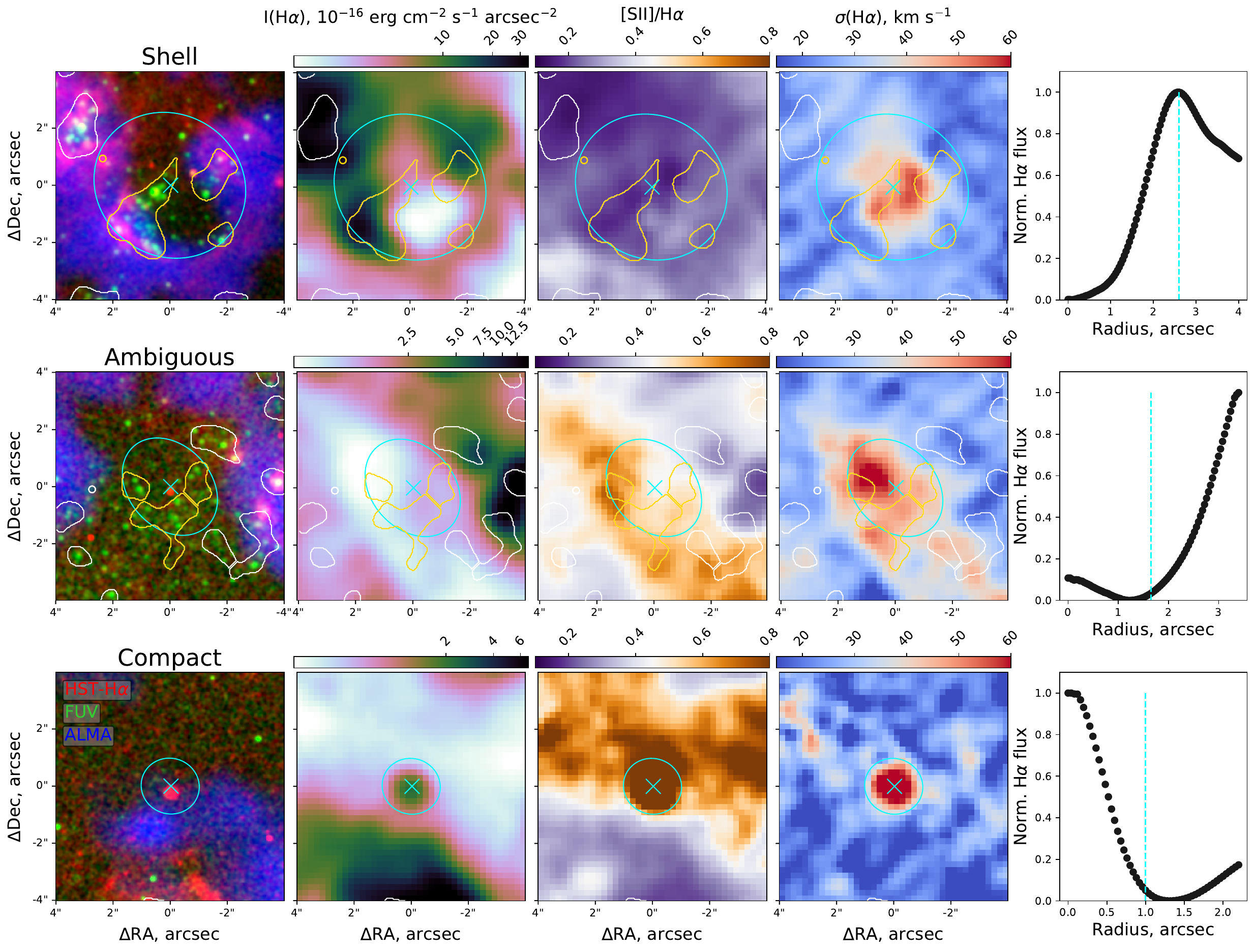}
    \caption{Examples of regions classified as shells (top row), ambiguous (middle row) and compact (bottom row). Images from left to right correspond to a three-colour multiband image of the region (\Ha\ image from HST (PI: R. Chandar, GO-10402) is in red, FUV from AstroSat (Hassani et al., in prep.) in green, CO from ALMA \citep{Leroy2021} in blue), the \Ha\ flux distribution, the \SIIHa\ line ratio, and the \Ha\ velocity dispersion. Cyan ellipses show the adopted borders of the region and overlaid contours correspond to the stellar associations' masks. The HST stellar associations shown in yellow are selected as associated with the region. The right-hand plots show the  radial \Ha\ flux distribution relative to the centre of each region (marked by crosses) }
    \label{fig:regions_example}
\end{figure*}

\begin{figure*}
    \centering
    \includegraphics[width=\linewidth]{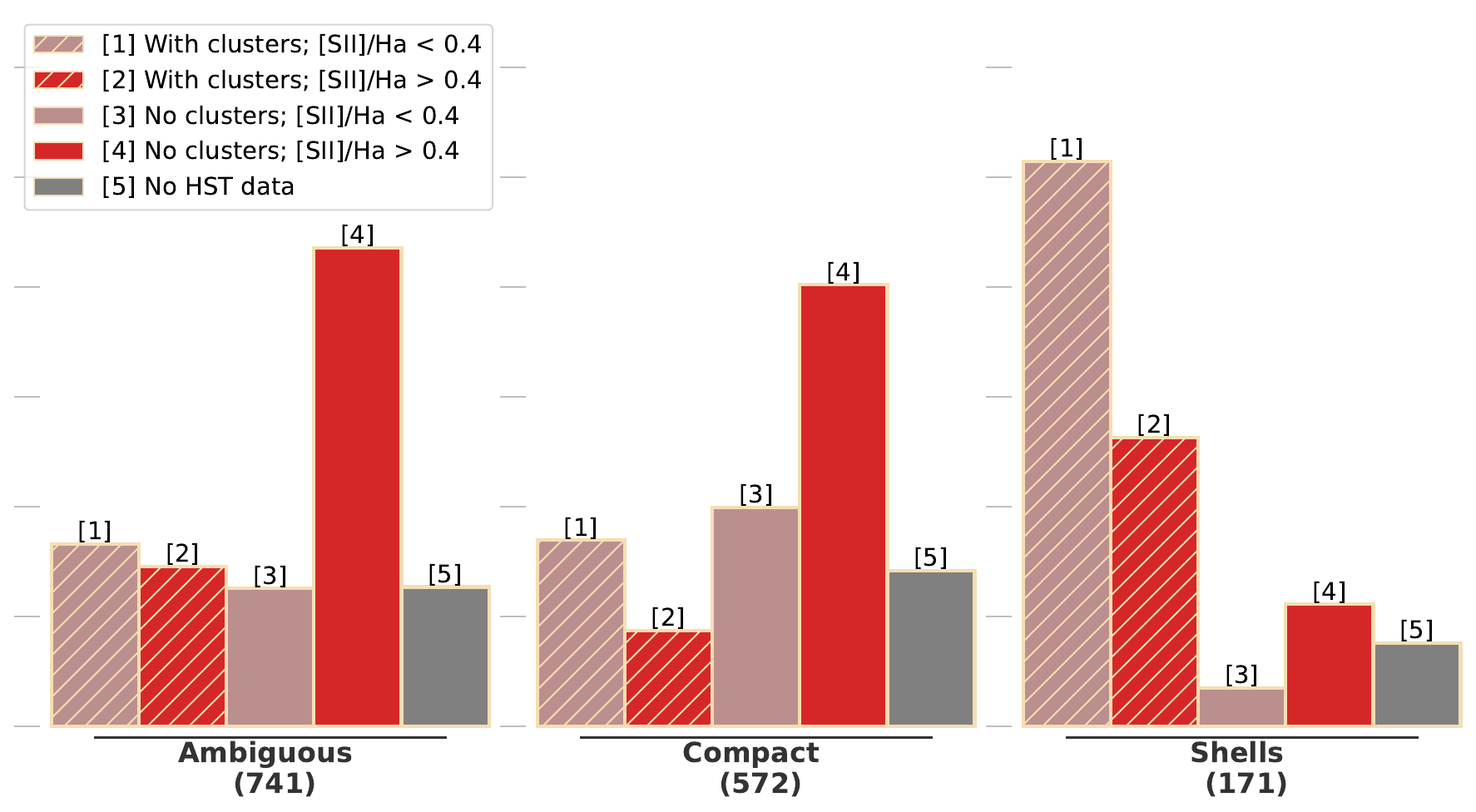}
    \caption{Relative distribution of the number of regions with different morphological types, ionisation states and associated young stellar associations. Three groups are shown, corresponding to the three morphological types (see text in Sec.~
    \ref{sec:regions_morphology}), and their sizes correlate with the relative number of corresponding regions. Different colours correspond to 
    select properties: red and orange colours correspond to regions with \SIIHa~$>0.4$, while blue and green -- to those with \SIIHa~$<0.4$; regions that have at least one stellar association within their borders are shown by blue and orange colours, while green and red correspond to those without identified young stars). Violet colour corresponds to the regions outside the \HST\ footprint (and thus without information about star clusters there). 
    Grey lines mark each 10\% level.}
    \label{fig:reg_stat}
\end{figure*}

Assuming that the measured elevated \Ha\ velocity dispersion relative to the mean value in a galaxy (or in the immediate surroundings in the more general case) is a proxy for turbulent motions or is associated with gas expansion, we can estimate the kinetic energy of these supersonic motions in the ionised gas as: 
 \begin{equation}
 \label{eq:ekin}
     E_{\rm kin} \simeq E_{\rm turb} \simeq \frac{3}{2} M_{\rm ion} (\sigma(\mathrm{H}\alpha)^2 - \sigma(\mathrm{H}\alpha)_m^2).
 \end{equation}

 To estimate the energy input from massive stars, we rely on the parameters of the stellar associations,  described in Sec.~\ref{sec:stars_identifications}. The age and total mass of the stars were taken from the PHANGS-HST catalogue of stellar associations (see Sec.~\ref{sec:obs:hst}). Then we used models produced by \textsc{starburst99} (v.7.0.1, \citealt{Leitherer1999, Leitherer2014}) to calculate the cumulative mechanical energy produced by the winds and supernovae from a star cluster with parameters (mass, age, metallicity) matched to each of the selected stellar associations. Namely, we calculated a grid of models assuming instantaneous star formation, a Kroupa \citep{Kroupa2001} initial mass function (IMF)\footnote{Note that the parameters of the star clusters have been determined using a Chabrier IMF \citep{Chabrier2003}. This is very similar to the Kroupa IMF, but there are small differences in their mass-to-light ratio (by a factor of $\sim 1.08$ according to \citealt{Madau2014}). Therefore, we may expect a negligible systematic offset ($\log(\Delta M) \sim 0.03$~dex) between the measured star cluster masses and those in the \textsc{starburst99} models.} and Geneva stellar evolution tracks which include rotation with two metallicities ($Z=0.014 \sim Z_\odot$, \citealt{Ekstrom2012}, and $Z=0.002 \sim 0.15 Z_\odot$, \citealt{Georgy2013}), for clusters with total stellar mass $\log(M_*/\mathrm{M}_\odot)=(3.0...7.0)$ with steps of 0.25 dex and age varying from 0.2 to 60~Myr with steps of 0.2~Myr. The output parameters from this grid are the total mechanical energies produced by both stellar winds and supernovae ($E_{\rm WSN}$) and by winds only ($E_{\rm W}$) during the whole lifetime of a cluster. The evolution of these cumulative energies is dependent on the input parameters and is shown in Fig.~\ref{fig:sb99} for several models as examples. All the regions analysed in this paper have metallicity significantly higher than the value used for the low-metallicity grid, and only 86 regions (4\%) have $\mathrm{12+\log(O/H) < 8.35}$ ($Z < 0.45 Z_\odot$). Therefore we use a grid of models computed for solar metallicity to evaluate the energy input for all the clusters. Given the differences in the energy input between the high- and low-metallicity models (Fig.~\ref{fig:sb99}), we can expect \revone{an overestimation} of the cumulative $E_\mathrm{mech}$ for some of the young clusters by a factor of few. The results do not change significantly if using the interpolated values between these two available grids.   
 The exact values of $E_{\rm WSN}$ and $E_{\rm W}$ were obtained by interpolating  the model values to the measured mass and age of each association. As we explained in the previous Section, for further analysis we take into account only values of $E_{\rm WSN}$ and $E_{\rm W}$ injected by star clusters during the last 10~Myr of their evolution. Accounting for the total energy does not significantly change the results as the majority (77\%) of the star clusters in our analysis are younger than 10~Myr.

\section{Properties of the ionised gas in  regions with high velocity dispersion}
\label{sec:statistics}

Following the analysis described in the previous section, we selected \nregs\, regions of locally elevated velocity dispersion from all 19 galaxies. Here we consider the properties of these regions and classify them according to their most likely powering mechanism.

\subsection{Morphology, ionisation state, and links with  stellar associations}
\label{sec:regions_morphology}

\begin{figure*}
    \centering
    \includegraphics[width=0.85\linewidth]{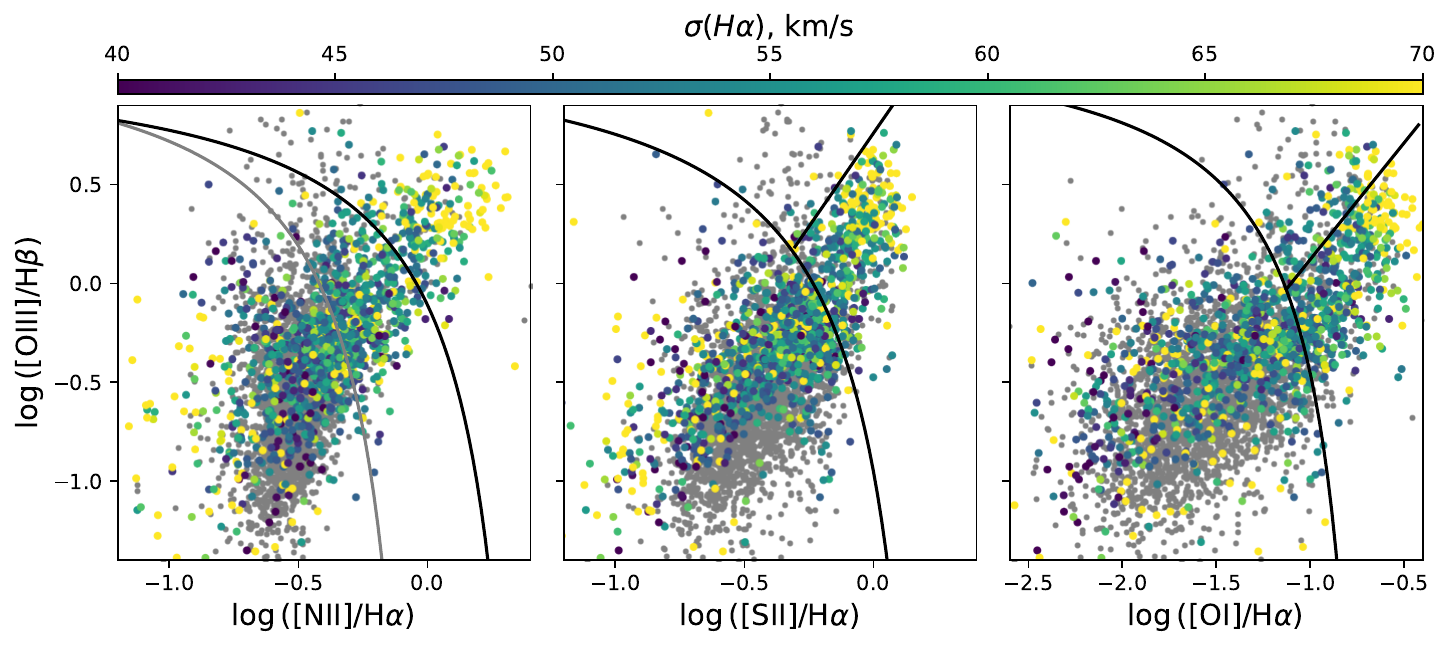}
    \includegraphics[width=0.85\linewidth]{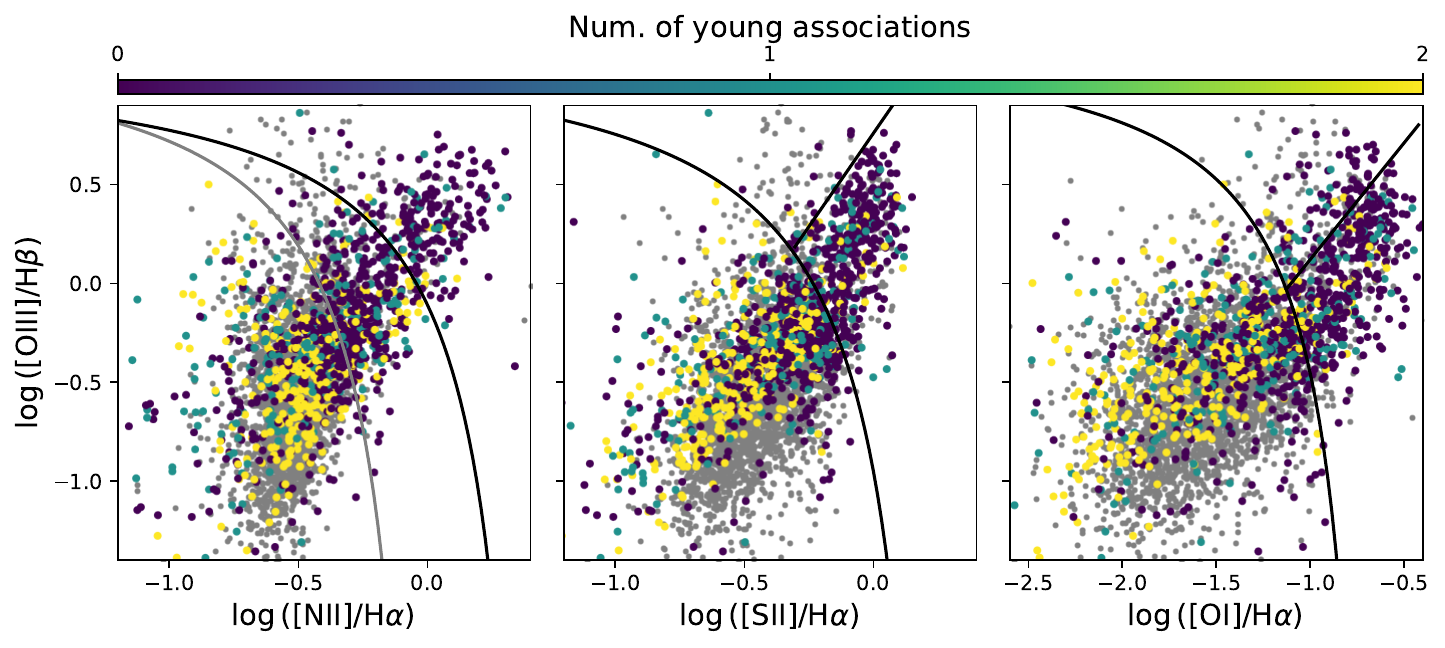}
    \includegraphics[width=0.85\linewidth]{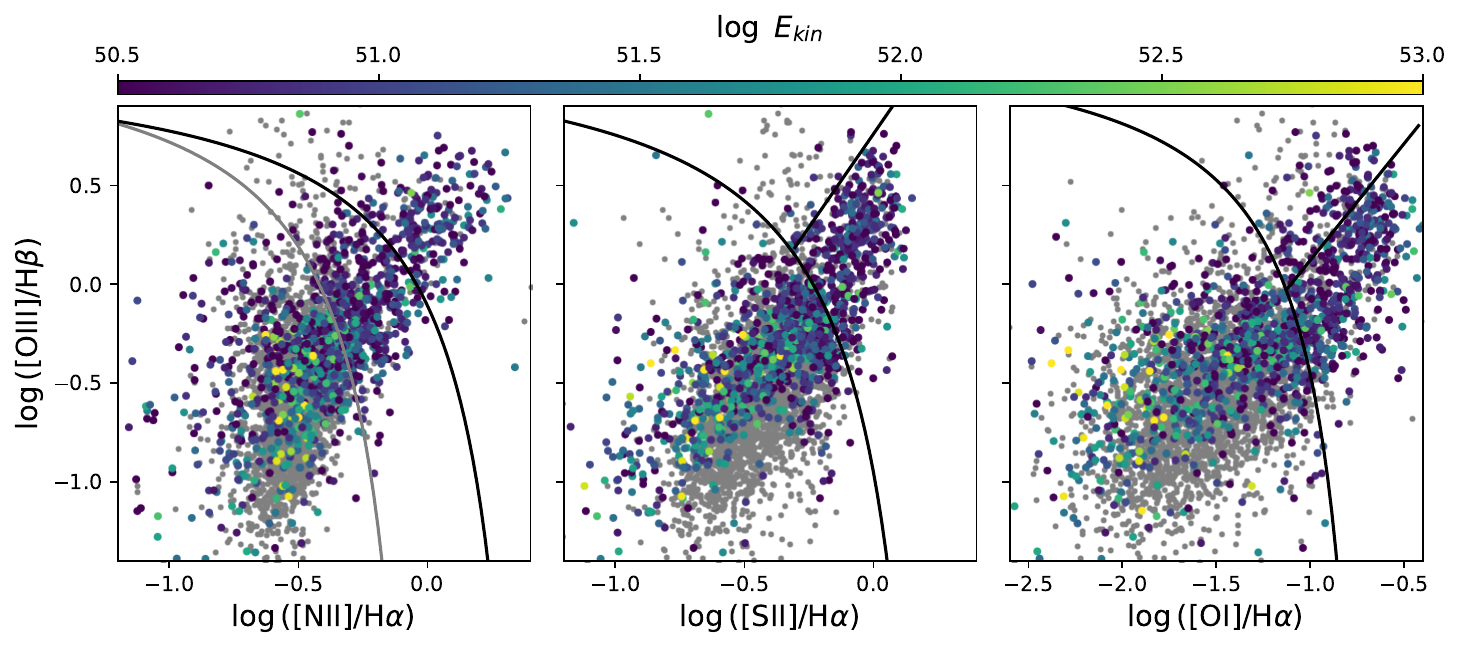}
    \caption{Diagnostic BPT diagrams showing the optical line ratios for all identified regions of locally elevated velocity dispersion (colour-coded circles). Colour denotes either \Ha\ velocity dispersion (top panels), number of young stellar associations and star clusters within the region (middle panels) or estimated kinetic energy of the gas (bottom panels). For reference, we show as grey circles nebulae from the PHANGS-MUSE catalogue (including \HII\ regions, supernovae remnants, planetary nebulae; see \citealt{Groves2023}) excluding those falling in the masked areas (see Sec.~\ref{sec:isigma}). Demarcation lines separate areas on the diagrams typically occupied by regions with different excitation mechanisms: the black solid curve from \citet{Kewley2001} separates \HII\ regions from those with a dominant contribution from other excitation mechanisms (e.g. DIG, shocks); composite mechanism of excitation is probable in the area between this and the grey curve from \citet{Kauffmann2003}; AGNs usually occupy the area above the black straight line from \citet{Kewley2006}.}
    \label{fig:bpt}
\end{figure*}

Based on the results of our automated technique for refining the size of each region (by searching for the peak in the radial \Ha\ flux distribution, see Appendix.~\ref{app:isigma}) we can divide the whole sample into three categories. In Fig.~\ref{fig:regions_example} we show three regions exemplifying each of these classes. The first are classified as probable expanding superbubbles, as their radial \Ha\ flux distribution shows weak flux in the centre increasing to a peak at the edges, corresponding to a shell-like morphology. We identified \nshells\ such regions. Many regions (\ncentralpeak) show a central peak in the \Ha\ flux distribution and thus are classified as compact objects, with both \Ha\ flux and velocity dispersion elevated in their centres. This can be due to the presence of a small unresolved expanding bubble, and in particular of SNRs. Finally, those regions without identified peaks in their radial \Ha\ flux distribution are classified as ambiguous. They still could be superbubbles where the rim is not clearly identified, or they could also represent regions of more complex morphology with elevated velocity dispersion due to locally induced turbulence by stellar feedback or gravity. Note that sometimes superbubbles are more easily identifiable by their kinematics than by their morphology, especially in an inhomogeneous environment \citep[e.g.][]{Naze2001}.

In Fig.~\ref{fig:reg_stat} we show the statistical distribution of these three classes of regions, considering their link with the young stellar associations and the \SIIHa\, line ratio, which is a common tracer for the presence of shocks \citep[e.g.][]{Allen2008}. We distinguish here the number of regions having \SIIHa\, \revone{above or below} 0.4, which is often used to separate supernovae remnants from photoionised nebulae \citep{Blair1981, Blair2004}. 

We identified \nregswithcls\ regions (less than half of the whole sample) with locally elevated velocity dispersion that are linked with at least one young stellar association or compact star cluster.  
However, this does not rule out that  stellar feedback is driving the high \sigmaHa\ in the rest of the regions -- 73\% of them have \SIIHa~$>0.4$, making them probable SNR candidates. Moreover, 74\% of all regions with \SIIHa~$>0.4$ and covered by \HST\ observations do not host stellar associations within their borders. 

Compact regions make up 39\% of our sample and most are not linked to any stellar associations. Meanwhile, about 43\% of the compact regions without stellar associations have \SIIHa\ line ratios consistent with photoionisation. We suggest that at least some of these regions could be poorly resolved bubbles around individual massive O stars. Indeed, $\sim86$\% of them have an \Ha\ luminosity $L(\mathrm{Ha}) < 2.5\times10^{37}\ergs$ that can be produced by a single OV5 star (adopting the rate of \revone{production of} hydrogen-ionising photons from \citealt{Martins2005}). The nebulae surrounding such isolated stars would likely be observed as isolated compact \HII\ regions at the $\sim$50pc PHANG-MUSE resolution and contribute to the ionisation balance \citep{Scheuermann2023}, but these single stars will naturally not be detected as stellar associations. The brightest compact regions could originate from multiple stars with a single SNR -- in such systems the line flux ratios might be still dominated by photoionisation. PNe could also be identified as compact regions with low \SIIHa\ and would not be linked with the young stellar associations, although we find that only two such regions overlap with PNe identified in PHANGS-MUSE galaxies \citep{Scheuermann2022}. 

In contrast to the compact and ambiguous group, regions with shell-like morphologies are generally associated with at least one stellar association or star cluster (84\% of the regions covered by \HST\ observations). We consider these regions as our best candidates for expanding superbubbles, and discuss them in Sec.~\ref{sec:disc_superbubbles}.

To determine the excitation state of regions with high \sigmaHa, we plot them on diagnostic BPT diagrams (Fig.~\ref{fig:bpt}) together with other types of nebulae (including \HII\ regions, SNRs and PNe) from the PHANGS-MUSE nebular catalogue (also applying the same environmental masks). 
Our regions tend to be shifted towards the higher ionising radiation domain (right-hand side of the diagrams). 
This is in agreement with the BPT-$\sigma$ relation observed in nearby galaxies of different types -- the \Ha\ velocity dispersion usually increases towards the top-right side of the diagrams, which is typically explained by a growing contribution of shocks to the emission line excitation \citep{Oparin2018, DAgostino2019, LopezCoba2020, Law2021}. Interestingly, as seen in the central row of Fig.~\ref{fig:bpt}, regions without corresponding star clusters preferentially occupy the upper-right part of the BPT diagrams, while regions linked to star clusters are more similar to \HII\ regions. Furthermore, regions without star clusters have (in general) lower kinetic energy, with values close to $10^{51}$~erg,  typical of the energy ejected during a SN explosion. Altogether, these findings support our claim that most of the regions not linked to young stellar associations are SNRs, and thus their high velocity dispersion results from the influence of stellar feedback.

\begin{figure}
    \centering
    \includegraphics[width=\linewidth]{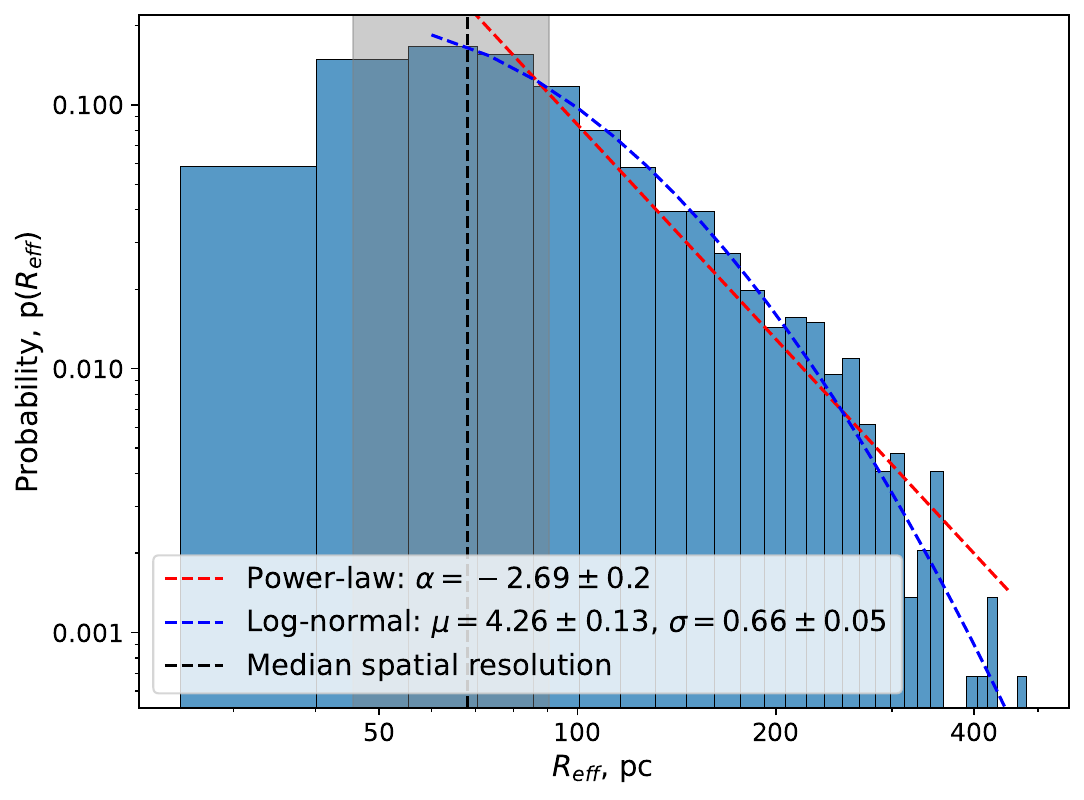}
    \caption{Distribution of sizes for regions of high \Ha\ velocity dispersion. The red line shows a result of a power-law fit to the data for $R_{\rm eff}=70-300$~pc. This fit agrees with a power-law approximation for the sizes of superbubbles according to hydrodynamical simulations ($\alpha = -2.7$ obtained by \citealt{Nath2020}). The distribution of larger and smaller regions is better approximated by log-normal distribution (the blue line). The dashed black line and the grey area show the median spatial resolution and its standard deviation across our sample.} 
    \label{fig:sizes}
\end{figure}

\subsection{Size distribution}
\label{sec:regions_size}

The distribution of the sizes (effective radii, $R_{\rm eff}$) of the identified regions of locally elevated velocity dispersion is shown in Fig.~\ref{fig:sizes}. As can be seen, the sizes of the large structures (above $\sim 70$~pc) are described very well by a power-law: $p(R_{\rm eff}) \propto R_{\rm eff}^{\alpha}$. Using the \textsc{powerlaw} package \citep{powerlaw}, we got a best-fit value of $\alpha = -2.71\pm0.19$. 
The power-law breaks at radii of $R_\mathrm{eff}\sim 60-80$~pc, and also at $R_\mathrm{eff} \gtrsim 300$~pc. We cannot properly resolve smaller structures because of the limited angular resolution. Most ($\sim86$\%) of the smallest regions (the first bin in Fig.~\ref{fig:sizes}) are detected in the four closest galaxies with $D<10$~Mpc. We note that a log-normal distribution describes the size distribution better than a power-law at its small and large ends. 

The measured slope of a power-law approximating our regions in Fig.~\ref{fig:sizes} is very similar to $\alpha = -2.7$, which is derived from hydrodynamical simulations based on the evolution of superbubbles by \cite{Nath2020}, as well as from the size distribution of the \HI\ holes in nearby galaxies \citep[$\alpha\simeq-2.9$;][]{Bagetakos2011, Oey1997}. The observed size distribution is also in good agreement with a power-law (with a slope $\sim -3$) defining the ISM density inhomogeneities due to supersonic turbulence \cite[see][for a review]{Burkhart2021}. 
\citet{Watkins2023} analysed the ALMA CO data for the same sample of galaxies and found a slightly steeper distribution ($\alpha\simeq-3.2\pm0.4$) for the molecular gas superbubbles. Molecular gas is hard to detect in the rims of the larger bubbles, which naturally explains their steeper slope. A shallower distribution ($\alpha\simeq-2.2\pm0.1$) was measured by \citet{Watkins2022jwst} for superbubbles in one galaxy from our sample (NGC~628) visually identified by the morphology of the dust in PHANGS-JWST data \citep{Lee2023}. Future analysis of superbubbles identified in JWST images for other galaxies from our sample will clarify the origin of the differences (if real) in their size distribution in different tracers.

\subsection{Expansion velocities and kinematic age}
\label{sec:tkin}

\begin{figure*}
    \includegraphics[width=0.49\linewidth]{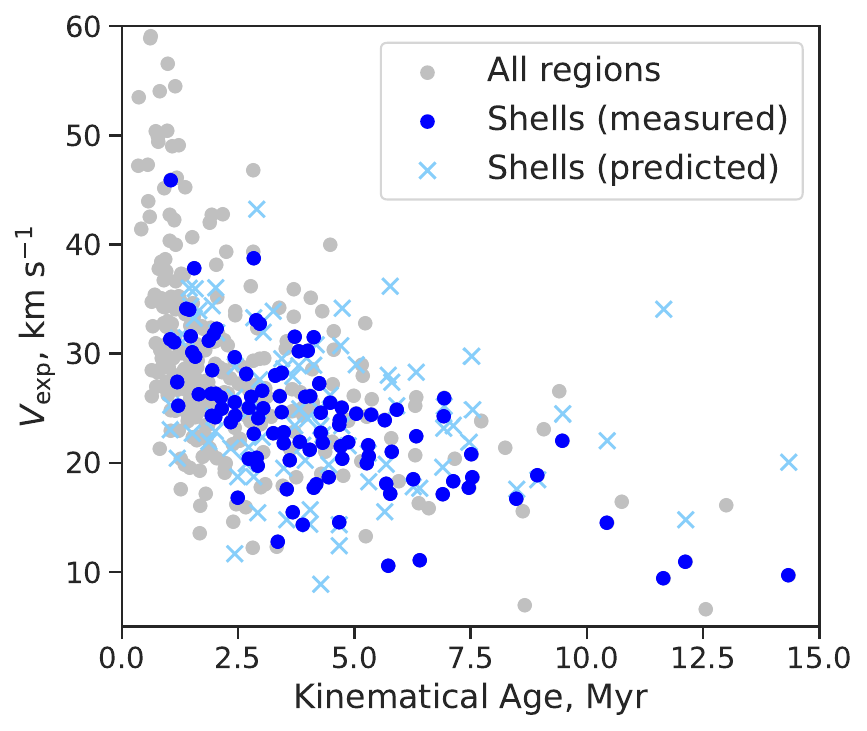}~
    \includegraphics[width=0.46\linewidth]{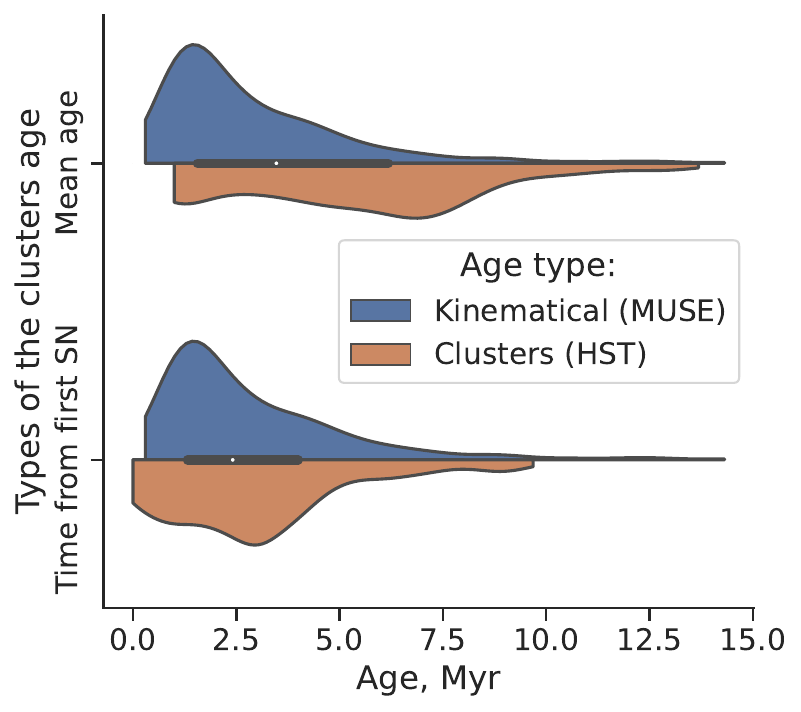}
    \caption{Kinematic properties of the regions with elevated velocity dispersion. Left panel shows the relation between the estimated expansion velocity ($V_\mathrm{exp}$) and kinematic age of each region (grey points). Blue points correspond to the regions classified as `Shells', and light-blue crosses -- to the values of $V_\mathrm{exp}$ for that sub-sample predicted from the analytical model (Eq.~\ref{eq:vexp_theor}) based on their estimated kinematic age and the properties of the star clusters. Top histograms in right panel shows the distribution of the  kinematic age (blue colour) and the mean age of the clusters (weighted by their total mechanical energy input during the last 10~Myr; orange colour). Orange colour on the bottom histograms considers only clusters older than 4~Myr and corresponds to the time passed since the first SN (assuming it is happening at 4~Myr according to Fig.~\ref{fig:sb99}). 
    Only the regions with at least one stellar association or star cluster are shown. }
    \label{fig:tkin}
\end{figure*}

Assuming that the elevated velocity dispersion has resulted from an expanding bubble, we can estimate its expansion velocity ($V_\mathrm{exp}$) and kinematic age ($t_\mathrm{kin}$). At the MUSE spectral resolution, we generally cannot resolve the \Ha\ line profile into kinematically distinct components corresponding to the approaching and receding sides (see, however, one example in Fig.~\ref{fig:data_example}), and thus direct measurements of the expansion velocity are not possible. However, we can use the information about spatial variation of \sigmaHa\ in order to estimate the corresponding $V_\mathrm{exp}$ following a recipe from \cite{SP2021} (their eqs. 7--10). For a Gaussian line profile and the MUSE spectral resolution, we obtain
\begin{equation}
    V_\mathrm{exp} \simeq 3.15 \times (\sigma(\mathrm{H\alpha})^2-\sigma(\mathrm{H\alpha})_\mathrm{m}^2)^{0.4} - 4.1 \kms,
    \label{eq:vexp}
\end{equation}
where the exact coefficients depend on the value of $\sigma(\mathrm{H\alpha})_\mathrm{m}$ (unperturbed velocity dispersion in the galaxy) and \revone{are parameterised} in that paper\footnote{The parameterisation was obtained relying on mock spectra of toy-models of expanding bubbles in the turbulent ISM with properties representative of those found in PHANGS-MUSE galaxies. The coefficients in Eq.\ref{eq:vexp} correspond to the median $\sigma(\mathrm{H\alpha})_\mathrm{m}$ in our sample.}, and \sigmaHa\ is measured from the integrated spectrum from the entire region. 

In the classical analytical solution by \citet{Weaver1977}, the kinematic age of a bubble with radius $R$ expanding in a homogeneous medium can be derived as $t_\mathrm{kin} \simeq 0.6R/V_\mathrm{exp}$ when the bubble is in the adiabatic stage of its evolution. Fig.~\ref{fig:tkin} (left) shows that the expansion \revone{velocities}, $V_\mathrm{exp}$, of the identified regions with elevated velocity dispersion decline with their kinematic age. That is expected given that the bubbles slow down once they have swept up sufficient mass during their evolution. Equation~\ref{eq:vexp} and estimates of $t_\mathrm{kin}$ are valid under the assumption of a spherical thin bubble expanding in a homogeneous medium, which is not necessarily true for all our regions (see Sec.~\ref{sec:regions_morphology}). We note, however, that \revone{in Fig.~\ref{fig:tkin}, all regions follow the same trends as those classified as `Shells'}.

The right panel of Fig.~\ref{fig:tkin} compares the \revone{distributions} of the derived kinematic age (blue histogram) and mean age of the young star clusters associated with each region, weighted by their total mechanical energy input over the last 10~Myr (orange histogram on the top plot). This plot demonstrates that the typical kinematic age of the observed supersonic ionised gas motions is generally lower than the mean age of the star clusters and stellar associations residing within the considered regions. This discrepancy can be understood if the development of these ionised gas motions (either turbulent or superbubble expansion) starts due to the influence of pre-SN feedback, but their main driver is energy and momentum from SN explosions. Therefore we expect that the kinematic age should correlate better with the time passed since the first SN explosion for older regions, or with the mean age of the clusters within the region if SNe have not exploded yet and the supersonic gas motions are mostly driven by pre-SN feedback. Thus, the bottom plot of Fig.~\ref{fig:tkin} (right) considers only clusters older than 4~Myr and shows their modified mean age $t_\mathrm{cls}^\mathrm{SN} = t_\mathrm{cls} - 4$~Myr (the threshold of 4~Myr approximately corresponds to the time when the first SN occurs according to the \textsc{starburst99} models shown in Fig.~\ref{fig:sb99}). We find that the estimated kinematic ages are in better agreement with such modified \revone{ages} of the clusters tracing the time passed from the first SN event.

To verify that the measured values of the expansion velocities are consistent with the properties of the young stellar population, we derive $V_{\rm exp}$ at a given $t_{\rm kin}$ from an analytical solution of a superbubble driven by multiple winds and SNe \citep{MacLow1988}: 
\begin{equation}
    V_\mathrm{exp}(t) \simeq 38.6 \left(\frac{L_{38}}{n_0}\right)^{1/5} t_6^{-2/5} \kms,
    \label{eq:vexp_theor}
\end{equation}
where $L_{38} = L_\mathrm{mech}/10^{38} \ergs$ is the mechanical luminosity of the stellar association, $n_0$ is the ambient density, and $t_6$ is $t_{\rm kin}$ expressed in Myr. Considering $n_\mathrm{e}^\mathrm{(min)}$ as a proxy of density and $L_\mathrm{mech} \simeq E_\mathrm{mech}^\mathrm{SN}/t_\mathrm{cls}^\mathrm{SN}$, we predict $V_\mathrm{exp}$ for `Shells' similar to what is observed.

While the velocity dispersion (and hence the derived expansion velocities) of the regions in this study is higher than it is in normal \HII\ regions (see Sec.~\ref{sec:compare_to_HII}), it does not vary  significantly between the regions and results in $V_{\rm exp} \sim 10-60 \kms$ (mostly within $20-35\kms$). Thus, the large spread of $E_{\mathrm{kin}}$ for these regions is driven mostly by the differences in their gas mass, not in their kinematics. We also note that the derived kinematic age is typically below 10~Myr, which supports our assumption made in Sec.~\ref{sec:stars_identifications} that only energy input from the star clusters during the last 10~Myr is relevant for establishing the energy balance between ionised gas and stars.

\subsection{Kinetic energy of the ionised gas and its dependence on metallicity}
\label{sec:regions_metallicity}

\begin{figure*}
    \centering
    \includegraphics[width=0.5\linewidth]{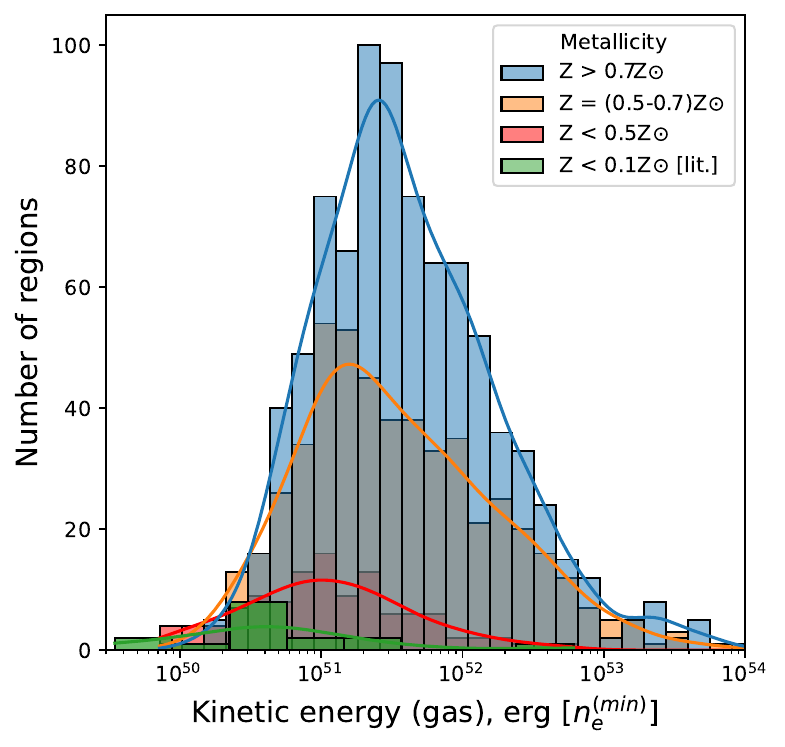}
    \includegraphics[width=0.48\linewidth]{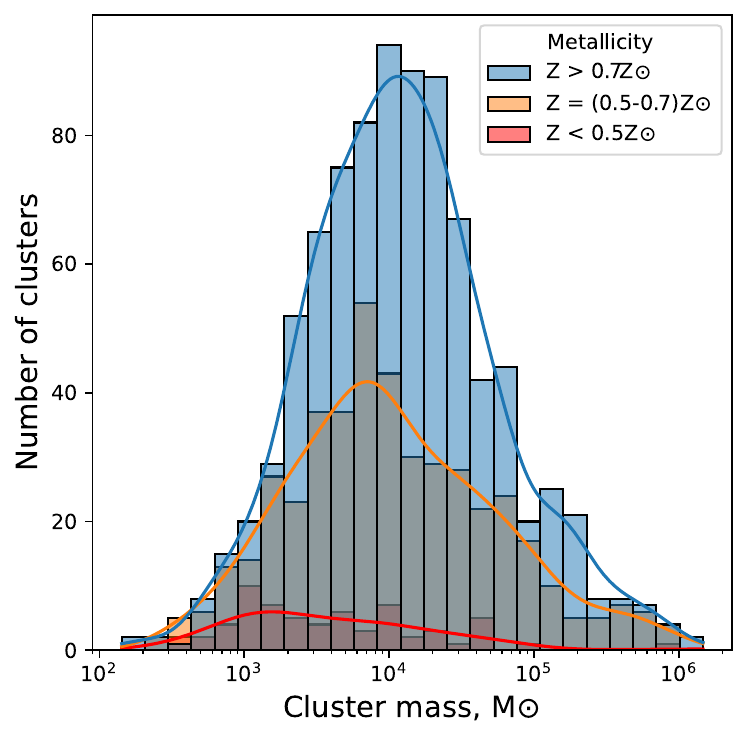}
    \caption{Distribution of the measured kinetic energy of the ionised gas in the identified regions of locally elevated \Ha\ velocity dispersion (left panel) and the masses of the star clusters in these regions (right panel) shown for different metallicity cuts. These regions tend to have lower kinetic energy in the low-metallicity environment, but also the typical masses of clusters decline. The bin for $Z < 0.1 Z_\odot$ corresponds to the estimates of kinetic energy for 16 ionised gas superbubbles identified in the dwarf galaxies Sextans~A \citep{Gerasimov2022} and DDO~53 \citep{Egorov2021} and is shown here for comparison.}
    \label{fig:energies_met}
\end{figure*}

\begin{figure*}
    \centering
    \includegraphics[width=\linewidth]{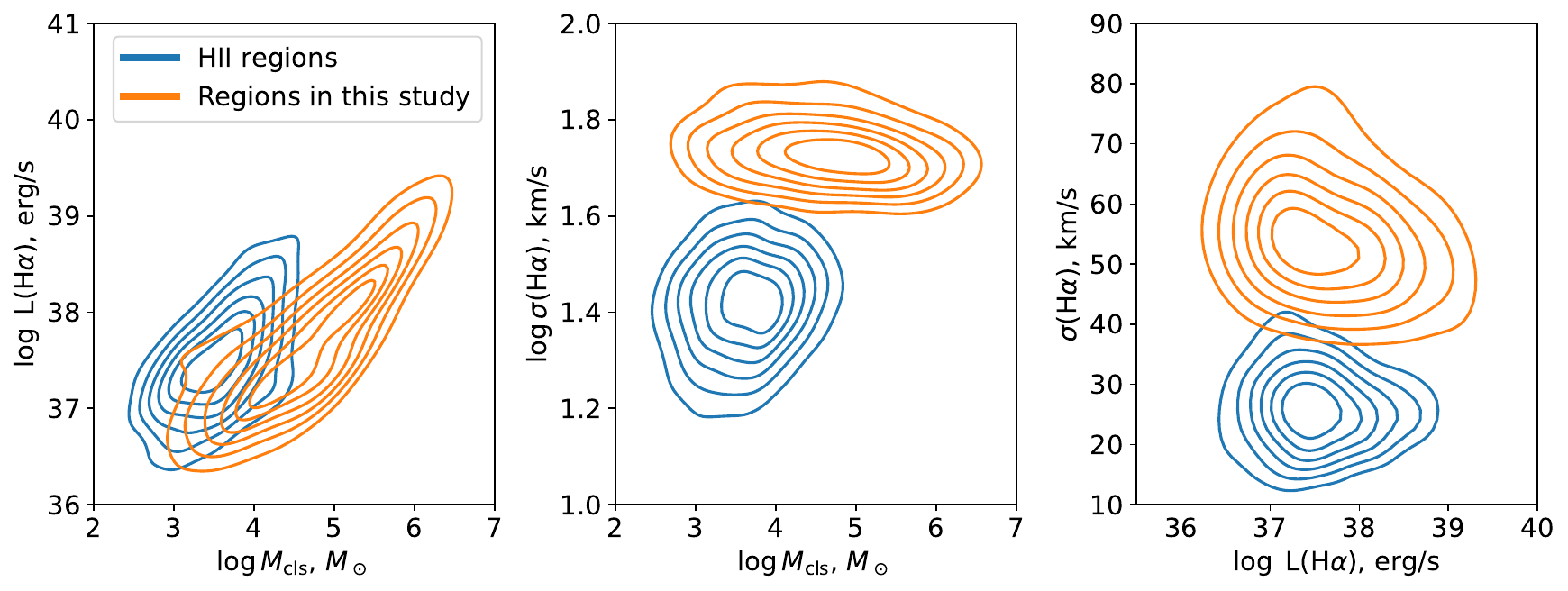}
    \caption{Properties of the regions of elevated velocity dispersion in comparison to those of normal \HII\ regions. Blue and orange contours show the statistical density distribution of the \HII\ regions (associated with \revone{a single star cluster, from} the catalogue compiled by \citealt{Scheuermann2023}) and the regions with locally elevated \sigmaHa, respectively. The \Ha\ luminosity ($L$(\Ha)) correlates with the total mass of the clusters ($M_\mathrm{cls}$) for both samples (left panel), but the slope is different implying differences in the escape fraction of the ionising photons and/or contribution of the additional ionising sources. \sigmaHa\ for the regions in this study slightly decreases with the $M_\mathrm{cls}$ (middle panel) and $L$(\Ha) (right panel) contrary to what is observed in the reference \HII\ regions.} 
    \label{fig:compare_to_HII}
\end{figure*}

The estimated kinetic energy $E_{\rm kin}$ of the ionised gas in the identified regions spans about three orders of magnitude -- from $\sim5\times10^{50}$~erg to $\sim5\times10^{53}$~erg. We consider how the measured values of $E_{\rm kin}$ change with the metallicity of a region in Fig.~\ref{fig:energies_met} (left panel). Regions with lower oxygen abundance tend to have lower kinetic energy, though the number of regions with $Z<0.5Z_\odot$ is rather low for drawing definitive conclusions. Note, however, that measurements of energetics of low-metallicity ($Z<0.1Z_\odot$) superbubbles available in the literature \citep{Egorov2021, Gerasimov2022} follow the same trend as in our data. 

Such a metallicity trend can be a consequence of either a lower mechanical energy input injected by young stars into the ISM, or higher energy losses, or it can result from differences in the distribution of cluster masses in low- and high-metallicity regions. The right panel of Fig.~\ref{fig:energies_met} shows that the latter is \revone{a likely explanation} of what we see in the PHANGS-MUSE data. Indeed, clusters at $Z<0.5Z_\odot$ tend to be less massive than at higher metallicities, which naturally leads to lower mechanical energy input from these clusters. We do not see any noticeable differences in the efficiency of stellar feedback (considering the ratio of $E_{\rm kin}$ and cumulative energy input produced by clusters, see Sec.~\ref{sec:disc_turbulence}), thus there is no evidence of differences in energy losses. We cannot, however, completely rule out the probable effect of a declining mechanical energy input per star cluster.
A decrease of the mechanical energy input from the star clusters at low-metallicity is present in the results of \textsc{starburst99} simulations in Fig.~\ref{fig:sb99}. Theoretical models suggest that the energy contributed by stellar winds \citep{Vink2001} declines in the low-metallicity regime. The observed number of SNe type II in low metallicity environments is also low \citep{Anderson2016}, however the mass range and fraction of stars ending as SNe do not vary significantly with metallicity \citep{Heger2003}. 
Note that the observed trend of $E_{\rm kin}$ with metallicity in Fig.~\ref{fig:energies_met} is derived from directly measurable values in the MUSE spectra only, it does not depend on any model assumptions (except the assumption of spherical geometry), and thus provides a purely observational indication for variations in mechanical stellar feedback as a function of metallicity.

\subsection{Ionised gas properties compared to normal \HII\ regions}
\label{sec:compare_to_HII}

The regions analysed in the current study are selected by their locally elevated \Ha\ velocity dispersion almost independently of their morphology, so they do not necessarily have the same properties as typical \HII\ regions. In Fig.~\ref{fig:compare_to_HII} we \revone{examine} some of the relations between the \Ha\ luminosity, velocity dispersion and mass of the associated star \revone{clusters. We compare these relations in the regions} of elevated velocity dispersion with what is found for \HII\ regions. We use a subsample of \HII\ regions \revone{from the PHANGS-MUSE nebulae catalogue, compiled by \citet{Scheuermann2023}, representing regions associated with a single star cluster. We can thus easily} establish the link between the ionised gas and young stars for this reference sample. 

Massive stars in the young stellar associations and compact clusters produce the hydrogen-ionising photons that are responsible for the observed \Ha\ flux from the \revone{ionised} gas surrounding the clusters. Therefore one naturally expects to see a correlation between the mass of the star clusters and the \Ha\ luminosity of the \HII\ regions. This is however not necessarily true for our regions of elevated velocity dispersion as they are not classical \HII\ regions. Given their kinematically disturbed nature, shock heating could 
contribute to their observed \Ha\ emission. Also given that these regions are likely superbubbles or turbulent volumes of gas, leakage of ionising photons into and out of the region can also play a role. 
The left panel of Fig.~\ref{fig:compare_to_HII} shows that the \Ha\ luminosity of the regions with high-velocity ionised gas motions still correlates with the total mass of the young stellar associations within the regions. The slope of this correlation is, however, shallower than for bona-fide \HII\ regions. At the same time, these regions span the same range of \Ha\ luminosities as the \HII\ regions. 

By construction, the regions considered in this paper have higher \Ha\ velocity dispersion than normal \HII\ regions. The middle and right panels of Fig.~\ref{fig:compare_to_HII} demonstrate that the distributions of \sigmaHa\ for these objects \revone{have almost no overlap with the} \HII\ region sample. The MUSE spectral resolution is not sufficient to measure the well-known positive correlation between the \Ha\ luminosity or the mass of the ionising star clusters and the \Ha\ velocity dispersion for \HII\ regions \citep[e.g.][]{Terlevich1981, Melnick1999, MoiseevKlypin2015}. However, we see a negative shallow trend for the regions with elevated velocity dispersion -- \sigmaHa\ is, in general, lower for the brighter regions and/or \revone{regions associated} with more massive star clusters. Therefore, from the distributions \revone{shown in } Fig.~\ref{fig:compare_to_HII} we conclude that the observed properties and, in particular, the \revone{correlation between properties of ionised gas} and massive stars slightly differ in these dynamically active regions from what is typically observed in \HII\ regions.

\section{Discussion}
\label{sec:discussion}

\subsection{Velocity dispersion at the location of the young stellar associations.}

\begin{figure*}
    \centering
    \includegraphics[width=0.49\linewidth]{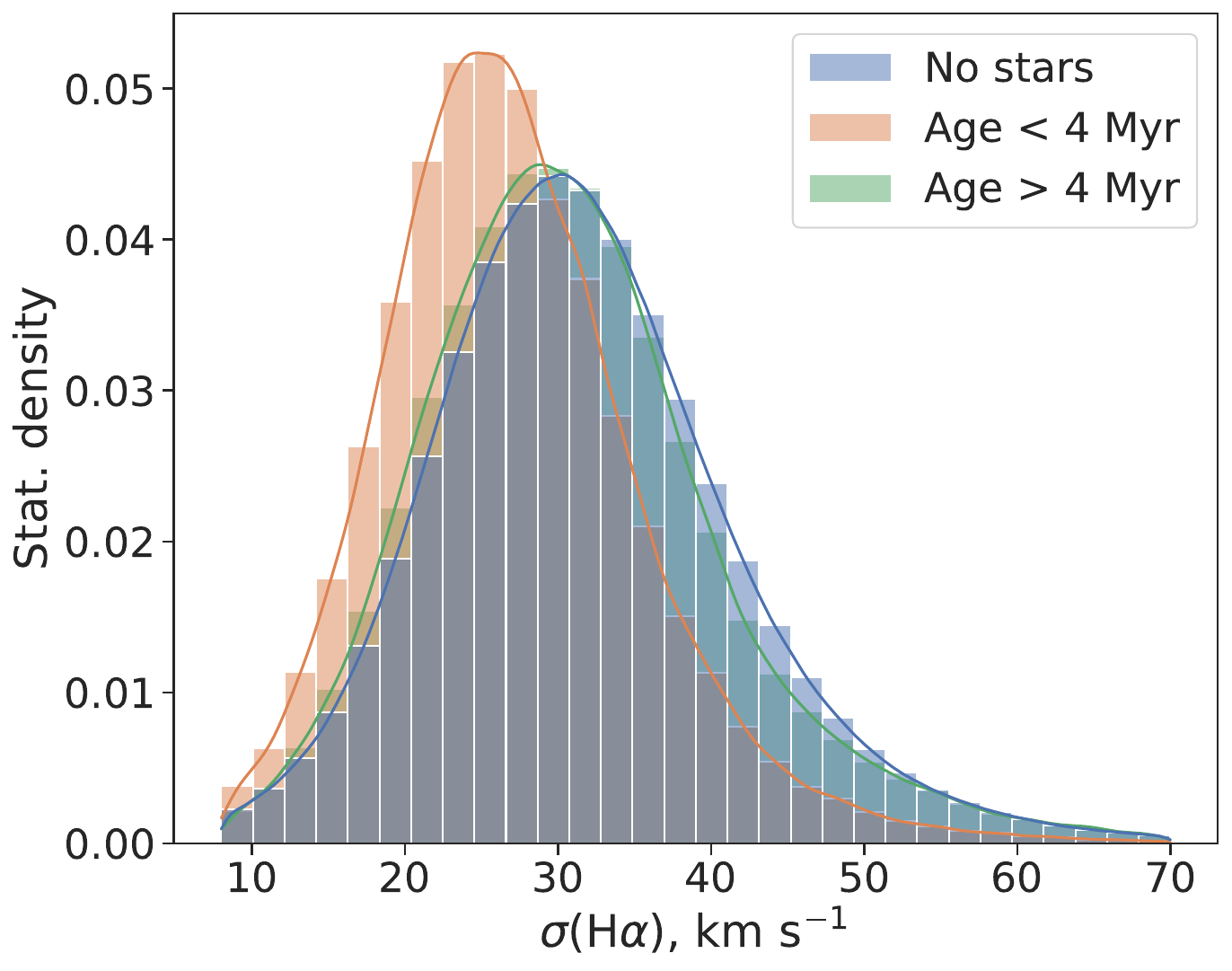}~\includegraphics[width=0.49\linewidth]{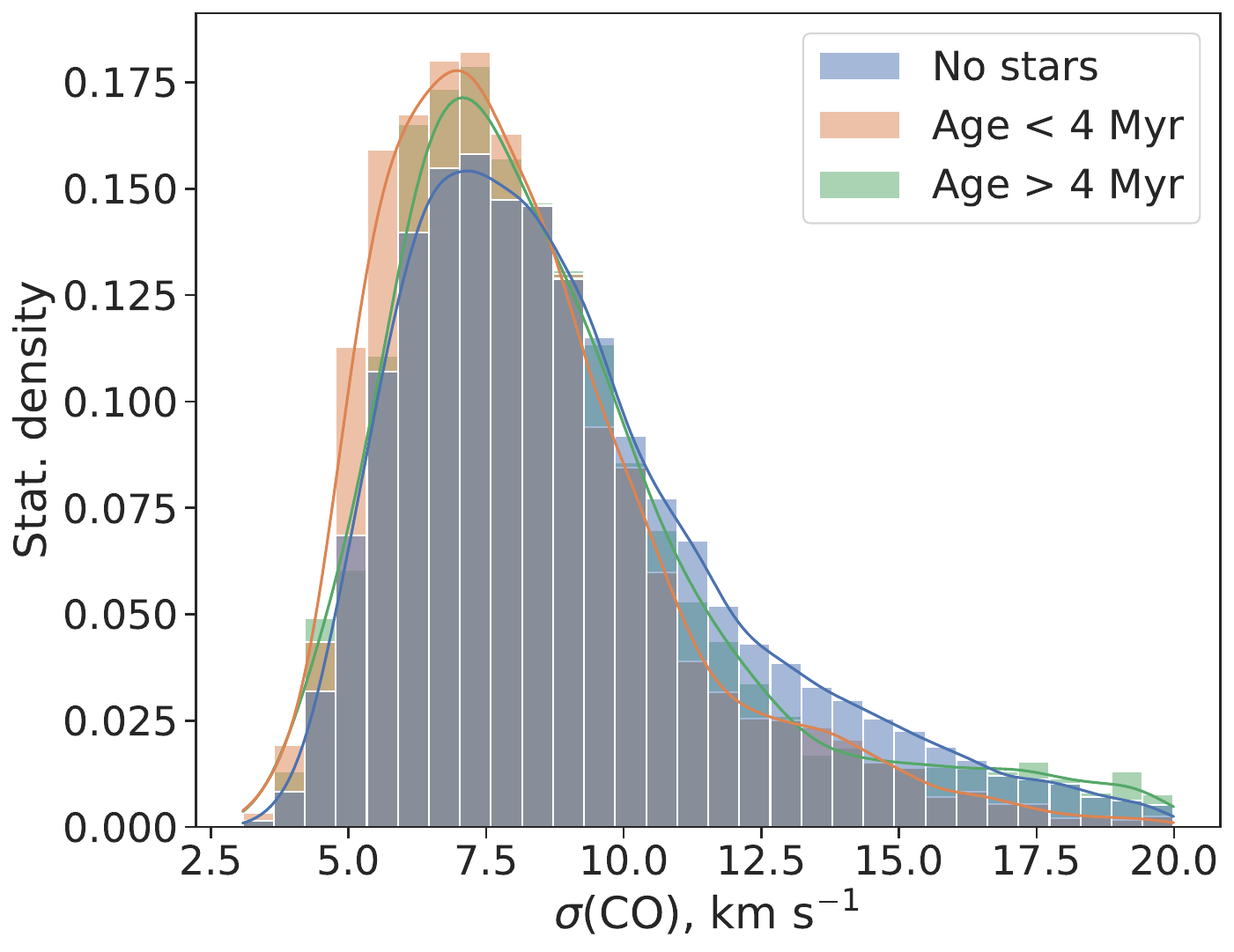}
    \caption{Distribution of the velocity dispersion of \Ha\ (left) and CO (right) emission within the borders of young stellar associations \revone{of} different ages (orange and blue colours for those younger and older than 4~Myr, respectively) and outside them (green colour). Only pixels with $S/N>30$ in the corresponding emission line are considered. Note that increasing the $S/N$ limit does not affect the result. }
    \label{fig:sigma_in_clusters}
\end{figure*}

\begin{figure*}
    \centering
    \includegraphics[width=0.5\linewidth]{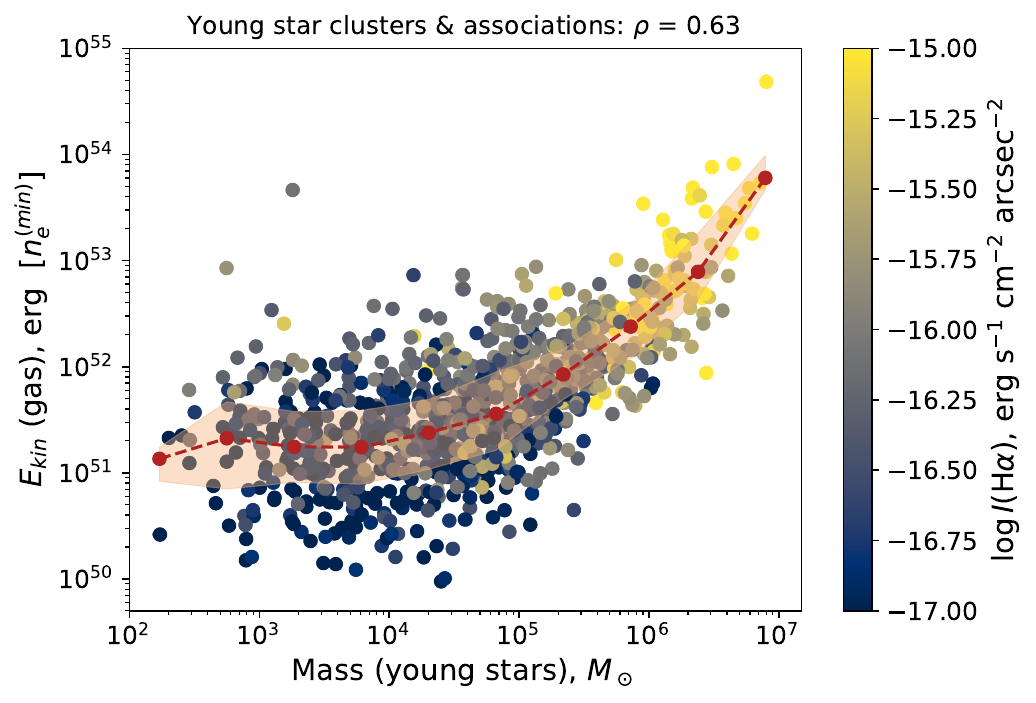}~\includegraphics[width=0.5\linewidth]{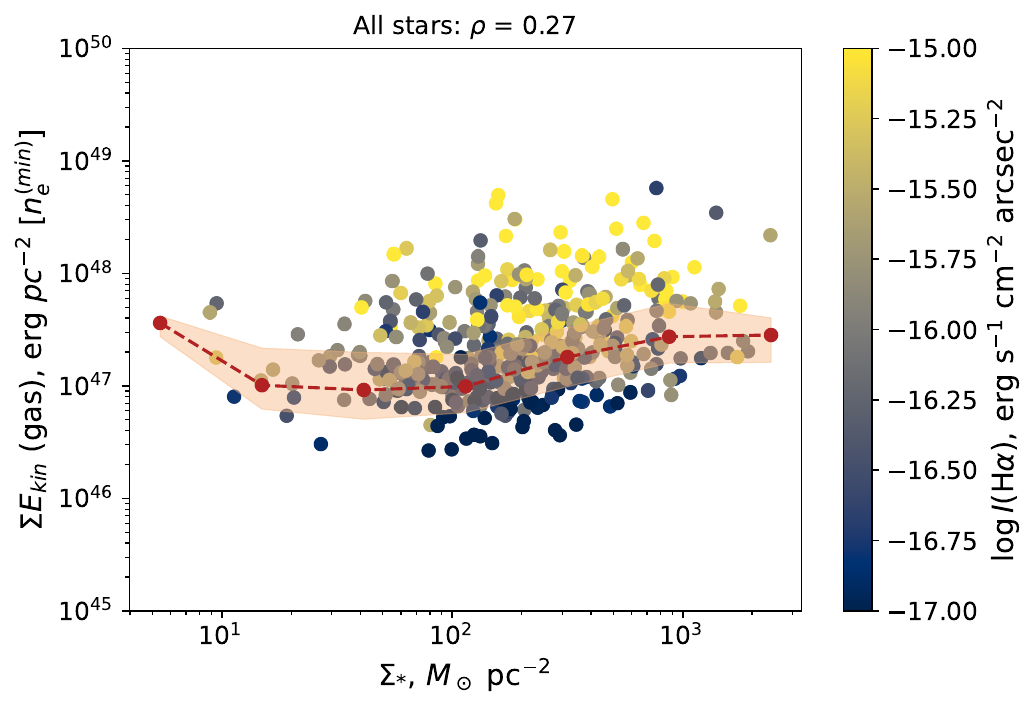}

    \includegraphics[width=0.5\linewidth]{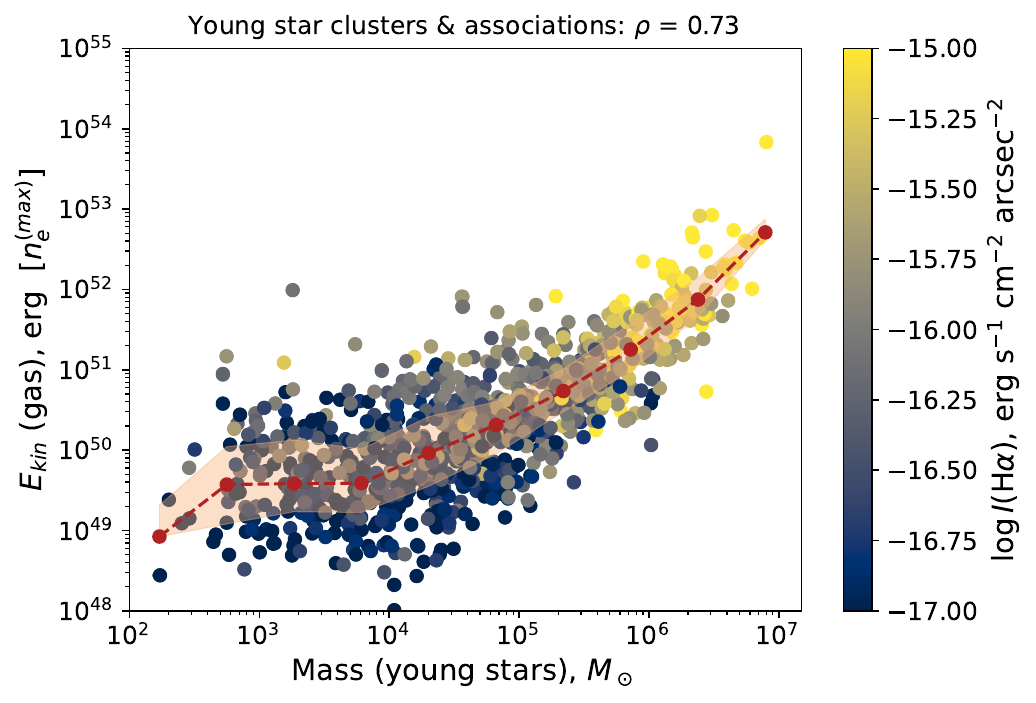}~\includegraphics[width=0.5\linewidth]{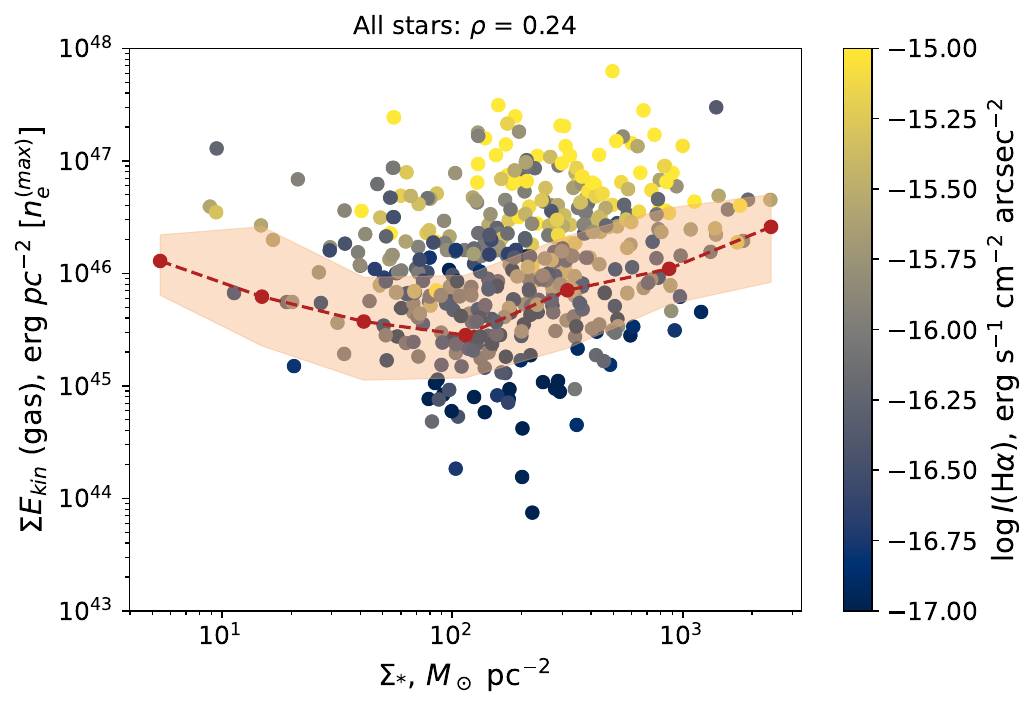}
    \caption{Dependence of the kinetic energy of ionised gas $E_{\rm kin}$ in the regions of locally elevated velocity dispersion on the mass of the stellar associations within them (left panels), and their gas kinetic energy per area, $\Sigma_{E_{\rm kin}}$, vs corresponding stellar mass surface density, $\Sigma_{\star}$ (right panels). Top and bottom panels correspond to minimal and maximal values of $n_{\rm e}$ underlying the $E_{\rm kin}$ measurements.  Colour encodes the \Ha\ flux surface brightness. The red line shows the median values of $E_{\rm kin}$ or $\Sigma_{E_{\rm kin}}$ in several bins, and the red shaded area corresponds to the 25--75 percentile interval. Spearman correlation coefficients $\rho$ are given above the plots.}
    \label{fig:energies_mass}
\end{figure*}

If the dynamics of the ionised gas is strongly affected by stellar feedback, we expect to see the velocity dispersion correlate with the presence of young stars. Using the positions of the young stellar associations (a dominant source of stellar feedback) identified in the \HST\ images, we check if an elevated velocity dispersion towards these stellar associations is observed. Fig.~\ref{fig:sigma_in_clusters} (left-hand panel) shows the distribution of \sigmaHa\ in the individual pixels \revone{residing within the borders of} 
young stellar associations (younger and older than 4~Myr shown by orange and blue colour, respectively) and outside the stellar associations. There is no measurable difference in \sigmaHa\ between the pixels \revone{associated with the young stellar associations older than 4~Myr age and those not overlapping with any stellar associations}. Surprisingly, \sigmaHa\ is systematically lower for stellar associations younger than 4~Myr.
 Applying a t-test to the bootstrapped sample of 1000 MC realizations of these distributions (accounting for the uncertainties of the measurements of \sigmaHa), we find the separation between the two distributions is $5-11 \kms$, and the probability that they are equal is negligible. 
 
It is not obvious why pixels corresponding to the locations of the youngest stellar associations show lower \Ha\ velocity dispersion. If this \revone{results} from lower turbulence towards the young stellar associations, then one would expect to see the same trend in the cold gas too. However, a similar comparison with the PHANGS-ALMA data \citep{Leroy2021}, using the CO line kinematics for the same sample of galaxies (Fig.~\ref{fig:sigma_in_clusters}, right-hand panel), shows no obvious difference\footnote{\revone{Note that the difference in absolute values of \sigmaHa\ and $\rm \sigma(CO)$ is consistent with what measured in other objects \citep[e.g.][]{Girard2021}}}. The elevated \sigmaHa\ towards stellar \revone{associations} older than 4~Myr (and the regions without any young stars) could be related to an increased contribution of the DIG to the total \Ha\ emission. Namely, we suggest that \revone{the \Ha\ emission around all} young stellar associations is tracing the kinematics of the thin star-forming gas disc, \revone{but} the diffuse \Ha\ outside the youngest \revone{of them} is more impacted by turbulence and outflows. A higher mean \sigmaHa\ relative to \HII\ regions is often reported for the DIG \citep[e.g.][]{Moiseev2015, Levy2019, DellaBruna2020, Law2022}, although it remains an open question whether stellar feedback or large-scale processes are responsible for this. The same effect, in principle, could be a consequence of superbubbles \revone{beeing powered by off-centre} stellar associations (see Sec.~\ref{sec:disc_superbubbles}), and thus a large kinematic separation between the approaching and receding walls is observed outside the young stellar associations. This implies that the influence of stellar feedback on the ionised gas kinematics can be observed better at some separation from the source of feedback.

\subsection{What drives the turbulent motions in star-forming galaxies?}
\label{sec:disc_turbulence}

\begin{figure*}
    \centering
    \includegraphics[width=0.49\linewidth]{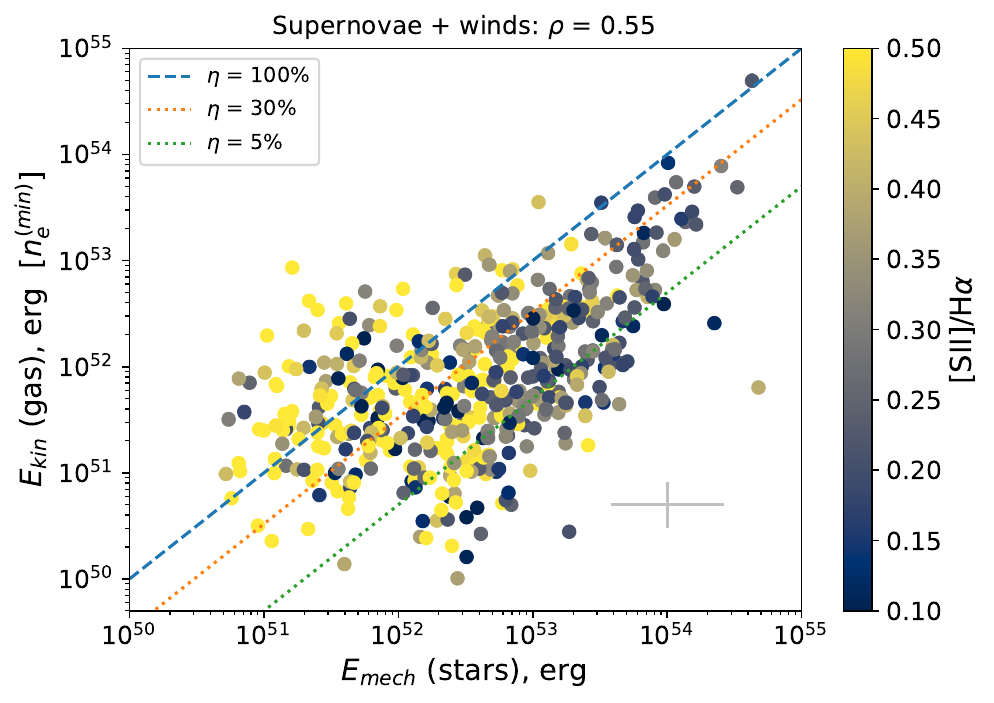}
    \includegraphics[width=0.49\linewidth]{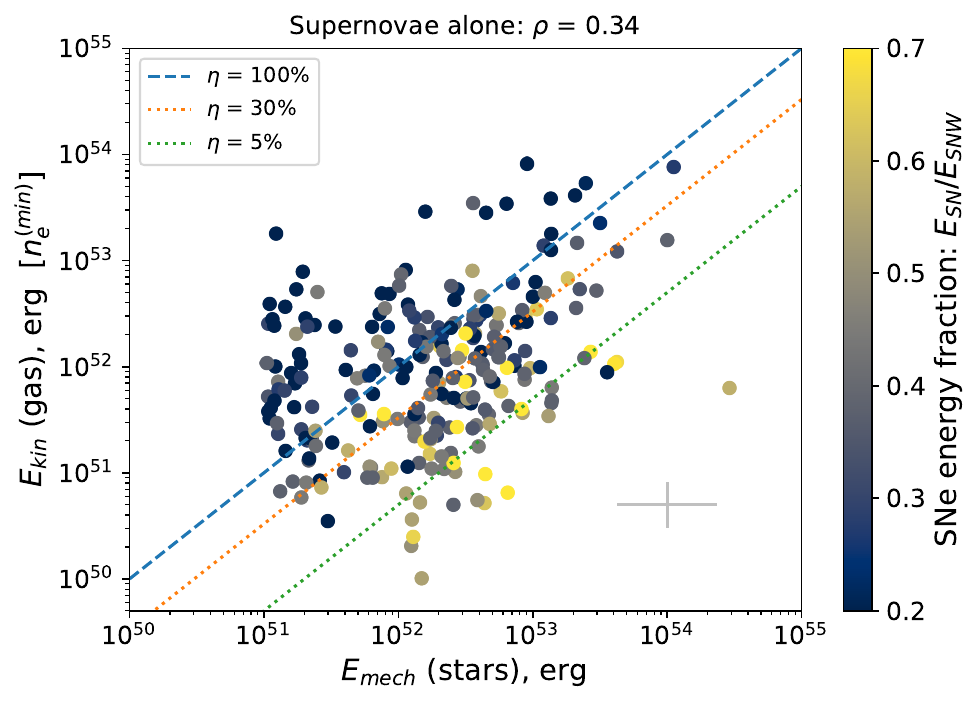}

    \includegraphics[width=0.49\linewidth]{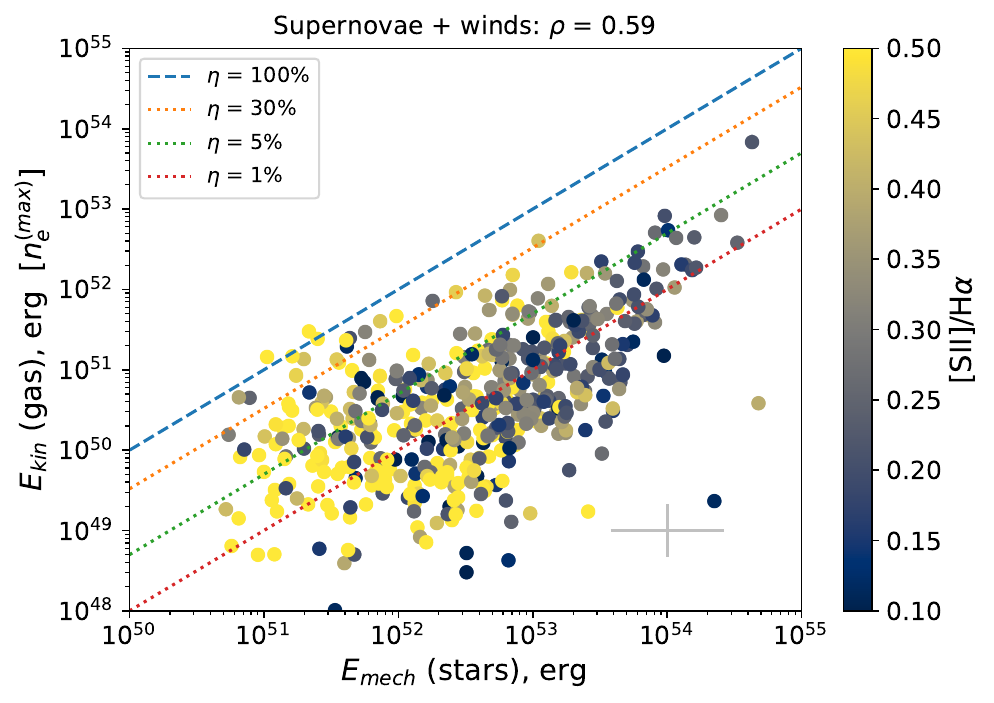}
    \includegraphics[width=0.49\linewidth]{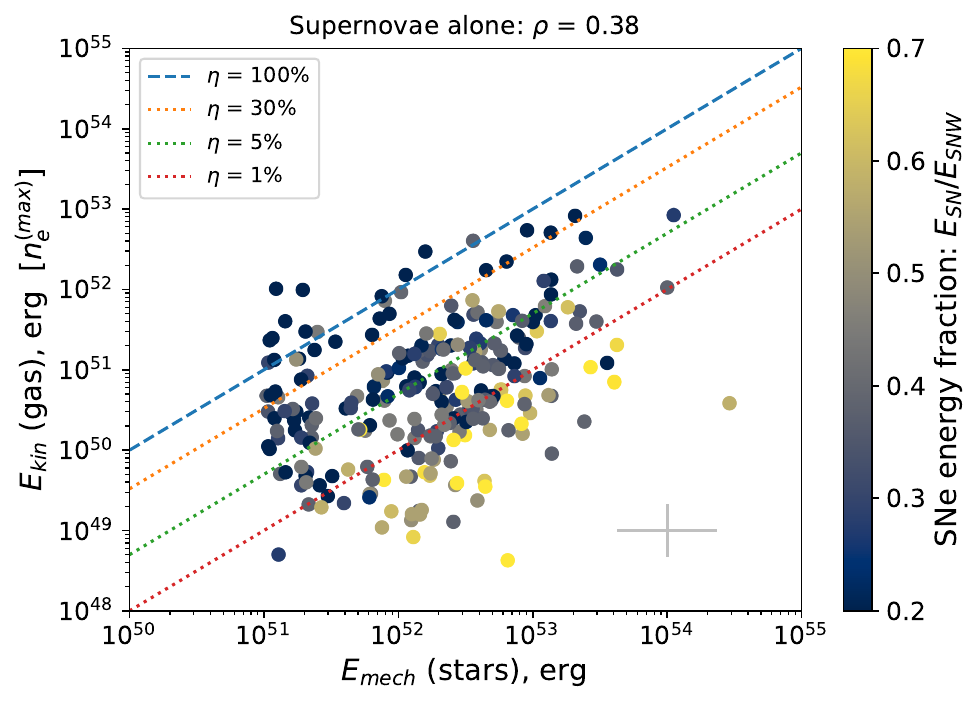}
    \caption{Dependence of the kinetic energy of ionised gas $E_{kin}$ in the regions of locally elevated velocity dispersion on the total mechanical energy input from the stellar associations in the form of supernovae and stellar winds (left panels), and supernovae \revone{alone} (right panels). Top and bottom panels correspond to the $n_e^{(min)}$ and $n_e^{(max)}$ density measurements (thus representing upper and lower limits of $E_{kin}$, respectively). Median value of logarithmic errors is shown in bottom-right corner of each panel. All regions with errors 2 times larger than presented are excluded from the plots. Colour encodes the \SIIHa\ lines ratio tracing the dominant gas excitation mechanism (\SIIHa$>0.4$ is likely produced by shocks; left panels), or the relative contribution of the SNe to the total mechanical energy input according to the \textsc{starburst99} models for the corresponding mass and age of each cluster \revone{right panels}. Blue, orange, green and red lines correspond to mechanical stellar feedback energy efficiency $\eta = 100, 30, 5, 1$\%, respectively. Spearman correlation coefficient $\rho$ is given above the plots.}
    \label{fig:energies}
\end{figure*}

As discussed in Section~\ref{sec:intro}, several processes have been proposed as drivers of high velocity dispersion: stellar feedback, gravitational instabilities, and external processes. In this paper we identified \nregs\, regions with locally enhanced \Ha\ velocity dispersion, and \revone{linked \nregswithcls\, of them} to young stellar associations or compact star clusters. Comparing the measured properties of the ionised gas with the modelled energy input from the stellar population (see Sec.~\ref{sec:physical_parameters}), we can determine which of these mechanisms (winds and SNe or gravity) is driving the observed supersonic gas motions.

We compare the measured kinetic energy $E_{\rm kin}$ of the ionised gas with the total mass of stars in the young stellar associations within the regions (Fig.~\ref{fig:energies_mass}, left panels) and the surface density of the kinetic energy $\Sigma E_{\rm kin}$ with the mean \revone{total stellar mass} surface density $\Sigma_\star$ (right-hand panels). The latter is taken from the PHANGS-MUSE \textsc{dap} maps, and measured from the local stellar continuum spectrum (see Sec.~\ref{sec:obs:muse}) \revone{and averaged over the area corresponding to each region}. The measured $\Sigma E_{\rm kin}$ is largely independent of $\Sigma_{\star}$, implying that gravitation alone\footnote{\revone{Where it was possible, we estimated the local gas mass fraction using the ALMA CO observations from \citet{Leroy2021}. Accounting for molecular and ionised gas, we found the median value of $\sim0.23$ implying that the additional contribution of gas to the total mass surface density will not affect significantly this relation.}} is unlikely to be the dominant powering source of the turbulence. 
In contrast, $E_{\rm kin}$ strongly correlates with the mass of the young stellar associations and clusters, pointing to  stellar feedback playing a critical role. We emphasize, however, that this is valid for discs of galaxies as bars and centres are excluded from our analysis. The result is independent of the adopted method for the $n_{\rm e}$ measurements except that $n_{\rm e}^{\rm (max)}$ leads to a more linear trend.

In Fig.~\ref{fig:energies} we compare the kinetic energy of the ionised gas with the total mechanical energy input produced by massive stars during the lifetime of the association (or the last 10~Myr if the star cluster is older than that) in the form of winds and supernovae (left panels), and supernovae alone (right panels). One can see that in most regions, the cumulative mechanical energy input from young stars is sufficient to explain the observed kinetic energy of the ionised gas. There are, however, $\sim16$\% of the regions residing in the top-left panel above the 1-to-1 line ($\sim4$\% if counting only those farther than the quoted uncertainties) -- in these cases young star clusters and associations do not provide sufficient mechanical energy input to explain the \revone{measured value of $E_{\rm kin}$. 64\% of them have} \SIIHa~$> 0.4$, indicative of shock-heated gas due to possible recent explosion of SNe, so we expect to underestimate the number of star clusters for these regions. Another explanation for these outliers is that we do not account for the pressure-driven expansion due to the presence of hot gas, which might be significant for these small regions \citep{Barnes2021}. Finally, we note that our measurements of $E_{\rm kin}$ are for the lower density limit $n_e^{\rm (min)}$ corresponding to an upper limit of $E_{\rm kin}$. All regions reside below the 1-to-1 line (within the errors) if we assume an upper limit estimate of $n_e = n_e^{\rm (max)}$ instead (bottom-left panel).

The situation for the energy balance changes if we account only for \revone{the SNe contribution} (right panels of Fig.~\ref{fig:energies}). In that case, 1/3 of the considered regions reside above the 1-to-1 relation assuming $n_e^{\rm (min)}$ (13\% if counting only those with higher $E_{\rm kin}$ than the quoted uncertainties) -- the star clusters do not produce enough mechanical energy via SNe alone to explain the observed kinetic energy of the ionised gas in these regions. The colours encode the relative contribution of supernovae to the total mechanical energy input from the star cluster during its lifetime, calculated based on \textsc{starburst99} models given the measured ages and masses. The mechanical energy input is mostly dominated by stellar winds for the outlier regions. This means that the contribution of pre-supernovae mechanical feedback (in the form of stellar winds) must be accounted for in order to understand what is driving the small-scale turbulence in the ISM. We note, however, that almost all regions reside below the 1-to-1 line if considering upper limit of the electron density ($n_e^{\rm (max)}$), but the correlation between $E_{\rm kin}$ of gas and $E_{\rm mech}$ from SNe is much weaker in both cases than that for the combined \revone{contribution of} SNe and winds. 

\revone{Significant contribution of pre-SNe feedback in total energy balance does not necessarily mean that it is also critical for reproducing the gas kinematics in simulations. As we demonstrated in Section~\ref{sec:tkin}, our estimates of the kinematic ages for our regions better agree with what is obtained from the SED fitting of star clusters when we consider only time passed since the first SN explosion. We suggest that SNe override the previous impacts of the star clusters onto the gas kinematics and that the latter is mostly regulated by SNe feedback (in agreement with simulations, e.g., \citealt{Walch2015}). However, pre-SNe feedback disperses molecular clouds ionising them and blowing-out gas away from the clusters \citep{Chevance2022}, and therefore the SNe explode in the pre-cleared medium. Also, the continuous impact of pre-SNe feedback during a few Myr can produce bigger superbubbles with more gas in their rims, which explains large differences in resulting $E_{kin}$ in Fig.~\ref{fig:energies}, in comparison with SNe contribution only. Therefore, accounting in simulations for the impact of pre-SNe feedback on the ISM can be essential for reproducing the realistic gas morphology and variation of SFR in the galaxies (in agreement with simulations by, e.g., \citealt{Haid2018, Keller2022}).}

\begin{figure}
    \centering
    \includegraphics[width=\linewidth]{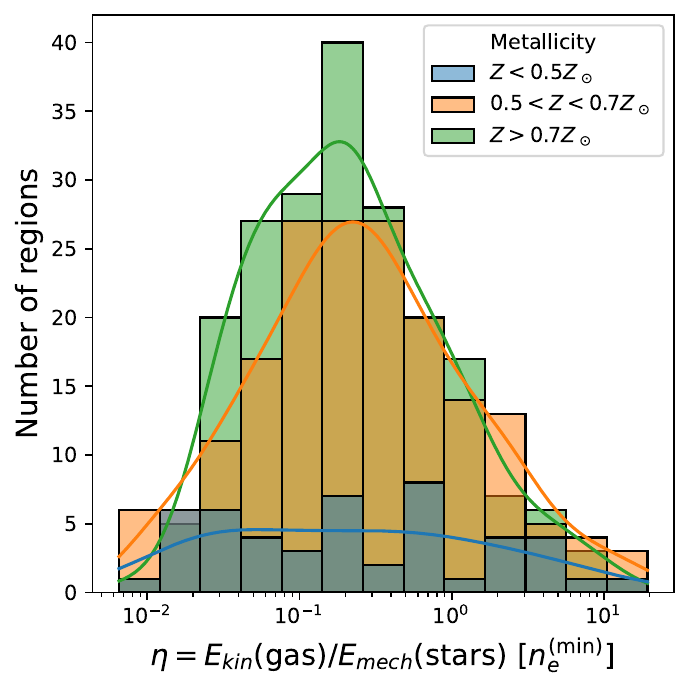}
    \caption{Distribution of the estimated mechanical stellar feedback energy efficiency for the regions shown in top-left panel of Fig.~\ref{fig:energies} grouped by metallicity.}
    \label{fig:efficiency_met}
\end{figure}

In general, there must be an energy balance between the young stellar associations and the turbulent ionised ISM. Indeed, in Fig.~\ref{fig:energies_mass} (left panel) the \Ha\ surface brightness (colour-coded) correlates with the stellar mass in the associations (see also Fig.~\ref{fig:compare_to_HII}). Given that the total \Ha\ flux is also a parameter in the calculations of the \revone{ionised gas mass} (see Eq.~\ref{eq:mass_flux}),
the observed correlations in Figs.~\ref{fig:energies_mass} and \ref{fig:energies} are likely driven by the fact that more massive associations produce more ionised gas around them. As a consequence of this, the total \Ha\ luminosity (and thus the observed \Ha\ flux) increases as the stellar association mass increases. This is often predicted by photoionisation models (e.g., in \textsc{starburst99} among many others), and  
this correlation is also observed on the scale of individual \HII\ regions \citep{Scheuermann2023}.
Meanwhile, our results imply that this correlation is preserved also for the turbulent ionised gas around \HII\ regions and for superbubbles, whose properties are not necessarily the same as \revone{those of} normal \HII\ regions (see Sec.~\ref{sec:compare_to_HII}). 
\revone{Given that the mass of the young stars and of the ionised gas (as well as $E_{\rm mech}$ from the stars and $E_{\rm kin}$ of the gas, respectively) are measured from unrelated data in a way independent of each other, we are confident that the correlation between these parameters is real and not a statistical effect.} 
Thus, we can use our measurements to estimate observationally what fraction of the mechanical energy produced by stars is converted into \revone{gas} kinetic energy in order to support the supersonic gas motions: $\eta = E_{\rm kin}(\mathrm{gas}) / E_{\rm mech}(\mathrm{stars})$. Our results imply that $\eta \sim 10-20$\% (see Fig.~\ref{fig:efficiency_met}), although there are regions with $\eta > 100$\% where the measured mechanical feedback alone cannot explain the gas kinetic energy (assuming the lower limit of electron density as discussed above). Note that in our measurement of $E_{\rm kin}$ we \revone{consider} the energy excess in each region with respect to that in the less perturbed ISM, so the energy coupling should be closer to $\sim 10$\% \revone{if we also account for the surrounding gas}. This is in good agreement with simulations by \cite{Ejdetjarn2022}, who found that $\sim10$\% of the energy from stellar feedback is necessary to drive the supersonic velocity dispersion of the ionised gas. We do not find any obvious change in the energy coupling with  metallicity, which suggests that the lower measured kinetic energy of the ionised gas in the \revone{low-metallicity turbulent regions} (see Sec.~\ref{sec:regions_metallicity} and Fig.~\ref{fig:energies_met}) results from a change in the cumulative mechanical energy input by stellar feedback and not the energy losses. However, the number of regions with metallicity $Z<0.5Z_\odot$ is very low, and more data are required to confirm this finding.

\subsection{Ionised superbubbles: energy balance and possible signs of star formation triggering}
\label{sec:disc_superbubbles}

\begin{figure}
    \centering
    \includegraphics[width=\linewidth]{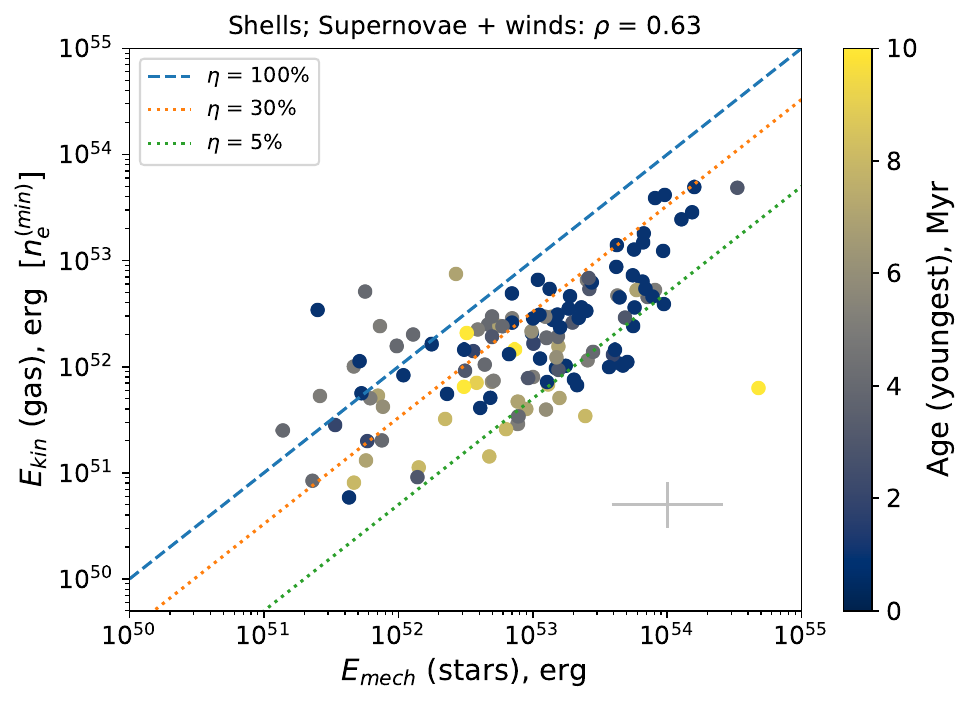}
    \includegraphics[width=\linewidth]{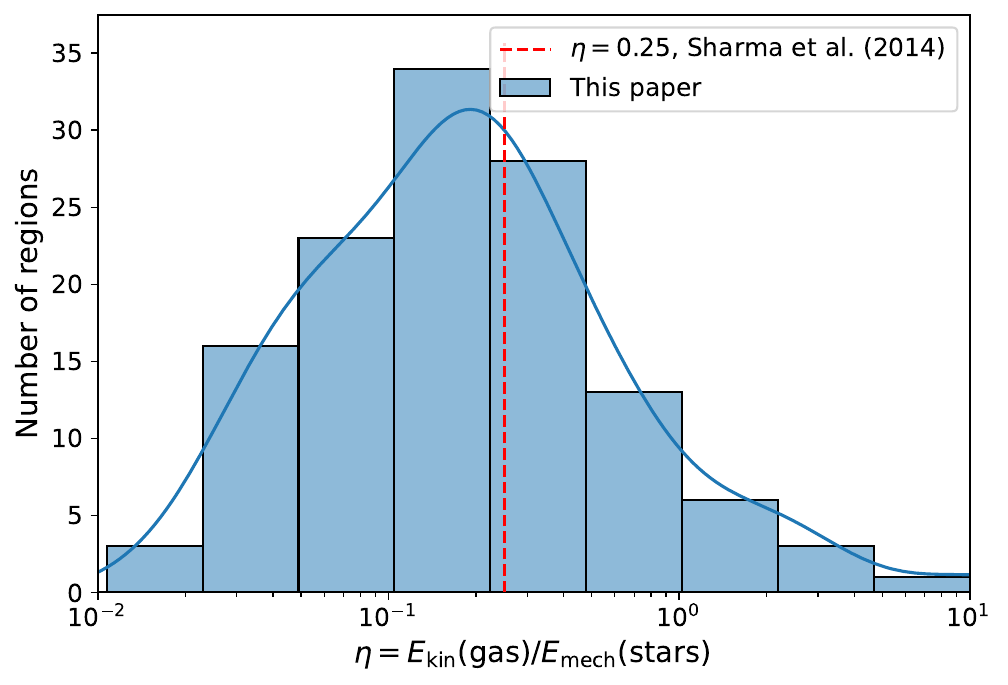}
    \caption{Comparison of the kinetic energy of ionised gas and the mechanical energy input by young stars for our best candidate superbubbles (as exhibiting shell-like morphology). Top plot: same comparison of energies as in Fig.~\ref{fig:energies}, superbubbles follow the same trend as the other regions. 
    Colour encodes the age of the youngest association in the superbubble. 
    Bottom plot: distribution of the estimated efficiency of the energy consumption in superbubbles. Its peak is slightly below $\eta = 0.25$ that was obtained from the modelling by \cite{Sharma2014} as an asymptotic value for the expanding superbubbles driven by multiple (up to a few hundred) OB stars in normal star-forming galaxies.}
    \label{fig:energies_superbubbles}
\end{figure}

As mentioned in Sec.~\ref{sec:isigma}, we identified a peak in the \Ha\ flux radial distribution corresponding to a swept-up shell for \nshells\, of our regions. These regions are our best candidates for expanding superbubbles, and it is interesting to compare their properties with the remaining regions. 

In Fig.~\ref{fig:energies_superbubbles} we show the same comparison of the kinetic energy of ionised gas and the mechanical energy input from stellar associations as in Fig.~\ref{fig:energies}, but in this figure, we include only those regions with shell-like morphologies. They follow the same trend as the rest of the regions. 
We measure similar mechanical energy efficiencies \revone{to those found} for the whole sample --  $\eta \sim 10-20$\%. This value is consistent with predictions from numerical simulations, where \cite{Sharma2014} followed the evolution of superbubbles powered by multiple (from tens to a few hundred) OB stars and obtained an asymptotic value of $\eta \sim 20-25$\%. We note that other simulations predict lower values -- e.g., \cite{Yadav2017} give $\eta \sim 6-10$\%, which is also close to the peak in Fig~\ref{fig:energies_superbubbles}. Our conclusion is valid for our adopted estimate of the electron density $n_e^{\rm (min)}$ generally corresponding to the upper limit of measured $\eta$. Using the upper limit on electron density $n_e^{\rm (max)}$ results in a much lower value of $\eta \sim 1$\%.  We would need larger statistics and more precise measurements of the electron densities to obtain strict observational limits on the parameter $\eta$, or to identify the main source of its variations. We note also that our measurements are related to ionised gas only. Based on observations of the same sample of galaxies as in this work, \cite{Watkins2023} found a coupling efficiency of $\sim5-12$\% required to put into agreement the derived and predicted stellar masses and ages inside the molecular gas superbubbles, that is close to our estimate. 

For the sample of \nshells\ likely superbubbles, we can resolve their morphology and thus check if the stellar associations driving their expansion are located close to their centres. For that, we calculated the separation as the angular distance between the centres of the regions and the position of the stellar associations (the geometric centre of the corresponding stellar association mask), deprojected them based on the position angle and axes ratio of the regions and translated to  physical distance normalized to $R_{\rm eff}$ of each region. For separations less than the PSF, we assume a separation equal to zero. We excluded from our analysis any region where $R_{\rm eff} < 3 \times \mathrm{PSF}$ to ensure the reliability of our results. As a result, we are left with 40 superbubble-like regions where stellar associations or star clusters \revone{younger than 40~Myr}\footnote{\revone{Further results do not change if we consider higher (50~Myr) or lower (e.g. 8~Myr) age limits}} are observed. In these regions, about 40\% of the star clusters are offset from the centre of the superbubble by at least $0.5R_{\rm eff}$. This implies that the stellar sources powering the expanding superbubbles are often located in their rims (see also \citealt{Egorov2017, Gerasimov2022, Barnes2022jwst}). We speculate that in such cases we observe the superbubbles (or outflows) driven by young stars at the edges of molecular clouds (`blister'-like regions). \revone{If such offsets are also common in other objects from our entire sample, this can} partially explain the difference between the \Ha\ velocity dispersion towards the youngest stellar associations and the regions without any young stars (see Sec.~\ref{sec:statistics} and Fig.~\ref{fig:sigma_in_clusters}). \revone{Namely,} more dynamically perturbed gas is observed in the lower density environment outside the powering stellar associations, which reside in the denser clouds.  

\begin{figure}
    \centering
    \includegraphics[width=\linewidth]{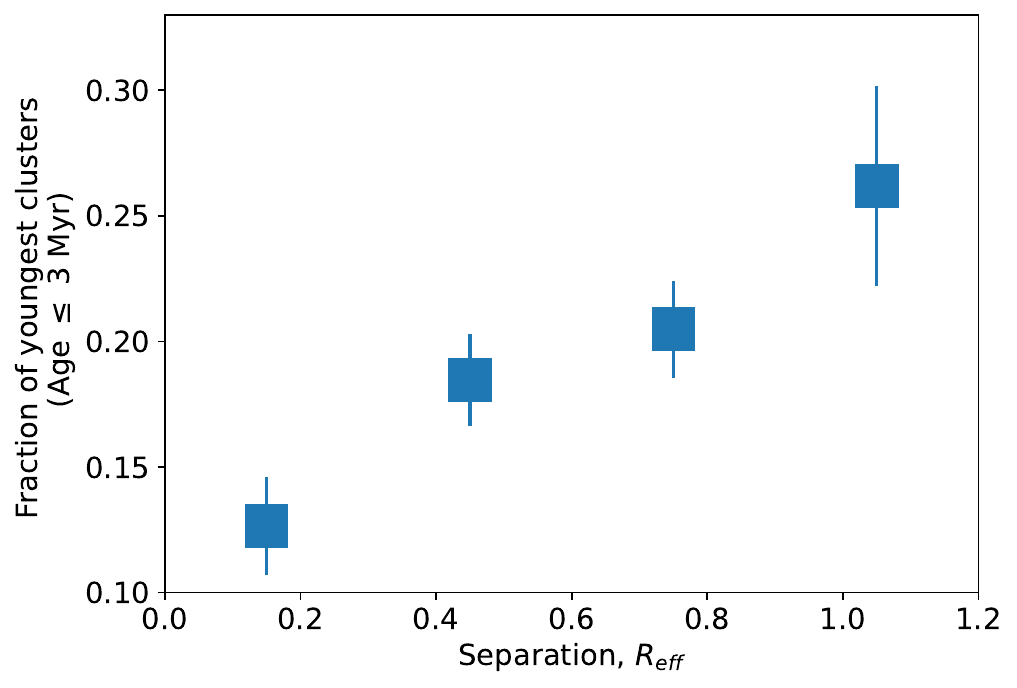}
    \caption{\revone{Relative fraction of the stellar associations younger than 3~Myr at different separations between clusters/associations and centres of the regions. Only a subsample of 40 well-resolved superbubbles and star clusters younger than 40~Myr are considered. Each separation bin corresponds to $0.3 R_{eff}$. Errors are calculated from 1000 Monte Carlo realizations with positions of each cluster randomly varying within the PSF of the MUSE observations.}}
    \label{fig:separation}
\end{figure}

We also find a weak trend with the age of the stellar association, \revone{shown in Fig.~\ref{fig:separation}: fraction of the associations younger than 3~Myr is higher (by $\sim15$\%) at the periphery of the regions (at $R > 0.5 R_{\rm eff}$) than close to their centres ($R<0.5R_{\rm eff}$)}. 
A higher fraction of the younger associations on the edge of superbubbles points towards sequential star formation \citep{Elmegreen1977}, or triggering of star formation as a result of the evolution of the superbubble \citep[e.g.][]{Hartmann2001}, though it is usually hard to distinguish between these processes \citep{Dale2015}. The triggering of star formation in the periphery of expanding supershells as a result of their interaction with the surrounding ISM, or collision with each other, is predicted in simulations \citep[e.g.][]{Ntormousi2011, Inutsuka2015, Vasiliev2022}, and indications of this have been observed in very nearby dwarf galaxies \citep[below 4~Mpc, see, e.g.][]{Lozinskaya2002, Egorov2017, Egorov2018, Fujii2021}. \cite{Dawson2013} demonstrated that about $12-25$\% of the molecular gas in the supergiant shells in the LMC was formed as a result of their \revone{interaction with} the ISM. Our identification of systematically larger separations from the centre of superbubbles for the youngest stellar associations extends the observational signatures of triggered star formation in superbubbles to more massive spiral galaxies. However, the statistics are limited, in principle by the limited angular resolution of the underlying data. Investigation of superbubbles in the delivered and scheduled data for nearby galaxies from JWST imaging will allow for a more detailed investigation into the age distribution of star clusters and stellar associations, in relation to morphologically identified superbubbles \cite[see, e.g.,][as examples of the first results]{Barnes2022jwst, Watkins2022jwst}.

\section{Summary}
\label{sec:summary}
   Here we quantify the contribution \revone{of mechanical feedback from young massive stars to the regulation} of the kinematics and morphology of the ionised ISM. Namely, we identify areas of locally elevated \Ha\ velocity dispersion (likely a result of the presence of unresolved expanding superbubbles, supernovae remnants, or turbulent ionised gas motions), \revone{connect them to young stellar associations and} quantify the energy balance between the mechanical energy in the ionised gas and what is contributed by the observed stellar populations in these regions. Our analysis combines observational data from  MUSE/VLT integral-field spectroscopy with \HST\ multiwavelength imaging for 19 nearby spiral galaxies obtained as a part of the PHANGS survey. Our main results are summarized below:
   \begin{itemize}
       
       \item We identify \nregs\, regions of locally elevated \Ha\ velocity dispersion (\sigmaHa\ $>45\kms$),  tens in each galaxy. About half of them contain at least one young stellar association or compact star cluster within their borders. We argue that most of those regions not connected with star clusters are supernovae remnants -- they  typically have a centrally-concentrated \Ha\ flux distribution and emission line flux ratios consistent with shock excitation. 
       \item We find that the kinetic energy of the ionised gas in the regions of locally elevated \Ha\ velocity dispersion strongly correlates with the mass of the young stellar associations within these regions, and with their total mechanical energy input into the surrounding ISM (in the form of stellar winds and supernovae based on Starburst99 models). At the same time, the correlation with mechanical energy produced by young stars is significantly weaker if we only consider the impact of supernovae. This implies that pre-supernovae feedback is an essential contributor to the energy balance. 
      \item About $\sim10-20$\% of the total mechanical energy input is retained in the regions in \revone{the form of} kinetic turbulent energy in the ionised gas. This result is in agreement with the \revone{coupling} efficiencies obtained from numerical simulations for feedback-driven superbubbles \cite[e.g.][]{Sharma2014} or ISM turbulence \cite[e.g.][]{Ejdetjarn2022}.
      \item The measured kinetic energy of the ionised gas changes with metallicity, which can be a consequence of both the lower masses of the clusters and weaker stellar feedback in the low metallicity environments.
       \item Among the high velocity dispersion regions, only \nshells\ (12\%) exhibit a clearly distinguishable shell-like morphology in their \Ha\ flux distribution. These regions are probably expanding ionised superbubbles.  In general, the size distribution of the identified regions is well described by a power-law with slope $\sim -2.7$, in agreement with observations \cite[e.g.][]{Oey1997, Bagetakos2011, Watkins2022jwst} and simulations \citep{Nath2020} of superbubbles in star-forming galaxies. Thus, we cannot exclude that most of the identified regions of high velocity dispersion appear to be expanding superbubbles, in many cases the shell-like morphology is not resolved  or overlaps with other structures. 
       \item \revone{The estimated expansion velocity $V_{exp} = 10 - 60 \kms$ (mostly within $20 - 35 \kms$) for the superbubbles in our sample, and corresponding kinematic ages are typically less than 10~Myr (mostly within $1 - 6$~Myr). The derived ages agree better with the time passed since the first SN explosion than with the total lifetime of the star clusters. This implies that once SNe occur, their impact becomes dominant in the regulation of the current ionised gas kinematics in the analysed superbubbles.} 
       \item About 40\% of the \revone{well-resolved} superbubbles in our sample have young stellar associations located on their rims, not in their centres. The youngest associations are more often observed at the periphery of the superbubbles, while older stellar associations are mostly observed closer to their centres. This is an indication of sequential or triggered star formation occurring at the edges of superbubbles in our sample.

   \end{itemize}
   
   Overall, we establish a direct observational connection between the mechanical feedback from young massive stars onto the surrounding ionised gas for a large and homogeneous sample of regions, including expanding superbubbles. We demonstrate that accounting for pre-supernovae mechanical feedback in form of stellar winds is important, as supernovae alone fail to explain the measured kinetic energy of the ionised gas. 
   
   We also show that the mechanical stellar feedback energy efficiency does not change \revone{with metallicity. However,} our data are very limited for metallicities $Z < 0.5 Z_\odot$. 
   A further extension of this analysis to dwarf galaxies will allow us to establish a purely observational constraint on the efficiency of mechanical stellar feedback in various environments which is a critical ingredient in cosmological simulations.

\begin{acknowledgements}
This work was carried out as part of the PHANGS collaboration.
Based on observations collected at the European Southern Observatory under ESO programmes 094.C-0623 (PI: Kreckel), 095.C-0473,  098.C-0484 (PI: Blanc), 1100.B-0651 (PHANGS-MUSE; PI: Schinnerer), as well as 094.B-0321 (MAGNUM; PI: Marconi), 099.B-0242, 0100.B-0116, 098.B-0551 (MAD; PI: Carollo) and 097.B-0640 (TIMER; PI: Gadotti). 

This research is based on observations made with the NASA/ESA Hubble Space Telescope obtained from the Space Telescope Science Institute, which is operated by the Association of Universities for Research in Astronomy, Inc., under NASA contract  NAS 5–26555. These observations are associated with program 15654.

This research made use of astrodendro, a Python package to compute dendrograms of Astronomical data (http://www.dendrograms.org/).

KK, OVE, EJW, JEM-D gratefully acknowledge funding from the Deutsche Forschungsgemeinschaft (DFG, German Research Foundation) in the form of an Emmy Noether Research Group (grant number KR4598/2-1, PI Kreckel). 

RSK and SCOG acknowledge funding from the European Research Council via the ERC Synergy Grant ``ECOGAL'' (project ID 855130),  from the Heidelberg Cluster of Excellence (EXC 2181 - 390900948) ``STRUCTURES'', funded by the German Excellence Strategy, and from the German Ministry for Economic Affairs and Climate Action in project ``MAINN'' (funding ID 50OO2206). RSK and SCOG also acknowledge computing resources provided by {\em The L\"{a}nd} through bwHPC and DFG through grant INST 35/1134-1 FUGG and for data storage at SDS@hd through grant INST 35/1314-1 FUGG.

\end{acknowledgements}

%
%

\bibliographystyle{aa}
\bibliography{PHANGS_SB}

\begin{appendix}
\section{\IS\ diagrams for all PHANGS-MUSE galaxies}
\label{app:isigma}
\begin{figure*}
    \centering
    \includegraphics[width=0.95\linewidth]{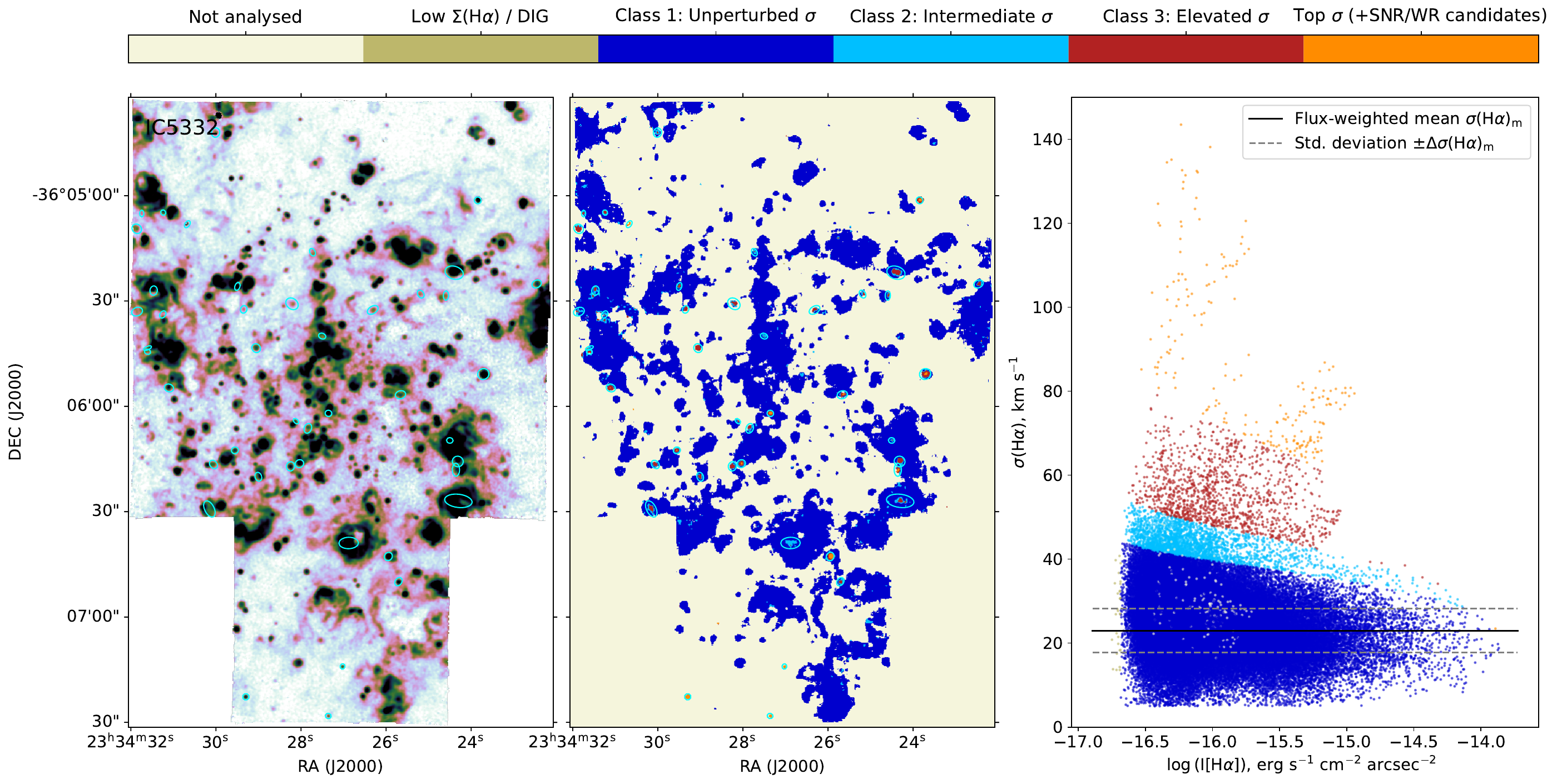}
    \includegraphics[width=0.95\linewidth]{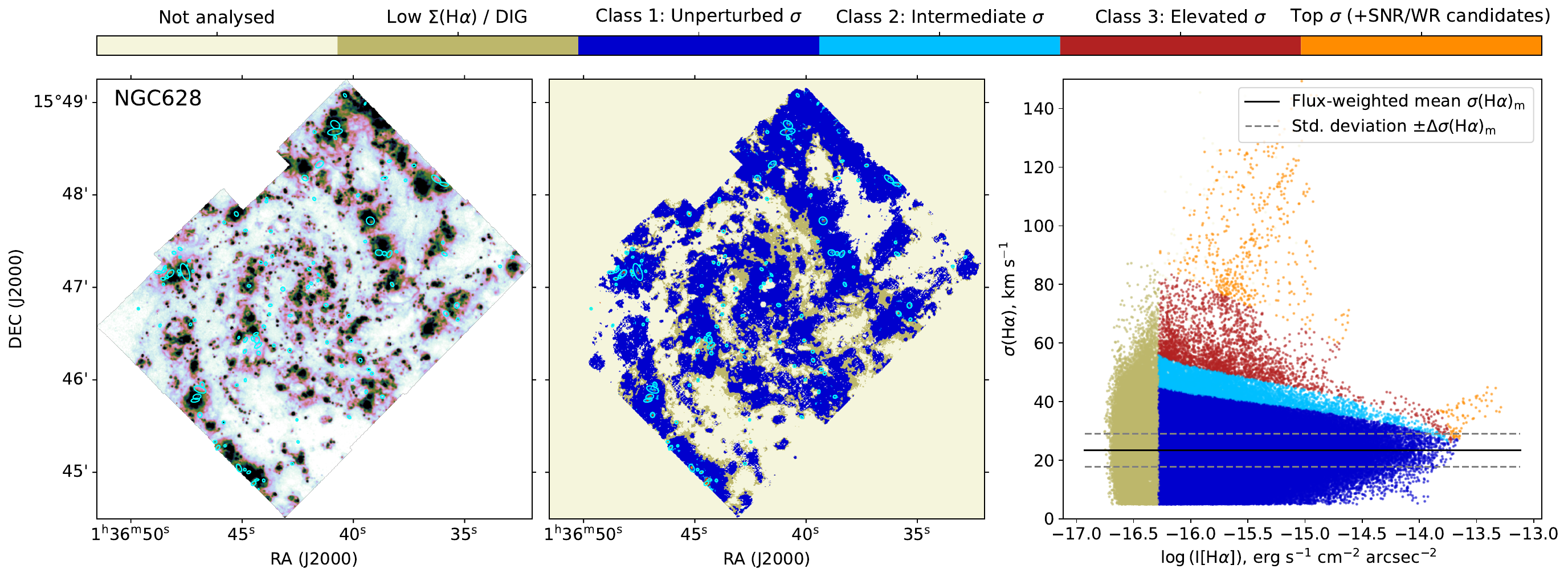}
    \includegraphics[width=0.95\linewidth]{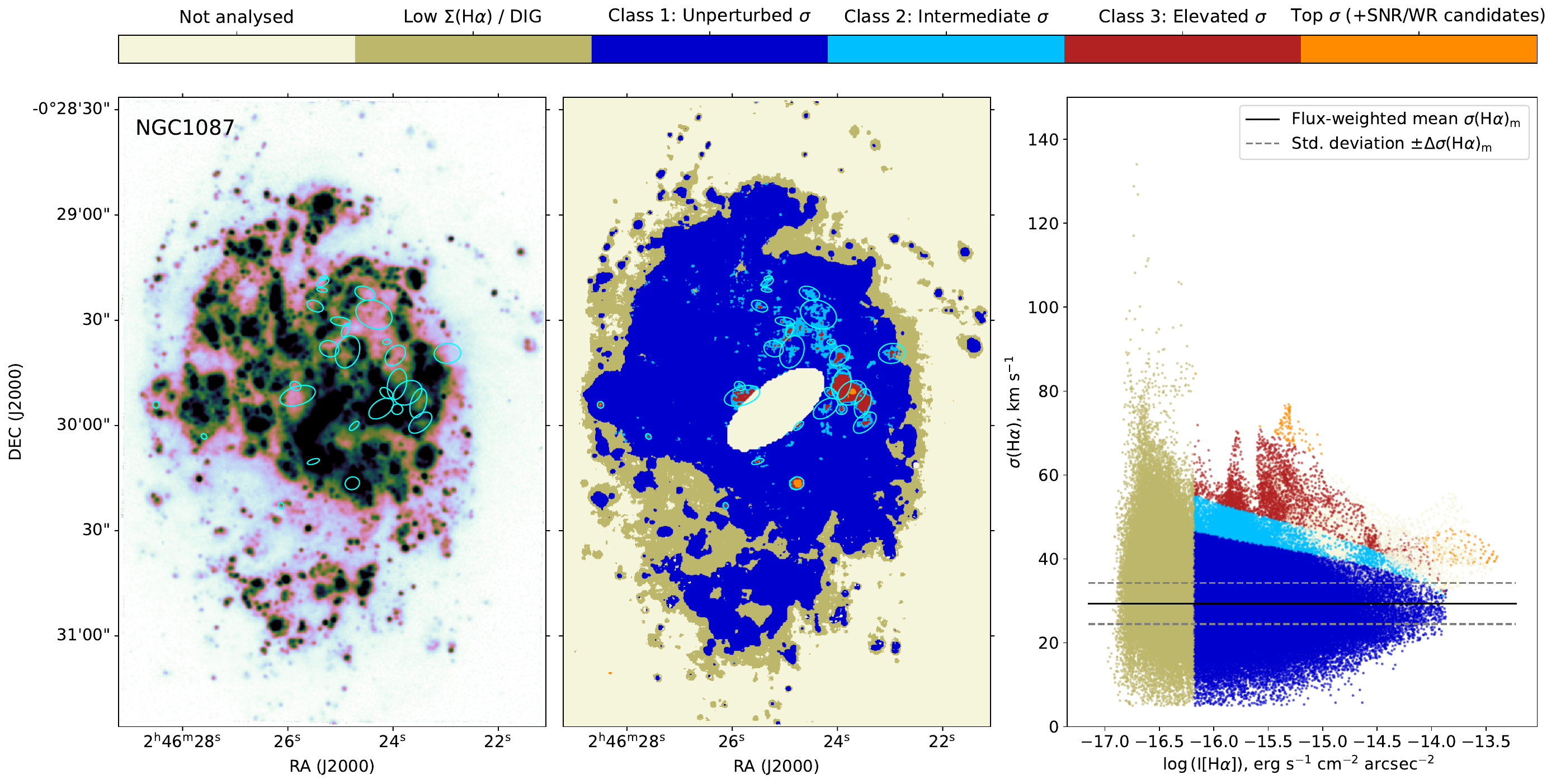}
    \caption{\IS\ diagram (right), classification map (centre) and the \Ha\ map (left) for all PHANGS-MUSE galaxies except NGC~4254 (shown in Fig.~\ref{fig:isigma}). Cyan ellipses encircle the identified regions of the locally elevated \Ha\ velocity dispersion.}
    \label{fig:isigma_all}
\end{figure*}

The regions analysed in this paper were selected based on the \IS\ diagrams introduced and described in Section~\ref{sec:isigma}. Here we describe the details of the performed analysis and how the selection criteria were determined. The \IS\ diagrams and classification maps for all PHANGS-MUSE galaxies are shown in Fig.~\ref{fig:isigma_all} (NGC~4254 is shown in Fig.~\ref{fig:isigma}).

A large scatter in \sigmaHa\ is usually observed at the low surface brightness end of the \IS\ diagrams \citep[e.g.][]{Moiseev2012}, and this is also the case for most of the PHANGS-MUSE galaxies. This scatter can result from both physical (e.g., the impact of shocks) and observational (low signal-to-noise ratio) reasons, though we remind that we consider only the pixels with $S/N > 30$. 
Here we exclude \sigmaHa\ peaks associated with the low surface brightness DIG. 
We tested two quantitative criteria for isolating the low surface brightness DIG. 
Both criteria suggest a constant level of surface brightness,  $\Sigma\mathrm{(H\alpha)_{DIG}}$, is suitable for each galaxy in order to separate the DIG from the rest of the emission structures. First, we adopted a median surface brightness for the DIG, $\Sigma\mathrm{(H\alpha)_{DIG}}$, as estimated for the PHANGS-MUSE galaxies in \citet{Belfiore2022}. 
In that paper, the DIG was isolated from \HII\ regions by a morphological analysis. Another criterion is based on the spectral properties of the DIG, as it exhibits elevated relative intensities in low-ionisation ions compared to \HII\ regions, and usually resides above the maximum starburst line from \cite{Kewley2001} on the \OIIIHb\ vs \SIIHa\ diagnostic \citep[e.g.,][]{Haffner2009, Belfiore2022}. We derived \SBDIG\, as the value corresponding to the 75th percentile of all pixels residing above the maximum starburst line. For half of the galaxies in our sample, both methods gave very similar values of \SBDIG. However, visually those taken from \citet{Belfiore2022} are better at separating the low brightness peak of \sigmaHa\ on \IS\ diagrams for the rest of the objects. Therefore, we used the values from \cite{Belfiore2022} to determine \SBDIG. For five galaxies (NGC~1087, NGC~1300, NGC~1385, NGC~3351, and NGC~7496) we additionally increased (by $0.1-0.5$~dex) the value of \SBDIG\, to better match the low-brightness peak in the \sigmaHa\ distribution. According to \cite{Belfiore2022}, the DIG emission in three of these galaxies has a very wide and shallow surface brightness distribution. The final adopted values of \SBDIG\, for each galaxy are given in Table~\ref{tab:sample}.

We estimated the mean \Ha\ velocity dispersion $\sigma\mathrm{(H\alpha)_m}_0$ for the \HII\ regions in each galaxy as the flux-weighted mean value across all pixels brighter than \SBDIG. The uncertainty of the mean velocity dispersion $\Delta(\sigma_\mathrm{m})_0$ was measured as the standard deviation of \sigmaHa\ around $\sigma\mathrm{(H\alpha)_m}_0$. The final adopted values of $\sigma\mathrm{(H\alpha)_m}$ and $\Delta(\sigma_\mathrm{m})$ were obtained on the second iteration, where we excluded all pixels with $\sigma\mathrm{(H\alpha)}>\sigma\mathrm{(H\alpha)_m}_0+\Delta(\sigma_\mathrm{m})_0$. The derived values of $\sigma\mathrm{(H\alpha)_m}$ characterize the mean velocity dispersion of the ionised ISM in each galaxy and are given in Table~\ref{tab:sample}. These values agree well with the empirical relations between $\sigma\mathrm{(H\alpha)_m}$ and SFR as reported in other works \citep[e.g.][]{Law2022}.

\begin{table*}
\caption{\revone{Summary of the criteria applied for classification and segmentation of the pixels in \IS\ diagrams }}
    \centering
    \begin{small}
    \begin{tabular}{cccc}
    \hline
        Classification & Observed \sigmaHa\ range & Clipped \sigmaHa\ for \textsc{astrodendro} & Physical interpretation\\
        \hline
        Low $\Sigma$(\Ha) & Any; $\rm I(H\alpha) < \Sigma(H\alpha)_{DIG}$ & $\rm \sigma_{ref} - \Delta(\sigma_m)$ & Diffuse ionised gas\\
        Class 1 & $\rm \sigma(H\alpha) < \sigma(H\alpha)_{norm}$  & $\rm \sigma_{ref} - \Delta(\sigma_m)$ &  Unperturbed \HII\ regions\\
        Class 2 & $\rm [\sigma(H\alpha)_{norm}, \sigma(H\alpha)_{1.5\times norm})$ & $\rm \sigma_{ref}$ & May be illusive, but likely periphery\\  
        & & &  of superbubbles; low-velocity shocks\\
        Class 3 & $\rm [\sigma(H\alpha)_{1.5\times norm}, \sigma(H\alpha)_{3\times norm})$ & $\rm max[\sigma_{ref}+2\Delta(\sigma_m), \sigma(H\alpha)]$ & Expanding superbubbles; shocks;\\
        & & &  other supersonic gas motions\\
        Top $\sigma$ 
 & $\rm \sigma(H\alpha) \geq min[\sigma(H\alpha)_{3\times norm}, 100\ km\ s^{-1}] $ & $\rm max[\sigma_{ref}+3\Delta(\sigma_m), \sigma(H\alpha)]$ & SNR/WR or other extreme source candidates \\
         \hline
    \end{tabular}
    \end{small}
    
\begin{footnotesize}\raggedright
\revone{
 -- $\rm \sigma(H\alpha)_{norm}$, $\rm \sigma(H\alpha)_{1.5 \times norm}$ and $ \rm \sigma(H\alpha)_{3\times norm}$ are the upper boundaries of velocity dispersion defined by Gaussians in $\rm \sigma(H\alpha)$ vs $\rm \log I(H\alpha)$ space with \texttt{mean} =  $\rm \sigma(H\alpha)_m$ and \texttt{std}~=~$\rm \Delta(\sigma_m)$, $\rm 1.5\Delta(\sigma_m)$ and $\rm 3\Delta(\sigma_m)$, respectively.\\
-- $\rm \sigma(H\alpha)_m$ and $\rm \Delta(\sigma_m)$ are flux-weighted mean velocity dispersion and its standard deviation (see measured values for each galaxy in Table~\ref{tab:sample}).\\
-- $\rm \sigma_{ref}$ is the median \sigmaHa\ derived from all pixels classified as Class 2, 3 or Top $\sigma$.\\
}
\end{footnotesize}

    \label{tab:method_summary}
\end{table*}

Based on $\sigma\mathrm{(H\alpha)_m}$ and $\Delta(\sigma_\mathrm{m})$, we performed a classification of all pixels brighter than \SBDIG\, into four classes (shown by different colours in Figs.~\ref{fig:isigma} and \ref{fig:isigma_all}) in a way similar to that in \cite{Egorov2021}. \revone{The particular criteria and physical meanings are summarized in Table~\ref{tab:method_summary} and described in detail below}:

\begin{itemize}
    \item Class 1: Unperturbed $\sigma\mathrm{(H\alpha)}$. \revone{Normal \HII\ regions tend to occupy the elongated area of relatively high surface brightness and low velocity dispersion on \IS\ diagrams. For brighter regions, a scatter of \sigmaHa\ measurements is usually smaller than for faint regions, which can also be seen for most galaxies in Fig.~\ref{fig:isigma}. The upper boundary of the distribution of \sigmaHa\ vs $\log(I[\mathrm{H\alpha}])$  for such regions can be parameterized by a Gaussian with the mean value and standard deviation equal to $\sigma\mathrm{(H\alpha)_m}$ and $\Delta(\sigma_\mathrm{m})$, respectively. We refer to velocity dispersion defined by such a normal distribution for each flux value as $\sigma\mathrm{(H\alpha)_{norm}}$.  
    We classify a pixel as related to Class~1 if $\sigma\mathrm{(H\alpha)}<\sigma\mathrm{(H\alpha)_{norm}}$ for its \Ha\ brightness. Note that} 
     all pixels with $\sigma\mathrm{(H\alpha)}<\sigma\mathrm{(H\alpha)_m}$ are also considered as belonging to this class. This class contains all bright \HII\ regions and surrounding ionised ISM that do not exhibit signs of perturbation in their kinematics. 
    \item Class 2: Intermediate $\sigma\mathrm{(H\alpha)}$. \revone{This includes all pixels with values of \sigmaHa\ between two distributions -- $\sigma\mathrm{(H\alpha)_{norm}}$ and $\sigma\mathrm{(H\alpha)_{1.5\times norm}}$. The latter is defined in the same way as $\sigma\mathrm{(H\alpha)_{norm}}$, but with} a standard deviation equal to $1.5 \Delta(\sigma_\mathrm{m})$. This is an intermediate class: \revone{the elevated \sigmaHa\ may still be illusive} (due to the contamination by pixels in the wings of the distribution defining Class 1 objects), but can reflect real physical changes in the velocity dispersion due to turbulence, shocks, the presence of slowly expanding superbubbles, or is observed at the edges of the more dynamically active Class 3 regions.
    \item Class 3: Elevated $\sigma\mathrm{(H\alpha)}$. This class is assigned to pixels with $\sigma\mathrm{(H\alpha)} > \sigma\mathrm{(H\alpha)_{1.5\times norm}}$ (thus the contamination by wrongly classified `unperturbed' pixels should not exceed 7\% assuming their normal distribution). The corresponding regions clearly show remarkable local peaks of velocity dispersion, that can be driven by turbulence, shocks or moderately expanding superbubbles (due to the presence of emission from both approaching and receding sides). In agreement with the toy models in \cite{Moiseev2012, MunozTunon1996}, and with the simulations in \cite{Vasiliev2015}, the regions corresponding to this class usually show triangular shapes on the \IS\ diagrams.
    \item Top \sigmaHa\ and SNR/WR candidates. This is a sub-group of the Class 3 pixels selected as having intrinsic $\sigma\mathrm{(H\alpha)} > 100\kms$, or \mbox{$\sigma\mathrm{(H\alpha)} > \sigma\mathrm{(H\alpha)_{3\times norm}}$} (where $\sigma\mathrm{(H\alpha)_{3\times norm}}$ is the normally distributed value with a standard deviation equal to $3\Delta(\sigma_\mathrm{m})$). The cores with the most extreme peaks in velocity dispersion are selected in this subclass. As was demonstrated by \cite{Moiseev2012},  these pixels might correspond to the presence of energetic stellar objects like SNRs, WR, LBV etc. Those pixels that pass the criteria for Classes 2 or 3 and have $\log(I[\mathrm{H\alpha}])$ above that for any Class 1 pixel are also considered as part of this subclass, as they are definitely outliers from the normal distribution of  $\sigma\mathrm{(H\alpha)}$. 
\end{itemize}

The regions of correlated high velocity dispersion are relatively easy to identify by eye in the constructed classification maps (central panel of Figs.~\ref{fig:isigma} and \ref{fig:isigma_all}). To automatically select these regions and derive their boundaries, we used the \textsc{astrodendro}\footnote{\url{http://www.dendrograms.org/}} package. However, the application of \textsc{astrodendro} directly to the classification map resulted in a very rough selection of the regions (many of the individual Class 3 regions were merged into a single region), and only the strongest peaks of velocity dispersion were often identified when we applied \textsc{astrodendro} to the $\sigma\mathrm{(H\alpha)}$ map directly. To overcome these issues, we created a synthetic map based on both the classification and velocity dispersion distribution \revone{by clipping the velocity dispersion range for each of the classes (as summarized in Table~\ref{tab:method_summary})}. Namely, for each galaxy, we derived the median \sigmaHa\ of all pixels within the galaxy classified as Class 2 or above (hereinafter $\sigma_\mathrm{ref}$), and set $\sigma\mathrm{(H\alpha)} = \sigma_\mathrm{ref} -  \Delta(\sigma_\mathrm{m})$ for all Class 1 and DIG regions, and $\sigma_\mathrm{ref}$ for Class 2 regions. For Class 3 and SNR/WR candidates, we left their observed \sigmaHa, but with the lower limit equal to $\sigma_\mathrm{ref} + 2\Delta(\sigma_\mathrm{m})$ and $\sigma_\mathrm{ref} + 3\Delta(\sigma_\mathrm{m})$, respectively (approximately equal to the mean thresholds used for the classification in the previous step).

Applying \textsc{astrodendro} to the constructed synthetic maps, we required the minimal area of adjoined pixels to be no less than $\pi( \mathrm{PSF}/2)^2$ (where the full width at half maximum (FWHM) size of the PSF -- point spread function -- characterises the angular resolution of the MUSE data and is given in Table~\ref{tab:sample}), and the minimal deviation of the identified region from the background should be $0.9\Delta(\sigma_\mathrm{m})$. Each of the identified regions was then approximated by an ellipse having minor and major diameters equal to $2.65\times\mathrm{FWHM}_{A}$, where $\mathrm{FWHM}_{A} = 2\sqrt{2\ln 2}\sigma_A$, and $\sigma_A$ is the standard deviation along the minor and major axes of the 2D distribution resulting from the statistics computed by \textsc{astrodendro}. The ellipses defined in such a way correspond to the baselines of the \sigmaHa\ distribution in the region and encircle the entire area of the elevated velocity dispersion. In the case of a bubble, this corresponds to its rim. To prove this, we used a toy model of a thin sphere expanding in a homogeneous medium. The velocity separation between the components from the approaching and receding sides of such a simulated bubble is $\Delta V \propto \sqrt{1-r^2/R^2}$, where $r$ is the distance in the sky plane from the centre of the bubble with radius $R$. We analysed the radial distribution of the velocity dispersion measured from the simulated spectra integrated along the line-of-sight and found that the weighted standard deviation of the \sigmaHa\ (calculated in the same way as by \textsc{astrodendro}) $\sigma_A \simeq 0.32\times R$ for various expansion velocities and spatial resolutions\footnote{We used \textsc{sdss5-lvmdatasimulator} package (\url{https://github.com/sdss/lvmdatasimulator}) for doing this analysis as it allows one to easily construct such toy models with various parameters and numbers of spatial resolution elements across the nebula and obtain the simulated IFU spectra for them.}.
We excluded all regions having an axis ratio less than 0.35 (chosen based on visual inspection of the results obtained with the different parameters). Even if real, such elongated regions may add significant scatter to the further analysis because of the uncertainties of their size measurements.
 
 \revone{Where possible}, we refined the sizes of the elliptical regions identified by \textsc{astrodendro}. We considered the oversampled radial \Ha\ flux distribution for each region. For those regions representing the expanding superbubbles we expect to see a peak in the \Ha\ flux radial distribution associated with the rims of the swept-up shell, and thus we define the radial distance to this peak as the radius of the superbubble (see example in Fig.~\ref{fig:regions_example}). We were able to clearly identify  peaks in the radial flux distribution for \nshells\ regions, which we use as our best candidates for expanding ionised superbubbles (see Section~\ref{sec:disc_superbubbles}). For most of these regions, the change in their size after this refinement was not significant (within 15\% for half of the regions, and $>30$\% for only 20 regions). On the other hand, we also identified many centrally concentrated regions, that could be SNR or other objects with unresolved morphology (see Section~\ref{sec:regions_morphology}). In such a case, the analysis of the classification maps described above can significantly overestimate the size of the regions, and their real radius is better approximated by the standard deviation of the \Ha\ radial flux distribution \citep{Barnes2022}. We left the size of a region unchanged in the cases when we found neither a central peak nor an associated shell in the radial \Ha\ flux distribution. 
 
 Finally, we checked for overlapping ellipses and excluded the smaller regions when the overlap was more than 20\%. As a result, we identified \nregs\, regions across all 19 galaxies. The cyan ellipses in Figs.~\ref{fig:isigma} and \ref{fig:isigma_all} show the locations of the identified regions of locally elevated velocity dispersion, and their edges correspond to the adopted sizes for each region.
 
 The number of regions with `normal' velocity dispersion erroneously selected with our criteria is negligible. This is estimated by simulating the synthetic maps of velocity dispersion randomly distributed around $\sigma\mathrm{(H\alpha)}$ with standard deviation equal to $\Delta(\sigma_\mathrm{m})$ (changing with the \Ha\ flux in the same way as observed, see Fig.~\ref{fig:isigma}) and applying the same criteria to these data. For each of the 100 MC realizations, the number of erroneously selected regions per galaxy does not exceed 3, and the size of such regions is close to the PSF. 
 
\setcounter{figure}{0}
\begin{figure*}
    \centering
    \includegraphics[width=0.95\linewidth]{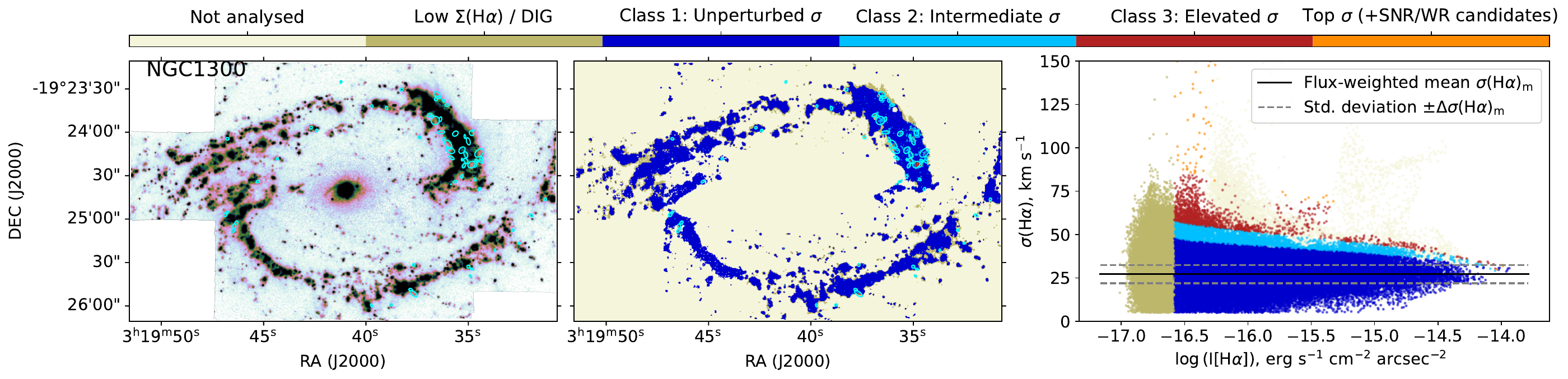}
    \includegraphics[width=0.95\linewidth]{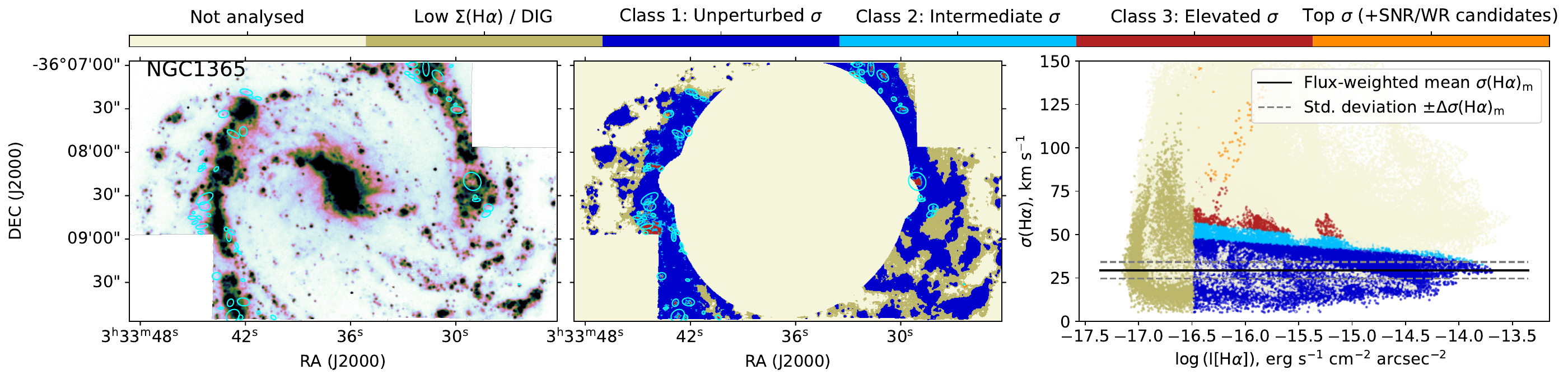}
    \includegraphics[width=0.95\linewidth]{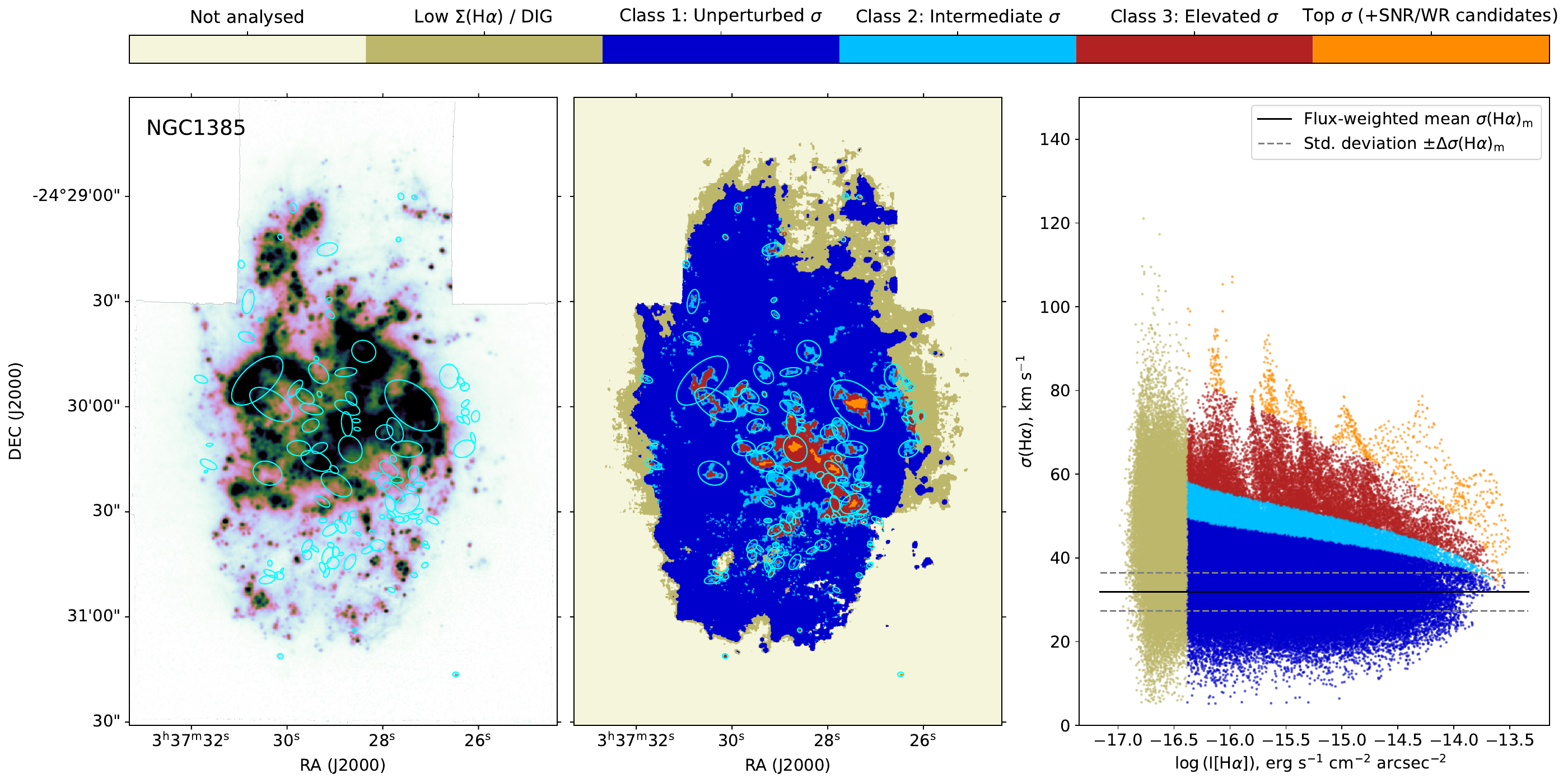}
    \includegraphics[width=0.95\linewidth]{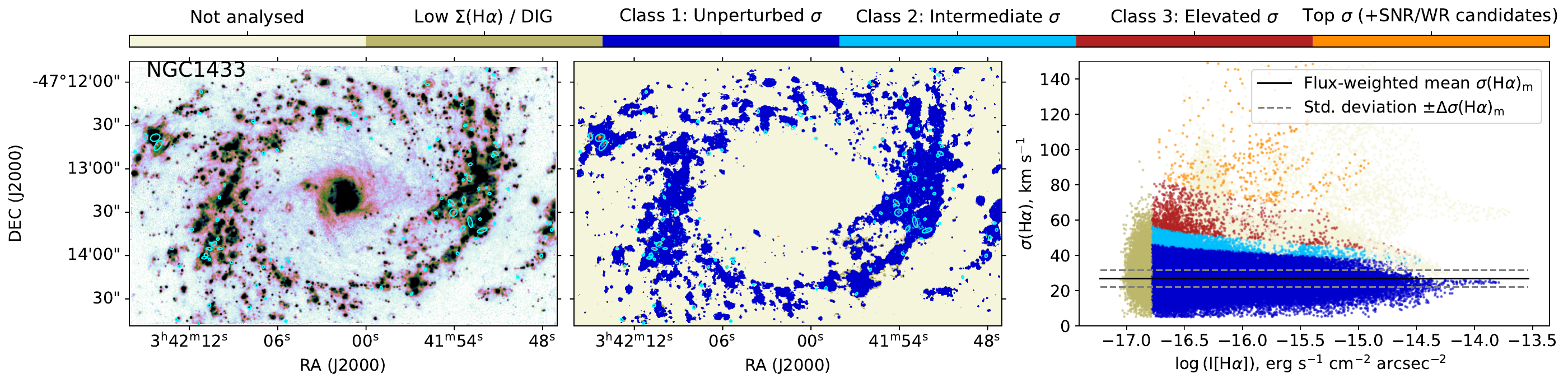}
    \caption{Continue.}
\end{figure*}

\setcounter{figure}{0}
\begin{figure*}
    \centering
    
    \includegraphics[width=0.95\linewidth]{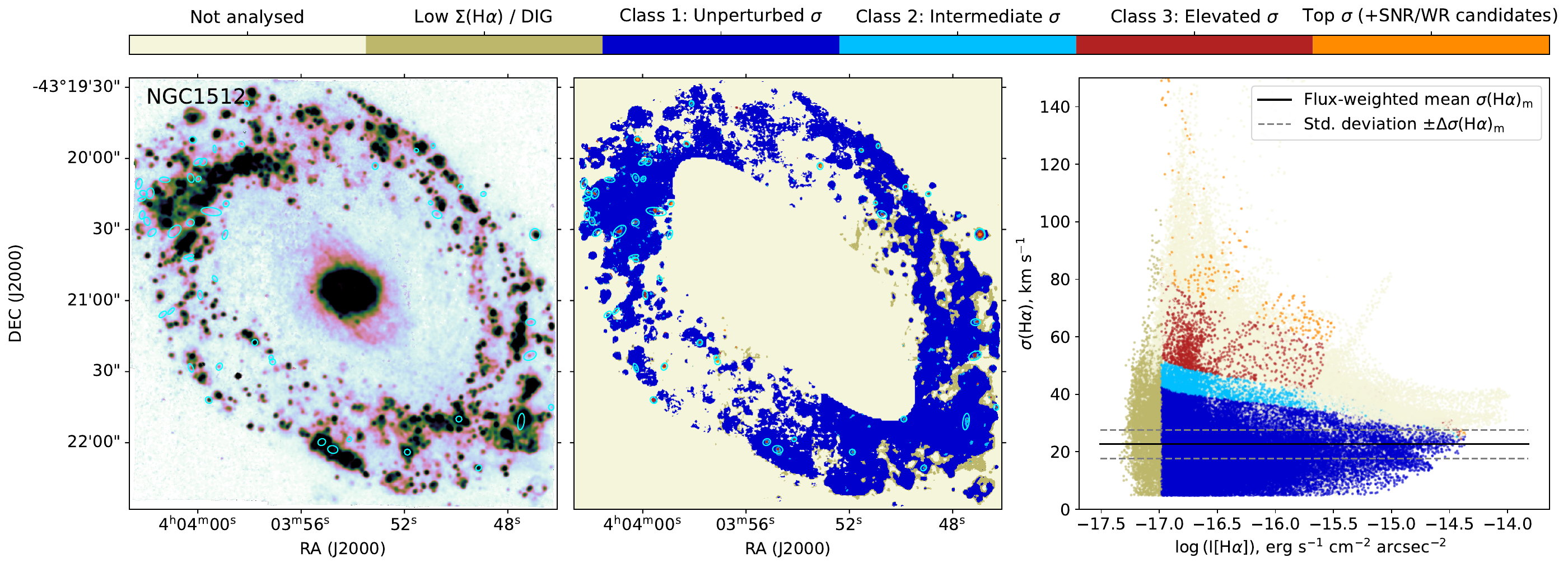}
    \includegraphics[width=0.95\linewidth]{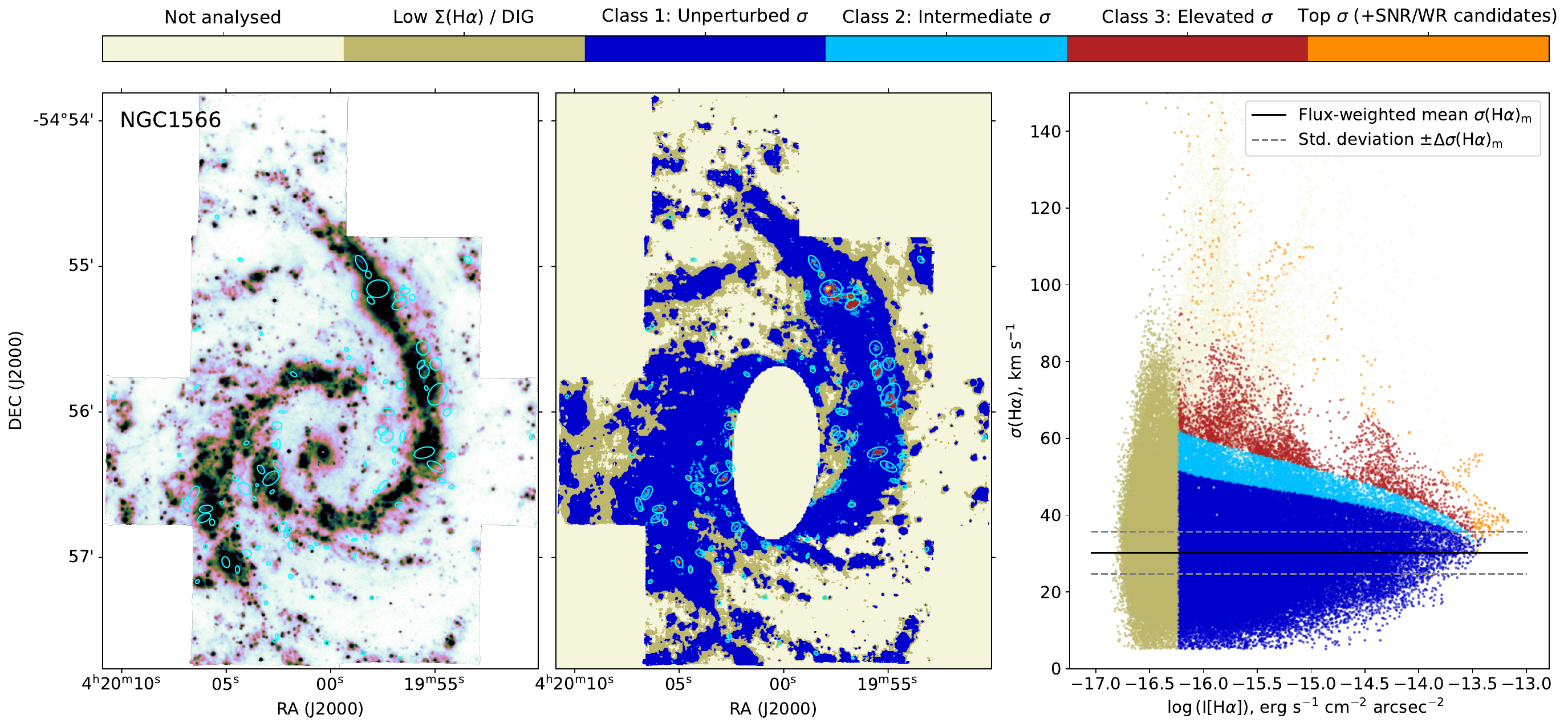}
    \includegraphics[width=0.95\linewidth]{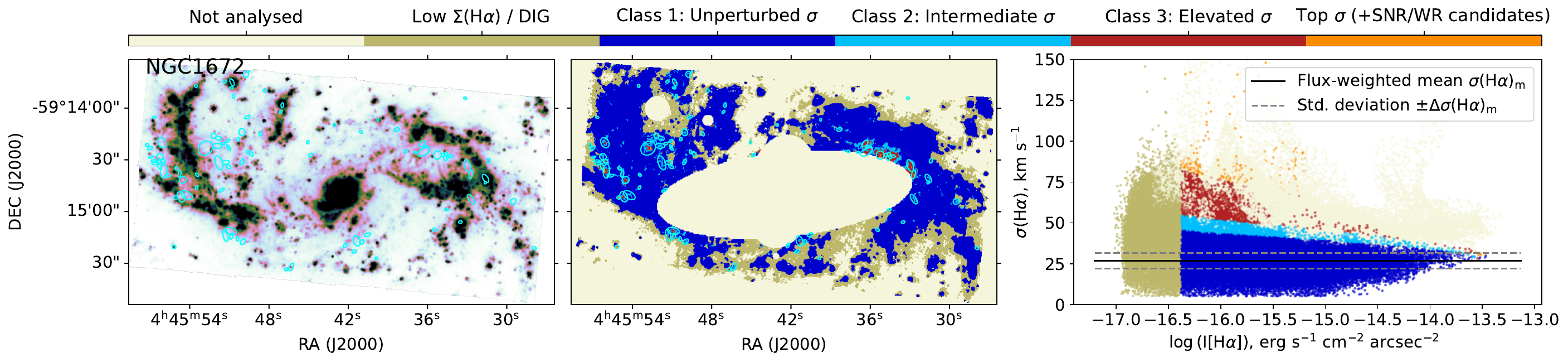}
    \caption{Continue.}
\end{figure*}

\setcounter{figure}{0}
\begin{figure*}
    \centering
    \includegraphics[width=0.95\linewidth]{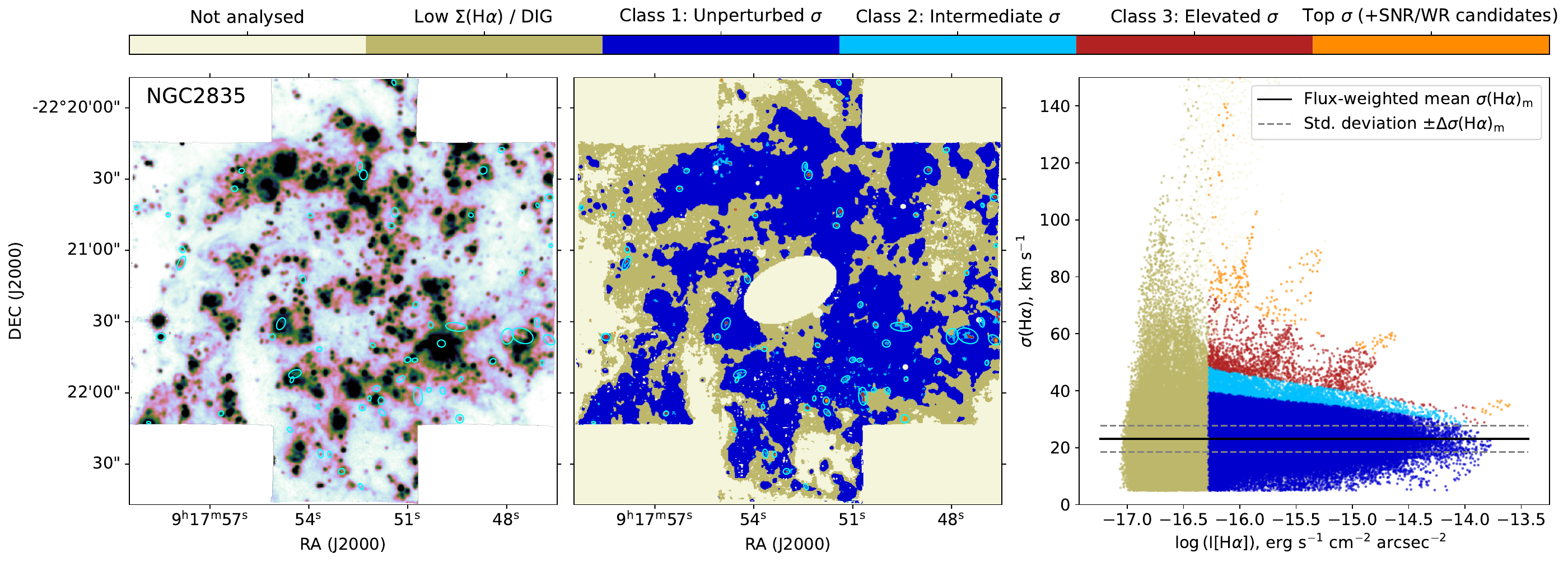}
    \includegraphics[width=0.95\linewidth]{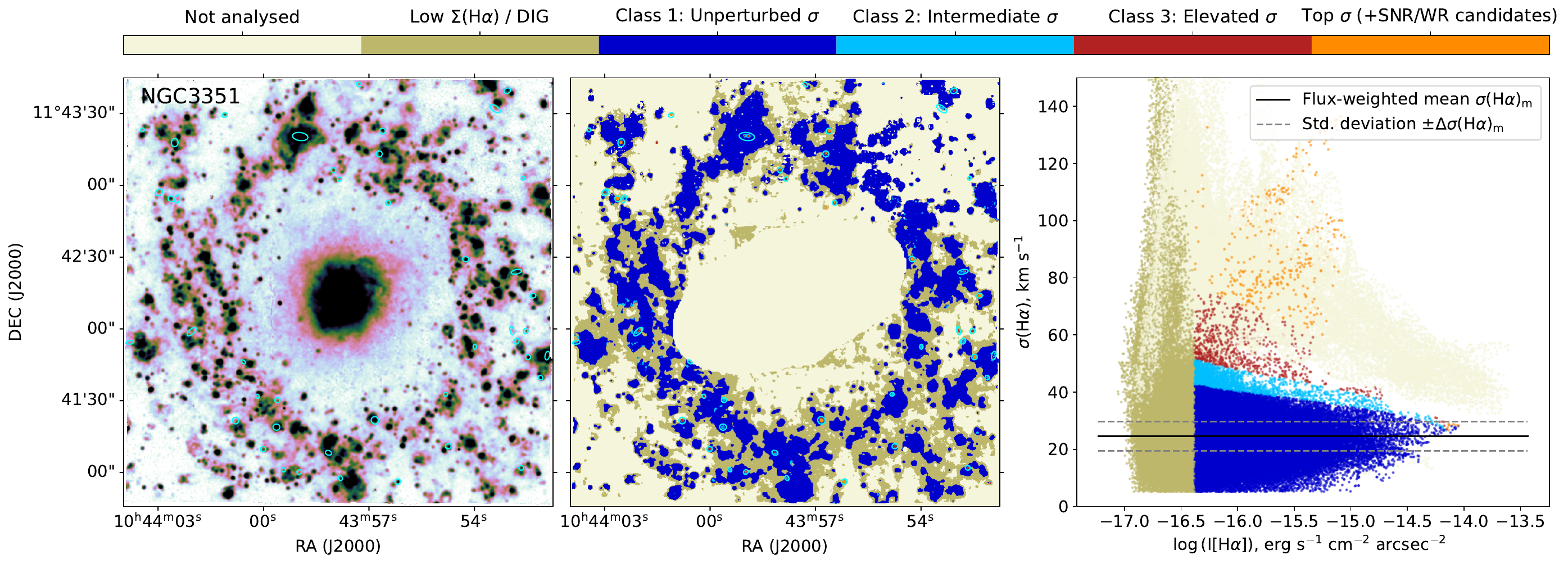}
    \includegraphics[width=0.95\linewidth]{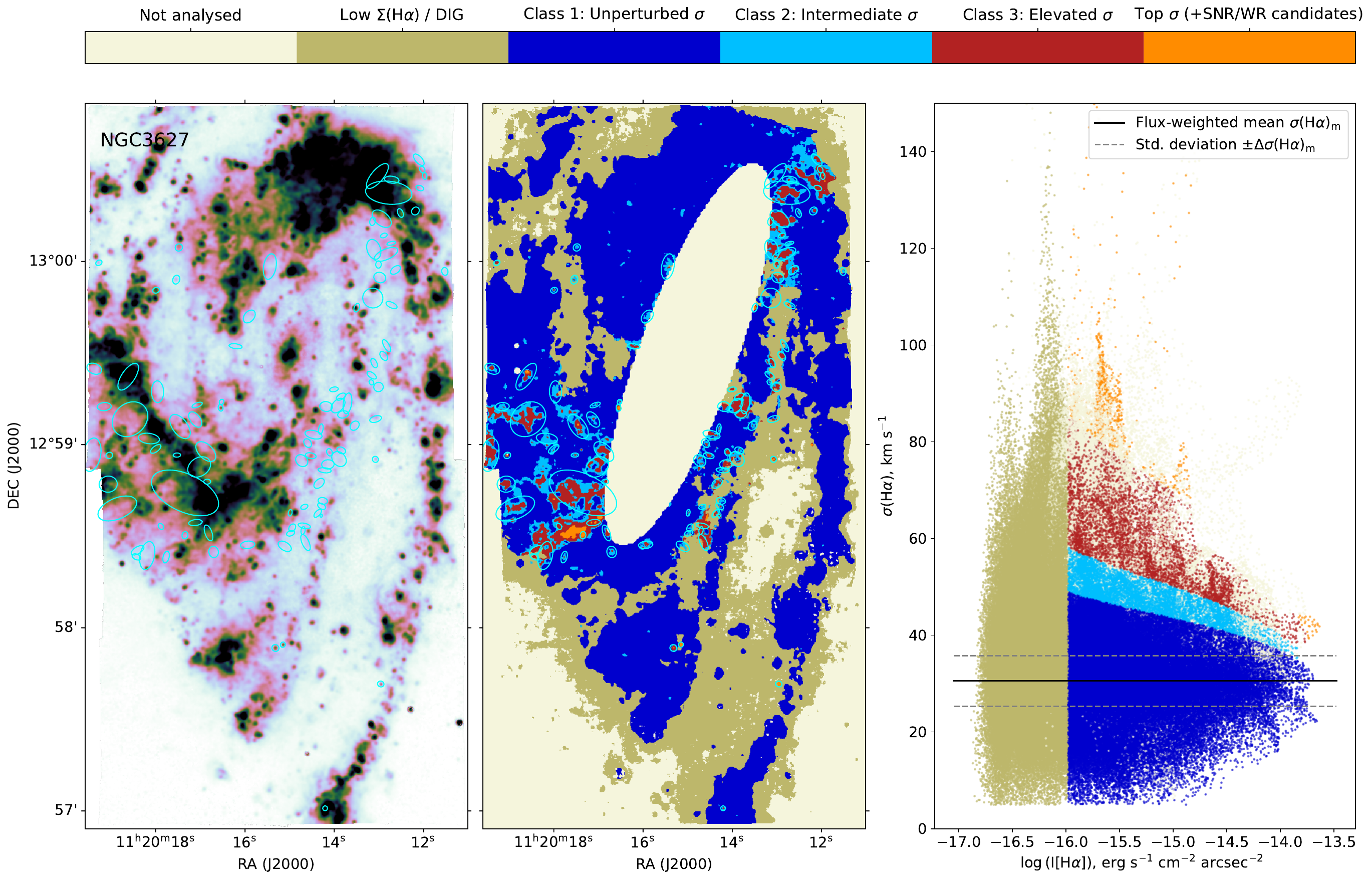}
    
    \caption{Continue.}
\end{figure*}

\setcounter{figure}{0}
\begin{figure*}
    \centering
    \includegraphics[width=0.95\linewidth]{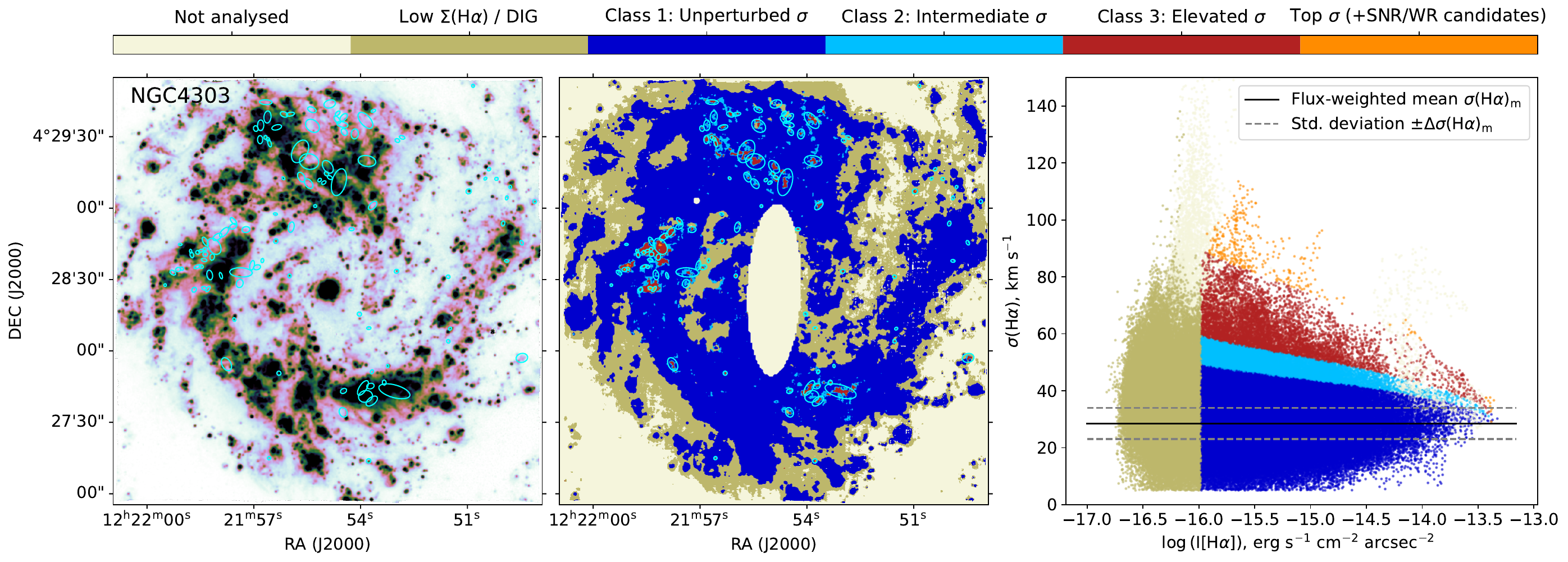}
    \includegraphics[width=0.95\linewidth]{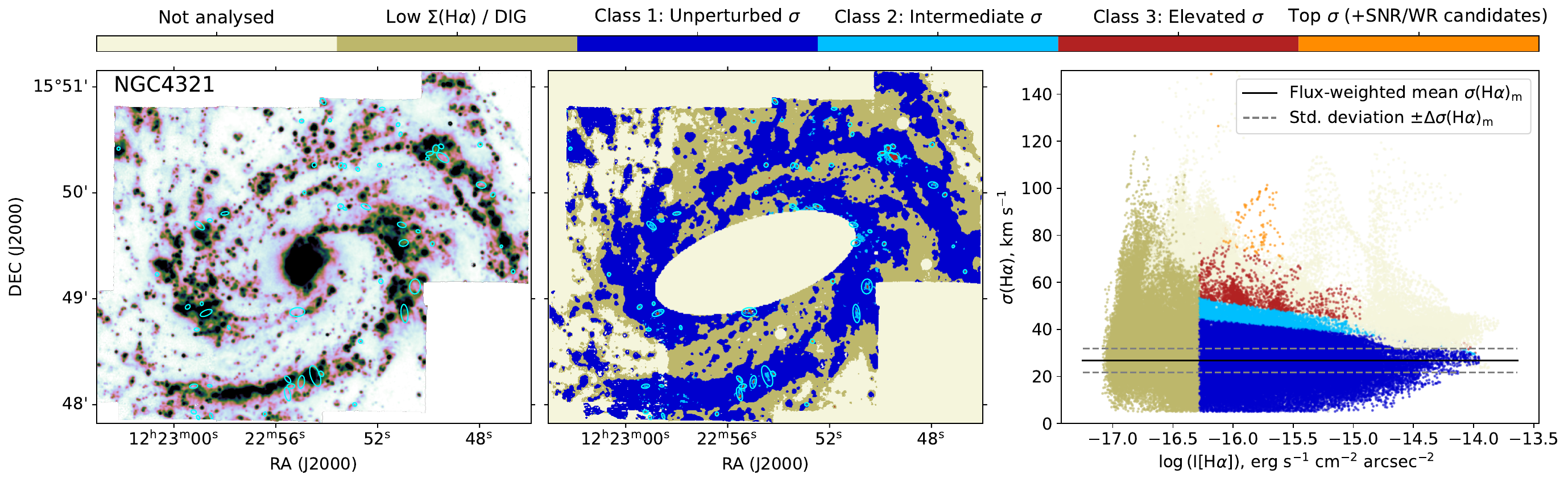}

    \includegraphics[width=\linewidth]{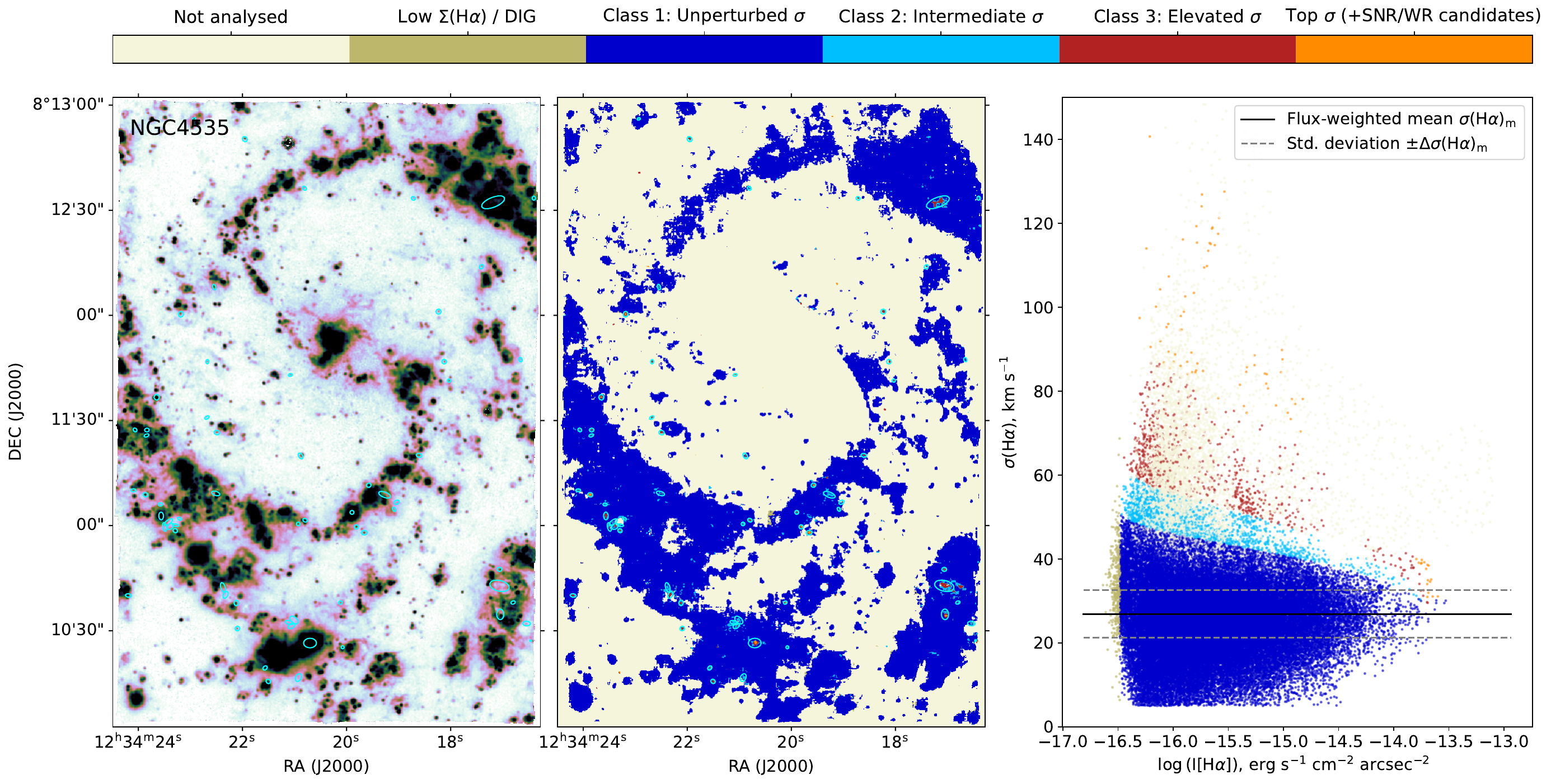}
    \caption{Continue.}
\end{figure*}

\setcounter{figure}{0}
\begin{figure*}
    \centering
    
    \includegraphics[width=0.95\linewidth]{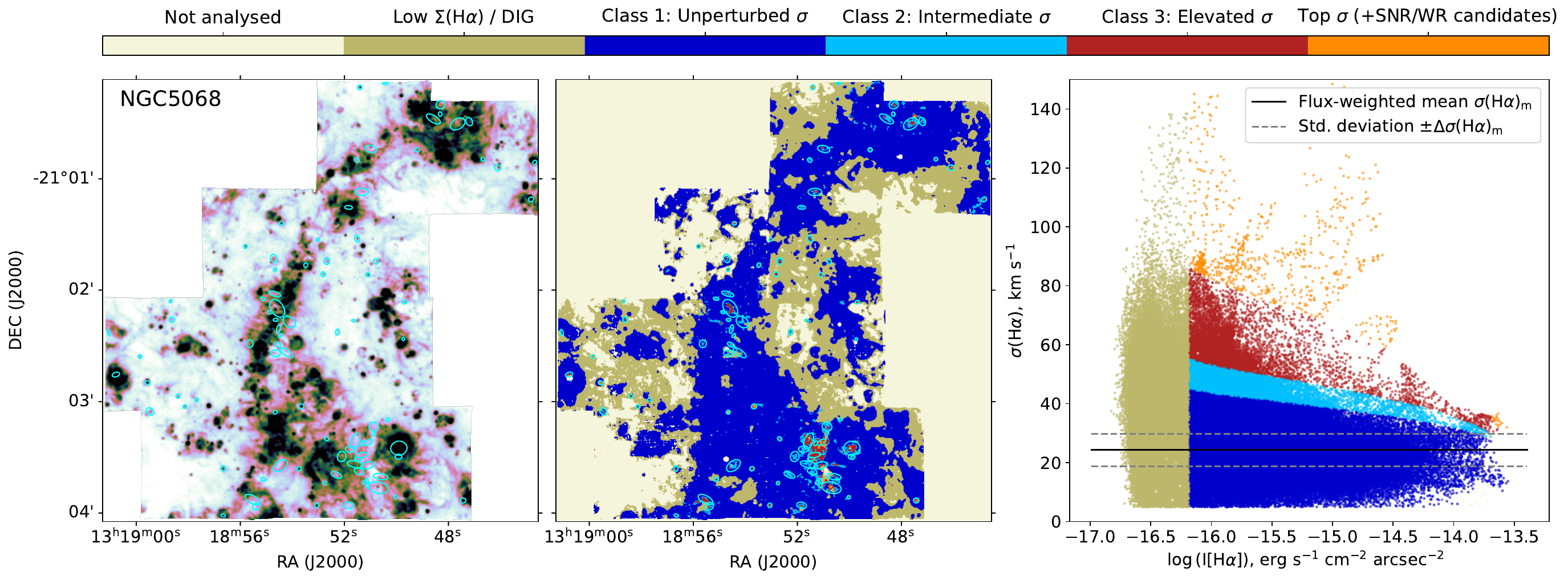}
    \includegraphics[width=\linewidth]{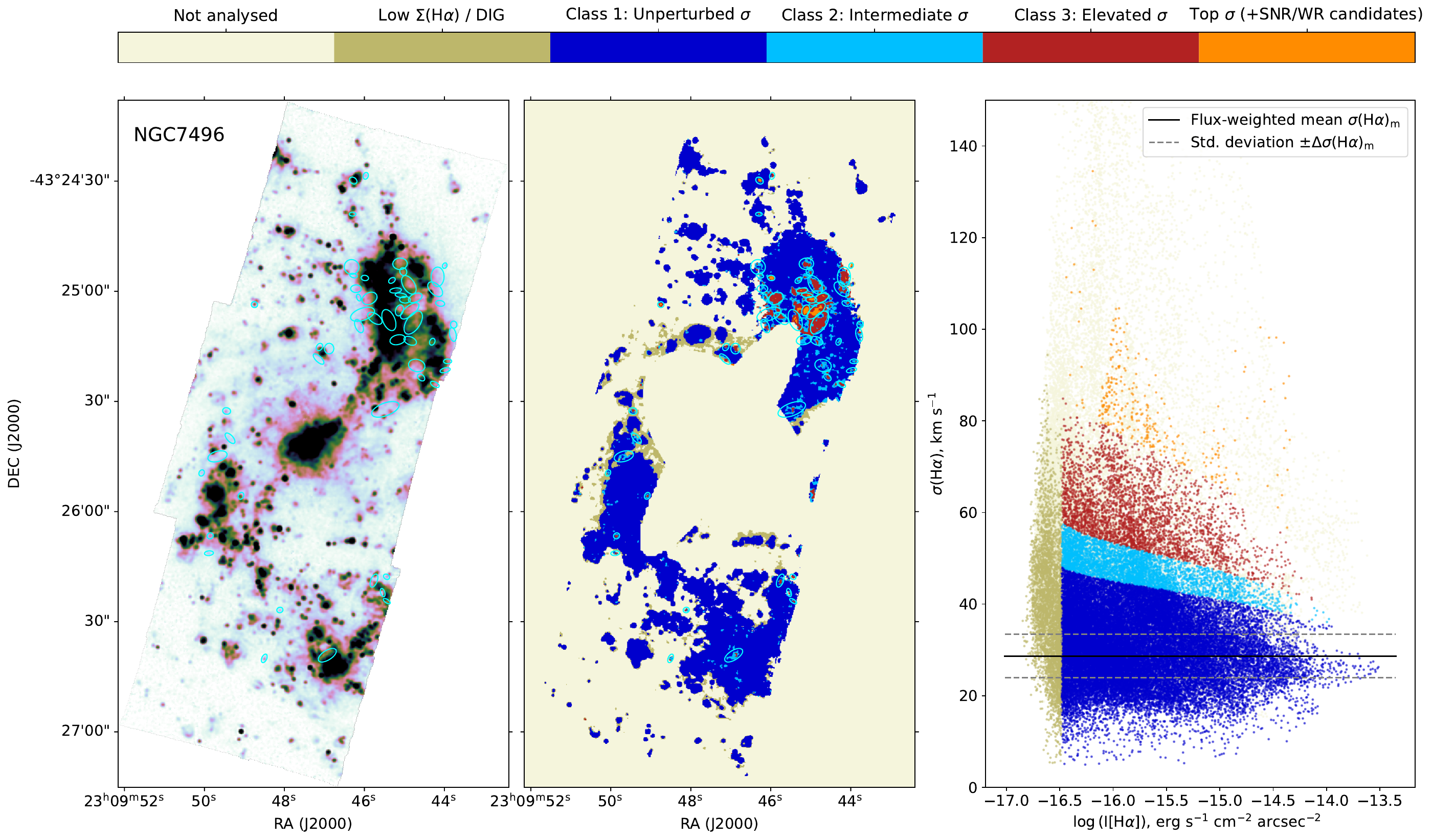}
    \caption{Continue.}
\end{figure*}
\end{appendix}
\end{document}